\documentclass[fleqn,usenatbib]{mnras}

\usepackage{mathptmx}

\usepackage[T1]{fontenc}

\newcommand\Tstrut{\rule{0pt}{2.6ex}}         
\newcommand\Bstrut{\rule[-0.9ex]{0pt}{0pt}}   

\usepackage{amssymb}	
\usepackage{graphicx}	
\usepackage{amsmath}	

\usepackage{hyperref}
\hypersetup{colorlinks=true,linkcolor=blue,citecolor=blue,filecolor=blue,urlcolor=blue}

\usepackage{siunitx}    

\usepackage[usenames, dvipsnames]{color}
\usepackage{soul}

\usepackage{subcaption}

\title[LERG LF evolution]{Cosmic evolution of low-excitation radio galaxies in the LOFAR Two-meter Sky Survey Deep Fields}

\author[R. Kondapally et al.]{Rohit~Kondapally,$^{1}$\thanks{E-mail: rohitk@roe.ac.uk}
Philip~N.~Best,$^{1}$
Rachel~K.~Cochrane,$^{2,3}$
Jos\'e~Sabater,$^{1}$
Kenneth~J.~Duncan,$^{1}$
\newauthor{Martin~J.~Hardcastle,$^{4}$
Paul~Haskell,$^{4}$
Beatriz~Mingo,$^{5}$
Huub~J.~A.~R\"{o}ttgering,$^{6}$
Daniel~J.~B.~Smith,$^{4}$}
\newauthor{Wendy~L.~Williams,$^{6}$
Matteo~Bonato,$^{7,8,9}$
Gabriela~Calistro~Rivera,$^{10}$
Fangyou~Gao,$^{11,12}$
Catherine~L.~Hale,$^{1}$}
\newauthor{Katarzyna~Ma{\l}ek,$^{13,14}$
George~K.~Miley,$^{6}$
Isabella~Prandoni,$^{7}$
and Lingyu~Wang$^{11,12}$
}
\\
$^{1}$SUPA, Institute for Astronomy, Royal Observatory, Blackford Hill, Edinburgh, EH9 3HJ, UK\\
$^{2}$Center for Computational Astrophysics, Flatiron Institute, 162 Fifth Avenue, New York, NY 10010, USA\\
$^{3}$Harvard-Smithsonian Center for Astrophysics, 60 Garden St, Cambridge, MA 02138, USA\\
$^{4}$Centre for Astrophysics Research, University of Hertfordshire, College Lane, Hatfield AL10 9AB, UK\\
$^{5}$School of Physical Sciences, The Open University, Walton Hall, Milton Keynes, MK7 6AA, UK\\
$^{6}$Leiden Observatory, Leiden University, PO Box 9513, NL-2300 RA Leiden, the Netherlands\\
$^{7}$INAF-Istituto di Radioastronomia, Via Gobetti 101, I-40129, Bologna, Italy\\
$^{8}$Italian ALMA Regional Centre, Via Gobetti 101, I-40129, Bologna, Italy\\
$^{9}$INAF-Osservatorio Astronomico di Padova, Vicolo dell'Osservatorio 5, I-35122, Padova, Italy\\
$^{10}$European Southern Observatory, Karl–Schwarzschild–Stra{\ss}e 2, D-85748 Garching bei M\"{u}nchen, Germany\\
$^{11}$SRON Netherlands Institute for Space Research, Landleven 12, 9747 AD, Groningen, The Netherlands\\
$^{12}$Kapteyn Astronomical Institute, University of Groningen, Postbus 800, 9700 AV Groningen, the Netherlands\\
$^{13}$National Centre for Nuclear Research, ul. Pasteura 7, 02-093 Warszawa, Poland\\
$^{14}$Aix Marseille Univ. CNRS, CNES, LAM, Marseille, France\\
}

\date{Accepted XXX. Received YYY; in original form ZZZ}

\pubyear{2022}

\begin{document}
\label{firstpage}
\pagerange{\pageref{firstpage}--\pageref{lastpage}}
\maketitle

\begin{abstract}
Feedback from low-excitation radio galaxies (LERGs) plays a key role in the lifecycle of massive galaxies in the local Universe; their evolution, and the impact of these active galactic nuclei on early galaxy evolution, however, remain poorly understood. We use a sample of 10\,481 LERGs from the first data release of the LOFAR Two-meter Sky Survey Deep Fields, covering $\sim$\,25\,deg$^2$, to present the first measurement of the evolution of the radio luminosity function (LF) of LERGs out to $z\sim2.5$; this shows relatively mild evolution. We split the LERGs into those hosted by quiescent and star-forming galaxies, finding a new dominant population of LERGs hosted by star-forming galaxies at high redshifts. The incidence of LERGs in quiescent galaxies shows a steep dependence on stellar-mass out to $z \sim1.5$, consistent with local Universe measurements of accretion occurring from cooling of hot gas haloes. The quiescent-LERGs dominate the LFs at $z<1$, showing a strong decline in space density with redshift, tracing that of the available host galaxies, while there is an increase in the characteristic luminosity. The star-forming LERG LF increases with redshift, such that this population dominates the space densities at most radio-luminosities by $z \sim 1$. The incidence of LERGs in star-forming galaxies shows a much weaker stellar-mass dependence, and increases with redshift, suggesting a different fuelling mechanism compared to their quiescent counterparts, potentially associated with the cold gas supply present in the star-forming galaxies.
\end{abstract}




\defcitealias{Kondapally2021}{K20}
\defcitealias{Williams2018}{W18}

\section{Introduction}\label{sec:intro}
It has been over two decades since the discovery of the tight correlation between the mass of the central supermassive black hole (SMBH) $M_{\bullet}$, and the velocity dispersion ($\sigma$), of the bulge component of a galaxy (the $M_{\bullet}-\sigma$ relation; \citealt{2000ApJ...539L...9F,2001MNRAS.320L..30M}), suggesting a co-evolution between the SMBH and its host galaxy. Accretion of matter onto a SMBH can power an active galactic nucleus (AGN), producing vast amounts of energy, which, when efficiently coupled to the surrounding gas, can disrupt star formation in the host galaxy or offset cluster cooling; this process is known as AGN feedback. It is widely accepted that almost all massive galaxies harbour a SMBH at their centres \citep[e.g.][]{2013ARA&A..51..511K}, with observations showing that feedback from AGN plays a key role in shaping the observed galaxy populations in the local Universe (see reviews by \citealt{2007ARA&A..45..117M,cattaneo2009_feedback_review,2012ARA&A..50..455F,2014ARA&A..52..589H}). This wealth of observational work is complemented by galaxy formation models \citep[see review by][]{Somerville2015} and cosmological simulations \citep[e.g.][]{2006MNRAS.370..645B,2006MNRAS.365...11C,Kaviraj2017horizon_agn,dave2019simba}, which often require feedback from AGN to successfully reproduce the observed local galaxy properties, such as the high-mass end of the galaxy luminosity function \citep[e.g.][]{2006MNRAS.370..645B} and the observed bi-modality in galaxy colours \citep[e.g.][]{Cattaneo2006}, and to prevent runaway cooling flows in galaxy clusters \citep[e.g.][]{Peterson2006,cattaneo2009_feedback_review}.

Detailed characterisation of the AGN in the local Universe has revealed that AGN can be split into two classes, which show differences in their host galaxy properties: `radiative-mode' and `jet-mode' AGN \citep[e.g.][]{Allen2006,Hardcastle2007,Janssen2012,2012MNRAS.421.1569B,2014ARA&A..52..589H,Mingo2014,Tadhunter2016,Ching2017b,Hardcastle2020}. Radiative-mode AGN accrete matter onto the black hole in a radiatively efficient manner; such an accretion flow results in the formation of an optically thick, geometrically thin, accretion disc and a dusty obscuring structure \citep[e.g.][]{1973A&A....24..337S}; these AGN can sometimes launch jets. The photoionisation of gas by photons from the accretion disc results in high-excitation emission lines in their optical spectra \citep[e.g.][]{2012MNRAS.421.1569B}. In contrast, jet-mode AGN are fuelled by radiatively inefficient accretion on to the black hole \citep[][]{1994ApJ...428L..13N,1995ApJ...452..710N} and do not display signs of an optically thick accretion disc or torus structure or AGN activity at other wavelengths but are efficient at producing bi-polar relativistic jets. 

Radio observations provide a powerful technique for identifying AGN based on the synchrotron emission from the jets that result from the acceleration of relativistic charged particles by the central engine; these sources are appropriately called `radio-loud' AGN. The radio regime offers the only way to identify and study the jet-mode AGN class, and also allows the selection of radiative-mode AGN in a manner that is unaffected by dust extinction. Due to the nature of the emission line properties of these sources, historically, the radio-detected radiative-mode AGN are known as high-excitation radio galaxies (HERGs), while the jet-mode AGN are referred to as low-excitation radio galaxies (LERGs).

Detailed characterisation of the host galaxy properties of these two AGN classes has revealed that LERGs are typically hosted in massive red galaxies with old stellar populations and very massive black holes \citep[e.g.][]{Tasse2008,Smolcic2009b,2012MNRAS.421.1569B,Janssen2012,Mingo2014,Ching2017,Whittam2018,Williams2018}, whereas the HERGs tend to be hosted in less massive, bluer galaxies with recent star-formation activity \citep{2014ARA&A..52..589H}. Compared to HERGs, LERGs are also found to have typically lower radio luminosities \citep[e.g.][]{2012MNRAS.421.1569B,2014MNRAS.445..955B,Pracy2016}, and are more often found in rich group or cluster environments \citep[e.g.][]{Best2004,Tasse2008,Gendre2013,Sabater2013,Croston2019}. These results are consistent with the idea that the radio-loud AGN activity in these two classes of AGN is triggered by different mechanisms related to the available fuel source; in this picture, LERGs are thought to be fuelled by the cooling hot gas from haloes present in their massive host galaxies, whereas the HERGs tend to accrete efficiently from cold gas, from either external (e.g. mergers or interactions) or internal (e.g. bars or instabilities) processes, with the more plentiful gas supply leading to the formation of an accretion disc \citep[e.g.][]{Allen2006,Hardcastle2007,2014ARA&A..52..589H}.

However, in reality, the picture is not quite so simple; in recent years, improved models for the nature of the accretion flow from the hot phase suggest that under certain conditions, thermal instabilities in the hot gas medium can result in the formation of cold filaments of gas that then `rain' down on to the black hole \citep[e.g.][]{Sharma2012,Gaspari2013,Gaspari2015,Gaspari2017}. This means that accretion on to the SMBHs in LERGs may also be from gas in a `cold' state. Indeed there is now growing observational support for this scenario, with observations of a handful of LERGs being fuelled by cold gas \citep[e.g.][]{Tremblay2016,Ruffa2019b}. 

It has been suggested that the Eddington-scaled accretion rate on to the black hole, rather than the gas origin alone, may determine the formation of a radiatively efficient/inefficient AGN \citep[e.g.][]{2012MNRAS.421.1569B,2014MNRAS.445..955B,Hardcastle2018,Hardcastle2020}. An analogy of this model can be drawn to the different spectral states of X-ray black hole binaries, where a switch from the `soft' (analogous to radiative-mode AGN) to the `hard' (analogous to the jet-mode AGN) state is governed by a change in the nature of the accretion flow occurring at a few per cent of the Eddington rate \citep[see reviews by][]{Remillard2006,Yuan2014}. In such a model, fuelling by cooling of hot gas generally occurs at low accretion rates in massive galaxies which host massive black holes, leading to a LERG, whereas low-mass galaxies with abundant cold gas supply are likely to result in higher (Eddington-scaled) accretion rates, and hence in the formation of a HERG. However, due to the stochastic nature of accretion processes, accretion from cold gas can occur at low Eddington rates leading to a LERG. Moreover, \citet{Whittam2018} found a difference in the Eddington-scaled accretion rate distribution of LERGs and HERGs out to $z \sim 0.7$, but with considerable overlap in these distributions in their deeper radio data compared to that found by \citet{2012MNRAS.421.1569B} in the local Universe. Therefore, the origin of the differences in these two populations and the precise physical mechanisms that trigger the different modes of AGN in different galaxies are not well understood.

In the local Universe, the incidence of LERGs shows a strong dependence on the stellar mass ($\propto M_{\star}^{2.5}$) and black hole mass ($\propto M_{\bullet}^{1.6}$; \citealt{Best2005fagn,Janssen2012}). In these massive systems, the time-averaged energetic output from the jets, in the form of mechanical energy, is found to balance the radiative cooling losses from the hot gas \citep[e.g.][]{Best2006,2007ARA&A..45..117M,Hardcastle2019}. Therefore, feedback from LERGs plays an important role in the evolution of massive galaxies as the medium into which the jet energy is deposited also forms the eventual fuel source for the AGN, thus providing the conditions for a self-regulating AGN feedback cycle. Understanding the cosmic evolution of the LERG population and the host galaxies in which they reside is crucial in marking their role in early galaxy evolution and testing this picture established from detailed observations in the local Universe.

\citet{2014MNRAS.445..955B} were the first to study the evolution of the HERG and LERG luminosity functions, separately, out to higher redshifts ($z \sim 1$) using a combination of various radio-AGN datasets and spectroscopic information. They found that for the LERGs, the space density decreases mildly with redshift at low-luminosities ($L_{\rm{1.4\,GHz}} \lesssim 10^{24}\,\rm{W\,Hz^{-1}}$) but increases with redshift at higher radio luminosities. They developed various models to explain the observed evolution but concluded that deeper, higher redshift data were required to distinguish between the models. Similarly, \citet{Pracy2016} constructed the evolving HERG and LERG luminosity functions (at $z < 0.75$) using a sample of 5000 radio-AGN derived from cross-matching the Faint Images of the Radio Sky at Twenty-cm \citep[FIRST;][]{becker1995first} survey at 1.4\,GHz and the Sloan Digital Sky Survey \citep[SDSS;][]{york2000sdss}. They also found that the LERG population shows little-to-no evolution over this redshift. More recently, \citet{Butler2019} investigated the evolution of a sample of 1729 LERGs using 2.1\,GHz observations of the XMM extragalactic survey field (XXL-S) using the Australia Telescope Compact Array (ATCA). They found that the space densities of the LERGs showed weak evolution out to $z \sim 1.3$. At low radio frequencies, \citet{Williams2018} studied the evolution of HERGs and LERGs, classified using photometry alone, between $0.5 < z \leq 2$ using 150\,MHz LOw Frequency ARray \citep[LOFAR;][]{2013A&A...556A...2V} observations of the Bo\"{o}tes field. Over this redshift range, they found that the LERG population shows a strong decline in space densities that is consistent with the decline in the space densities of the expected hosts, i.e. the massive quiescent galaxies.

In this paper, we present the first robust measurement of the LERG luminosity functions out to $z \sim 2.5$ and study the cosmic evolution of their host galaxy properties using a sample of over 10\,000 LERGs constructed by combining deep radio observations from LOFAR with deep, wide-area multi-wavelength photometry. The radio observations come from the first data release of the LOFAR Two-meter Sky Survey Deep Fields \citep[LoTSS-Deep;][]{Tasse2021,Sabater2021,Kondapally2021,Duncan2021}, covering three extragalactic fields and forming the deepest radio continuum survey to date at low frequencies. Our combination of deep and wide radio and multi-wavelength datasets, covering $\sim$ 25\,deg$^{2}$, is ideal for probing much fainter luminosities out to higher redshifts than previous studies \citep[e.g.][]{2014MNRAS.445..955B,Pracy2016,Williams2018,Butler2019} and also allows a better sampling of the bright end of the luminosity function compared to previous deep observations over small areas \citep[e.g.][]{smolcic2017_vla_cosmos_3G,smolcic2017agn_evol_vla}, while limiting the effects of cosmic variance. We focus on the evolution of the LERG population as our deep radio dataset is particularly well-suited to sample the low-luminosity AGN population, which is dominated by LERGs; this allows us to characterise the evolution of this population in unprecedented detail.

The paper is structured as follows. In Sect.~\ref{sec:data} we describe the radio and multi-wavelength datasets used, and the selection of the sample of LERGs. Sect.~\ref{sec:general_lfs} describes the construction of the luminosity functions, completeness simulations, and the evolution of the radio-AGN luminosity functions. Sect.~\ref{sec:cosevol_lerg_lf} presents the evolution of the LERG luminosity functions and comparisons with literature. Sect.~\ref{sec:agn_frac} investigates the incidence of LERGs as a function of stellar mass, split into those hosted by quiescent galaxies and star-forming galaxies. Sect.~\ref{sec:lerg_lfs_Q_SF} presents the LERG luminosity functions split into those hosted by star-forming and quiescent galaxies and models the evolution of the quiescent LERG population. We draw our conclusions in Sect.~\ref{sec:conclusions}.

Throughout this work, we use a flat $\Lambda$CDM cosmology with $\Omega_{\rm{M}}=0.3,~\Omega_{\rm{\Lambda}}=0.7$ and $H_{0} = 70~\mathrm{km~s^{-1}~Mpc^{-1}}$, and a radio spectral index $\alpha = -0.7$ (where $S_{\nu} \propto \nu^{\alpha}$). Where quoted, magnitudes are in the AB system \citep{1983ApJ...266..713O}, unless otherwise stated.

\section{Data}\label{sec:data}
In this paper, we utilise a large sample of radio-detected sources combined with other multi-wavelength datasets from LoTSS-Deep Data Release 1 in the European Large-Area ISO Northern-1 (ELAIS-N1), Lockman Hole, and Bo\"{o}tes fields.

\subsection{Radio data}\label{sec:radio_data}
The radio data in the three fields were obtained from observations taken with the LOFAR High Band Antenna (HBA) with high spatial resolution (6 arcsec) and a central observing frequency of 146~MHz for the ELAIS-N1 field and 144~MHz for the Lockman Hole and Bo\"{o}tes fields\footnote{Hereafter, we refer to the central frequencies for all fields as 150\,MHz, for simplicity.}. Multi-epoch observations were used to build deep radio images, which cover 68\,deg$^{2}$ in each field down to the 30 per cent power point of the primary beam. The radio images, with total integration times of 168, 112, and 80 hours, reach RMS noise levels of 20, 22, and 32$~\mu$Jy\,beam$^{-1}$ in the central regions of the ELAIS-N1, Lockman Hole, and Bo\"{o}tes fields, respectively. The full details of the radio calibration and imaging are described by \citet{Tasse2021} for Lockman Hole and Bo\"{o}tes, and by \citet{Sabater2021} for ELAIS-N1. Source extraction was performed on the Stokes I images in each field using Python Blob Detector and Source Finder (\textsc{PyBDSF}; \citealt{2015ascl.soft02007M}), with source catalogues extracted out to the 30 per cent power point of the primary beam.

\subsection{Multi-wavelength data}\label{sec:opt_ir_data}
The three LoTSS Deep Fields contain deep, wide-area panchromatic photometry; this rich multi-wavelength dataset makes these fields ideal for determining robust photometric redshifts and performing spectral energy distribution (SED) fitting.

The full details of the available multi-wavelength data, including the depth and areas covered by each survey, are given by \citet{Kondapally2021}. In summary, ultra-violet (UV) data in all fields come from the Galaxy and Evolution Explorer (\textit{GALEX}) space telescope \citep{2007ApJS..173..682M}. The optical data in the three fields comes from a combination of the Panoramic Survey Telescope and Rapid Response System (Pan-STARRS-1; \citealt{2010SPIE.7733E..0EK}) Medium Deep Survey, the Hyper-Suprime-Cam Subaru Strategic Program (HSC-SSP) survey data release 1 \citep{2018PASJ...70S...8A}, the \textit{Spitzer} Adaptation of the Red-sequence Cluster Survey (SpARCS; \citealt{2009ApJ...698.1943W,2009ApJ...698.1934M}), Red Cluster Sequence Lensing Survey \citep[RCSLenS][]{2016MNRAS.463..635H}, and the NOAO Deep Wide Field Survey (NDWFS; \citealt{1999ASPC..191..111J}). In both ELAIS-N1 and Lockman Hole, near-infrared (NIR) data comes from the UK Infrared Deep Sky Survey (UKIDSS) Deep Extragalactic Survey (DXS; \citealt{2007MNRAS.379.1599L}). The mid-infrared (MIR) data in all fields comes from the \textit{Spitzer} space telescope \citep{werner2004spitzer}. In ELAIS-N1 and Lockman Hole this consists of observations from the \textit{Spitzer} Wide-area Infra-Red Extragalactic (SWIRE; \citealt{2003PASP..115..897L}) survey and the \textit{Spitzer} Extragalactic Representative Volume Survey (SERVS; \citealt{2012PASP..124..714M}). In Bo\"{o}tes, the mid-infrared data come from \textit{Spitzer} observations of the NDWFS field (\citealt{eisenhardt2004iracshallow,2009ApJ...701..428A}; Decadal IRAC Bo\"{o}tes Survey).

In ELAIS-N1 and Lockman Hole, we make use of new forced, matched-aperture multi-wavelength catalogues with UV to mid-infrared photometry generated by \citet{Kondapally2021}, providing more robust and, at some wavelengths, deeper catalogues than those previously available in the literature. The full details of the catalogue creation process are provided by \citet{Kondapally2021}; in summary, sources were detected using deep $\chi^{2}$ images that incorporated information across optical-MIR wavelengths, with fluxes extracted in circular apertures and corrected to total fluxes by performing aperture corrections based on sources typical of distant galaxies. In the Bo\"{o}tes field, we made use of existing point spread function (PSF) matched I-band and 4.5\,$\mu$m band selected catalogues \citep{2007ApJ...654..858B,2008ApJ...682..937B} to generate a combined multi-wavelength catalogue, as described by \citet{Kondapally2021}.

Additional far-infrared (FIR) data come from the FIR-deblended catalogues generated by the \textit{Herschel} Extragalactic Legacy Project (HELP; \citealt{Shirley2021}) and McCheyne et al. (2021, in press). This dataset compiles observations from \textit{Spitzer}-MIPS \citep{rieke2004mips} at 24\,$\mu$m, and imaging from the Photodetector Array Camera and Spectrometer (PACS; \citealt{poglitsch2010pacs}; with photometry at 100\,$\mu$m and 160\,$\mu$m), and the Spectral and Photometric Imaging Receiver (SPIRE; \citealt{griffin2010spire}; with bands at 250\,$\mu$m, 360\,$\mu$m and 520\,$\mu$m) instruments on-board the \textit{Herschel} Space Observatory \citep{pilbratt2010herschel}. The observations from \textit{Herschel} come from the \textit{Herschel} Multi-tiered Extragalactic Survey (HerMES; \citealt{oliver2012hermes}).

\subsection{Multi-wavelength counterparts and value-added catalogues}
The new multi-wavelength catalogues generated in these fields were used to identify the counterparts of the LOFAR-detected sources in the LoTSS Deep Fields. The host-galaxy identification process was limited to the regions of each field with the best available multi-wavelength coverage\footnote{See \textsc{flag\_overlap} criteria described in \citet[][their table~5]{Kondapally2021}}, totalling $\sim$26\,deg$^{2}$.

The host-galaxy identification method is described in detail by \citet{Kondapally2021}. In summary, counterparts were identified using a combination of the colour-based statistical likelihood ratio (LR) method \citep{1977A&AS...28..211D,1992MNRAS.259..413S,2019A&A...622A...2W} and a visual classification scheme. The LR method is suitable for radio sources with a secure radio position; a decision tree was developed to select such sources. The LR method cross-matching achieved a reliability and completeness of $> 97.5$ per cent across the three fields. Sources not selected for LR cross-matching were identified using two main visual inspection workflows with any issues with source associations, for example, blending of distinct physical sources or association of core and lobe components of an AGN, also being corrected at this step. The first visual workflow used was LOFAR Galaxy Zoo (LGZ; \citealt{2019A&A...622A...2W}), a Zooniverse-based framework, where the consensus decision from five volunteers is used to form the source associations and identifications. Sources that required additional inspection after LGZ or were otherwise more complex, for example, radio blends, were finally classified by a single expert in a workflow designed with more functionality than LGZ. The cross-matching process resulted in the identification of the host-galaxies for $> 97$ per cent of the radio-detected sources across the three fields. Full details of the statistics and magnitude distributions of the counterparts are given in \citet{Kondapally2021}.

Building on the robust multi-wavelength photometric catalogues in each field, photometric redshifts for the full multi-wavelength catalogues were presented by \citet{Duncan2021}. These were generated using a hybrid approach, combining template fitting and machine learning methods, developed for the next generation of radio surveys \citep[see][]{duncan2018photz_templates,duncan2018photz_ml}. In this paper, we use spectroscopic redshifts where available and reliable, which is the case for a small fraction of the sources ($\sim$ 5, 5, and 22 per cent of all radio-sources in ELAIS-N1, Lockman Hole, and Bo\"{o}tes, respectively); otherwise the photometric redshifts are used.

\subsection{Spectral energy distribution fitting}\label{sec:sed_fits}
Spectral energy distribution (SED) fitting was performed using four different SED fitting codes for all LoTSS Deep Fields sources with counterparts that satisfy the quality cuts defined by \citet{Kondapally2021} and \citet{Duncan2021}; this process is described in detail by Best et al. (in prep.).

The SED fitting was performed using this same input catalogue using each of \textsc{AGNFitter} \citep{CalistroRivera2016}, Bayesian Analysis of Galaxies for Physical Inference and Parameter Estimation \citep[\textsc{bagpipes};][]{Carnall2018}, Code Investigating Galaxy Evolution \citep[\textsc{cigale};][]{Burgarella2005,Noll2009,Boquien2019}, and Multi-wavelength Analysis of Galaxy Physical Properties \citep[\textsc{magphys};][]{daCunha2008}. One of the key differences between the SED fitting codes employed by Best et al (in prep.) is that \textsc{AGNFitter} and \textsc{cigale}, unlike the other two routines, are able to model emission from AGN which imprint features in the mid-infrared regime in particular. This is done by incorporating models for the accretion disc and dusty torus surrounding the AGN in both SED fitting codes; for \textsc{cigale}, this includes a run based on the \citet{Fritz2006} AGN models, and a separate run based on the \textsc{skirtor} \citep{Stalevski2012,Stalevski2016} prescription which assumes a clumpy two-medium torus model \citep[rather than a smooth torus structure; see][]{Boquien2019,Yang2020}; for \textsc{AGNFitter}, this includes two-component models for the AGN, torus models \citep{2004MNRAS.355..973S}, and an accretion disk emission model \citep{Richards2006}. The other key difference between the different codes is that \textsc{bagpipes}, \textsc{magphys}, and \textsc{cigale} all enforce an energy balance between the UV-optical emission from starlight absorbed by dust and the re-radiated emission in the infrared; this ensures that the spectral distributions of galaxies are physically consistent. The version of \textsc{AGNFitter} used by Best et al. (in prep.) does not enforce this, for example to account for cases where the UV emission is spatially offset from the dust emission \citep[e.g.][]{CalistroRivera2018,Cochrane2021}; the most recent version allows users to enforce a strict energy balance. The inclusion of energy balance in the \textsc{magphys}, \textsc{bagpipes}, and \textsc{cigale} results  has implications when comparing our results with previous \textsc{AGNFitter}-based work from \citet{Williams2018}; these are discussed in Appendix~\ref{sec:ap_wendy_lerg}.

The outputs from the four different SED fitting routines were combined to generate \textit{consensus} estimates of physical galaxy properties; the key parameters relevant for analysis in this study are the stellar masses and star-formation rates (SFRs). This process is described in detail by Best et al. (in prep.) but in summary, for sources that showed no signatures of a radiative-mode AGN, the stellar masses and SFRs were estimated by taking the average of the \textsc{magphys} and \textsc{bagpipes} results (accounting for their goodness of fits); both stellar masses and SFRs agree very well with each other, with a scatter $< 0.15\,\rm{dex}$ (see Best et al. in prep.). For sources that showed signs of a radiative-mode AGN (see below; also referred to as `SED AGN' colloquially hereafter), the stellar masses and SFRs were taken by averaging the two \textsc{cigale} runs with the \citet{Fritz2006} and \textsc{skirtor} AGN models, provided that a good fit was found; \textsc{AGNFitter} was found to result in less reliable fits and was hence excluded from this step. We use these \textit{consensus} values determined by Best et al. (in prep.) throughout this paper unless otherwise stated. 

The `SED AGN' (i.e. AGN showing emission from the accretion disc or torus), were identified using the outputs of the four SED fitting routines. Firstly, Best et al. (in prep.) defined a diagnostic based on the $f_{\rm{AGN}}$ parameter, which corresponds to the ratio of MIR luminosity that arises from AGN components compared to that from the stellar galaxy component, as fitted by both \textsc{AGNFitter} and \textsc{cigale}. In particular, Best et al. used the 16\textsuperscript{th} percentile of the $f_{\rm{AGN}}$ parameter, to avoid bias from objects with large uncertainties on $f_{\rm{AGN}}$. Secondly, Best et al. also considered the goodness of fit estimates from \textsc{bagpipes} and \textsc{magphys} (which do not include AGN components) compared to those from \textsc{cigale} and \textsc{AGNFitter}; the latter two codes should find a better goodness of fit for sources with significant AGN contribution. Based on the combination of the $f_{\rm{AGN, 16\textsuperscript{th}}}$ parameters and the relative goodness of fit values, Best et al. identified the likely `SED AGN'. Finally, a small subset of the LOFAR-detected sources have either optical spectroscopy or X-ray detections, indicating the presence of a typical AGN \citep[see][]{Duncan2021}; most of these were already identified as `SED AGN' by the method above, but any additional sources were also added to this sample.

\subsection{Selection of LERGs and HERGs}\label{sec:agn_sel}
Radio continuum surveys detect both synchrotron emission associated with jets from AGN, and also emission from star-formation activity. The faint radio source population, especially at $S_{\rm{150\,MHz}} \lesssim 1\,\rm{mJy}$, is expected to be dominated by star-forming galaxies \citep{2008MNRAS.388.1335W,Padovani2016}. Source classification of the LOFAR-detected sources was therefore performed to separate star-forming galaxies from different classes of AGN using the outputs from the SED fitting process. This process is described in detail by Best et al. (in prep.) and summarised below.

As low-frequency radio observations trace cosmic ray electrons from supernovae from massive (recently formed) stars, there is a well-known relation between the radio luminosity and SFR for star-forming galaxies \citep[e.g.][and references therein]{CalistroRivera2016,gurkan2018lofar_sfr,Smith2021}, with a correlation also seen between the far-infrared and radio luminosities (far-infrared radio correlation; FIRC). We selected radio-AGN (also known as `radio-excess AGN') as sources that show excess radio emission ( $>0.7\,\rm{dex}$, $\approx 3\sigma$) compared with that expected from star-formation processes alone using a ridgeline analysis by Best et al. (in prep.). To this sample, we added the small fraction of sources that are resolved and show extended radio emission that is associated with jets from the AGN even if they show a radio-excess $< 0.7\,\rm{dex}$. The other sources without such radio-excess are largely star-forming galaxies (SFGs) but can also include `radio-quiet AGN' which are known to broadly follow the same radio luminosity - star formation rate relation \citep[e.g.][]{gurkan2018lofar_sfr}. Both of these groups are excluded from this analysis as we are focused on the radio-loud AGN population in this study. The luminosity functions of these populations are studied by \citep{Bonato2021} and Cochrane et al. (in prep.).

As discussed in the introduction, the total radio-AGN population consists of two classes of AGN: HERGs and LERGs. HERGs display strong optical emission lines indicating the presence of an accretion disc and dusty obscuring structure; these sources are consistent with being the radio-loud subset of the radiative-mode AGN population. The LERGs show powerful radio emission from the AGN but little to no evidence of having an accretion disc or obscuring structure in the IR through to the X-ray regime, and as such are not identified as AGN at other wavelengths. Using the above definitions of the two classes of AGN, in this study, we define LERGs as sources that host a `radio-excess AGN' but not an `SED AGN' (based on the SED fitting; see Sect.~\ref{sec:sed_fits}). Likewise, HERGs are defined as sources that are classed as both `radio-excess AGN' and `SED AGN'.

\begin{figure}
    \centering
    \begin{subfigure}{\columnwidth}
    \includegraphics[width=\columnwidth]{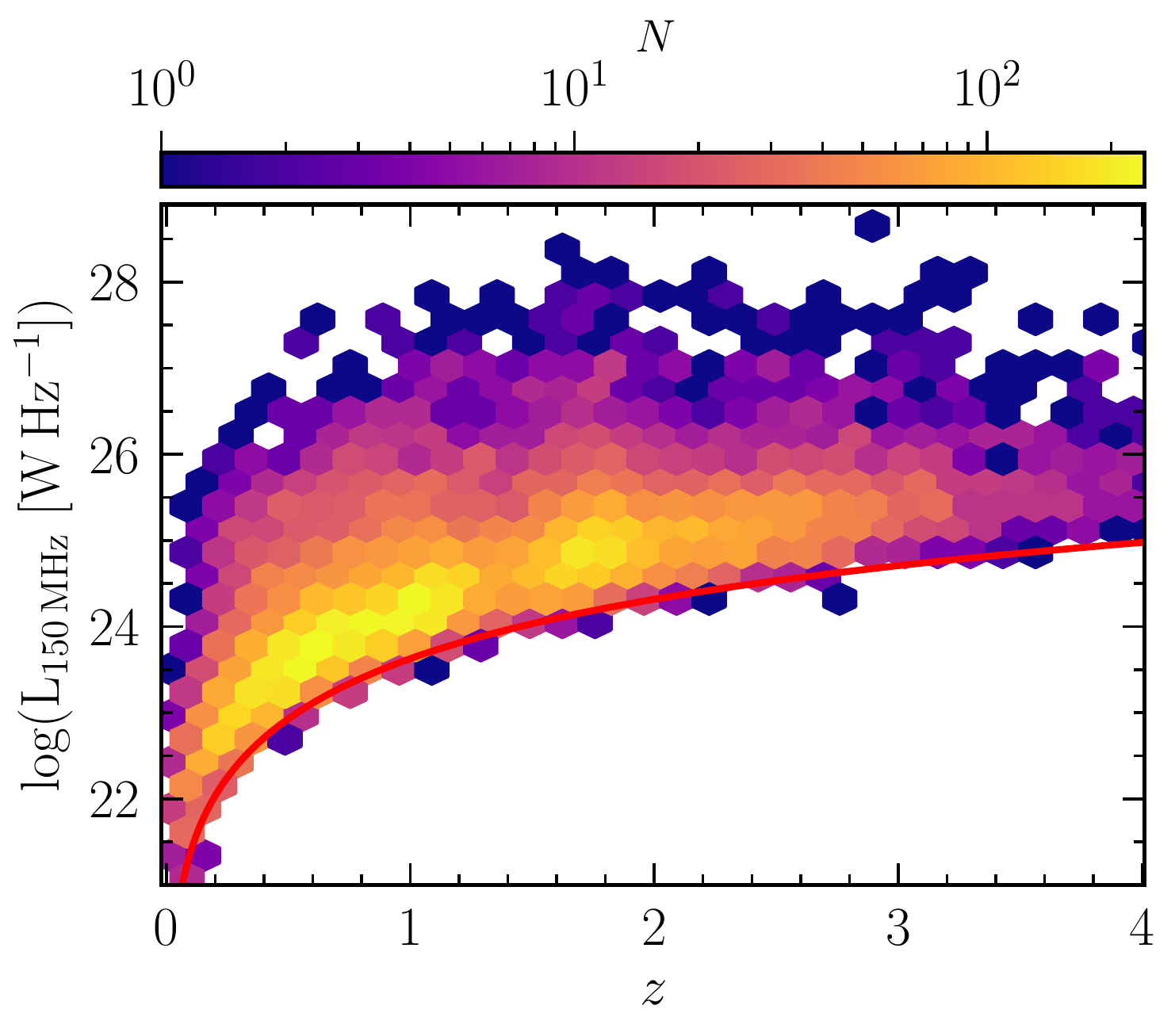}
    \end{subfigure}
    \begin{subfigure}{\columnwidth}
    \includegraphics[width=\columnwidth]{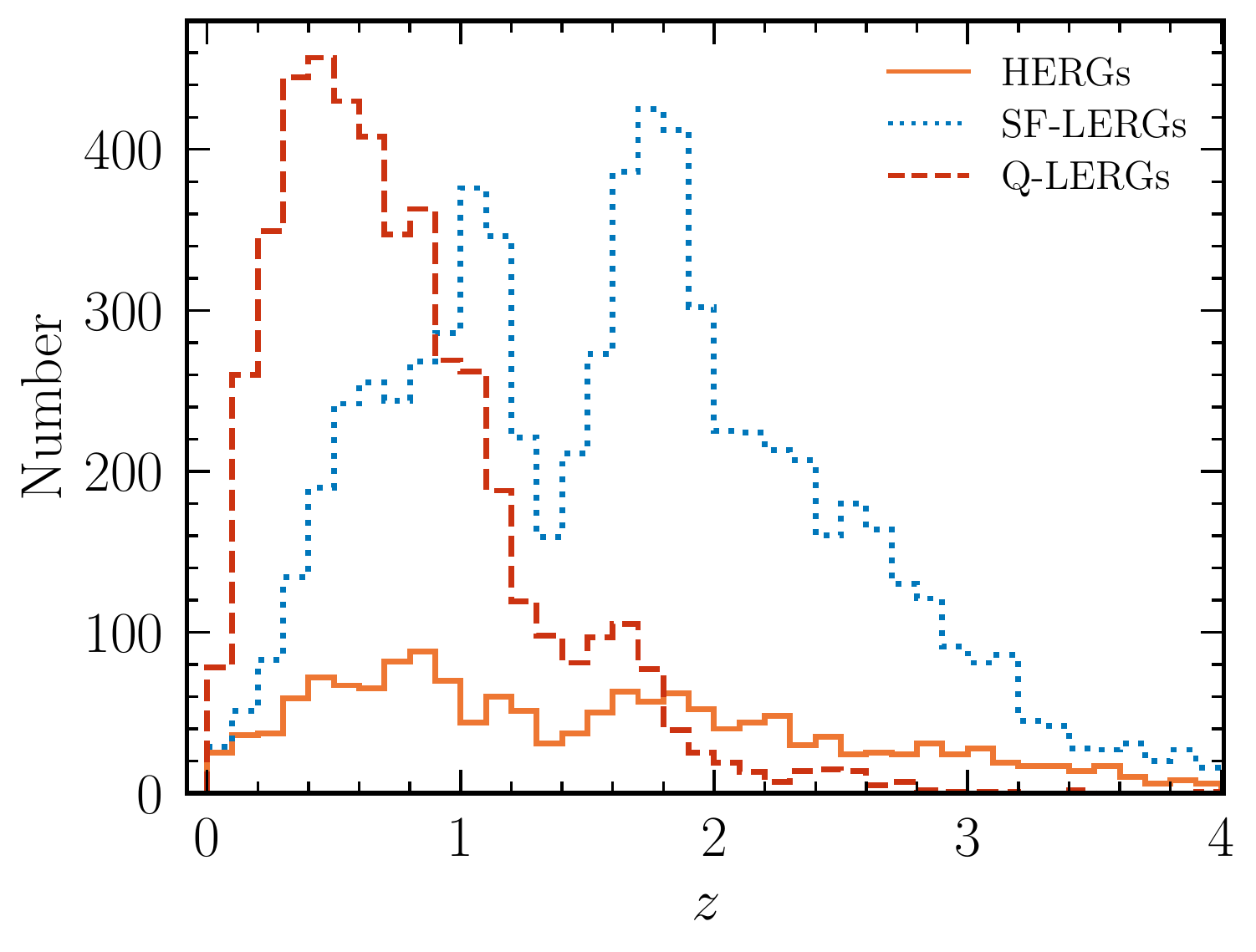}
    \end{subfigure}
    \caption{\label{fig:pz_hist} \textit{Top:} Radio luminosity as a function of redshift for the LERGs detected in the LoTSS Deep Fields. The colourbar with a logarithmic scale shows the number of sources across the parameter space. The red line shows the point source 5$\sigma$ detection limit using $\sigma=20\,\rm{\mu Jy}$ (achieved in the central region of ELAIS-N1) and assuming a spectral index $\alpha=-0.7$. \textit{Bottom:} Histogram of redshifts for the sample of LERGs split into those hosted by quiescent (red dashed line) and star-forming (blue dotted line) galaxies (see Sect.~\ref{sec:agn_frac_quies}), along with the HERGs (orange line), selected in the LoTSS Deep fields.}
\end{figure}

The final sample of LERGs and HERGs used in this paper and for the construction of the luminosity functions is limited to sources that have a radio detection with peak flux density at $>5\sigma$ level based on the local RMS, and excludes sources that are masked in optical bright-star masks (i.e. \textsc{flag\_clean} $= 1$; see \citealt{Kondapally2021} for details) to ensure a clean and robust sample. In addition, we limit our analysis of the luminosity functions to $0.03 < z \leq 2.5$; at lower redshifts the volume sampled by LoTSS-Deep is small and there may be incompleteness due to the larger angular size of the nearby sources, and at higher redshifts, we would reach the limits of the multi-wavelength datasets available in these fields, beyond which the source classifications of Best et al. from the SED fitting process become less secure. The above criteria result in a sample of 10\,481 LERGs (of which 4563 are hosted by quiescent galaxies and 5918 are hosted by star-forming galaxies; see Sect.~\ref{sec:lerg_lfs_Q_SF}), and 1302 HERGs, across the three deep fields.

Fig.~\ref{fig:pz_hist} (\textit{top}) shows the 150\,MHz radio luminosity as a function of redshift for the sample of LERGs from the LoTSS Deep Fields. The red line corresponds to the point source 5$\sigma$ detection limit calculated based on the noise in the central region of ELAIS-N1, $\sigma=20\,\rm{\mu}$Jy, and a radio spectral index $\alpha = -0.7$. Fig.~\ref{fig:pz_hist} (\textit{bottom}) shows the redshift distribution for the sample of LERGs (split into those hosted by quiescent and star-forming galaxies; see Sect.~\ref{sec:agn_frac_quies}) and HERGs. We note that the `peaks' seen in the redshift distribution, most prominent for the star-forming LERGs, are caused by the use of the median of the posterior photometric redshift distribution, which for very faint sources can suffer from aliasing due to gaps in the filter coverage. The full posterior distribution for such sources is smoother, and hence these `peaks' seen are not physical. Neither the SED fitting nor the source classifications account for uncertainties in the photometric redshifts, however, the impact of such features on the luminosity functions can be reduced by applying an optical magnitude completeness selection (see Sect.~\ref{sec:building_lfs}). Moreover, because of the wide redshift bins used for the analysis in this paper, this small aliasing has little effect on the derived luminosity functions.

\section{Total Radio-AGN luminosity functions}\label{sec:general_lfs}
\subsection{Building the luminosity functions}\label{sec:building_lfs}
We calculate the rest-frame radio luminosity of our sources by assuming that the radio spectrum is described by a simple power-law in frequency, $\nu$, with $S_{\nu} \propto \nu^{\alpha}$, where $S_{\nu}$ is the radio flux density at frequency $\nu$ and $\alpha = -0.7$ is the assumed spectral index throughout this study \citep[e.g.][]{CalistroRivera2017,Murphy2017}. The radio luminosity $L_{\nu}$ can be computed using the radio flux density $S_{\nu}$ as
\begin{equation}\label{eq:l150}
L_{\nu} = \frac{4{\pi}D_{\rm{L}}^{2}(z)}{\left(1 + z\right)^{1+\alpha}} S_{\nu},
\end{equation}
where $D_{\rm{L}}$ is the luminosity distance to the source and the $(1+z)^{-(1+\alpha)}$ term accounts for the radio spectrum \textit{k}-correction.

The luminosity functions (LFs) were built using the standard $1/V_{\rm{max}}$ technique \citep{schmidt1968lf,condon1989lf}, which weights each source in the sample by the maximum volume $V_{\rm{max}}$ that the source could be observed in, given the potential redshift range, and still satisfy all selection effects to be included in the sample. This is particularly important for surveys like the LoTSS Deep Fields where the RMS in the radio images (and hence the $5\sigma$ flux density limit) varies as a function of the position in the field due to primary beam effects, increased RMS around bright sources, and facet-to-facet variations in radio data calibration. The luminosity function $\rho(L,z)$ gives the number of sources per unit comoving volume observed per unit of log luminosity, and is given by
\begin{equation}\label{eq:lf_vmax}
    \rho(L,z) = \frac{1}{\Delta \log L} \sum_{i=1}^{N} \frac{1}{V_{\rm{max},i}}
\end{equation}
where $\Delta \log L = 0.3$ is the luminosity bin width in log-space, with the sum calculated over each source $i$ in a given luminosity and redshift bin.

\begin{figure}
    \centering
    \includegraphics[width=\columnwidth]{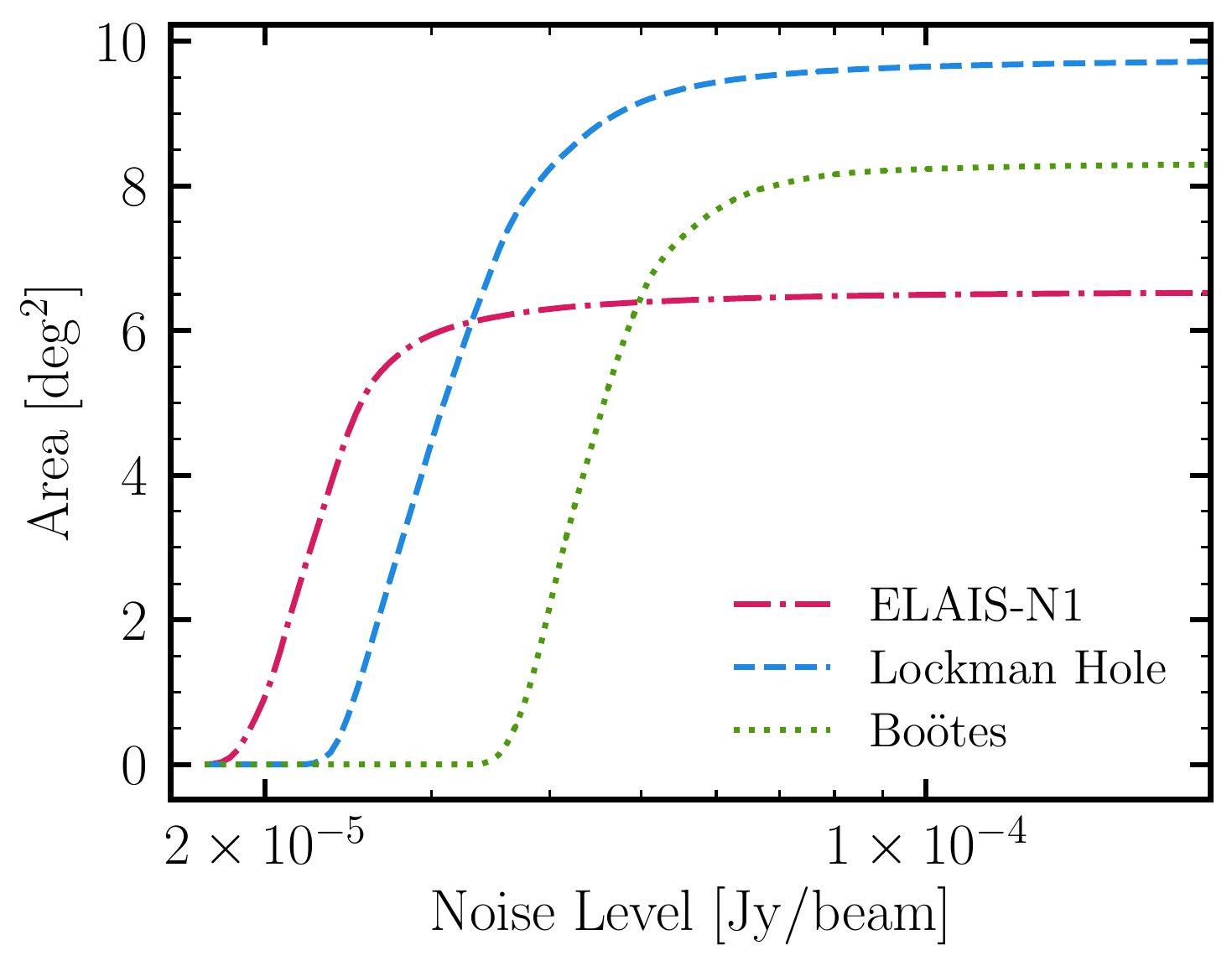}
    \caption{\label{fig:noise_area} Cumulative area covered down to a given noise level in the three LoTSS Deep Fields. The area is computed after applying the optical-based bright-star masks and only including the regions with coverage from the best available multi-wavelength surveys in each field \citep[see][]{Kondapally2021}. The total area across the three fields covers $\approx$ 24.6 deg$^{2}$.}
\end{figure}

The $V_{\rm{max},i}$ for a given source is then computed as 
\begin{equation}\label{eq:vmax_detailed}
    V_{\rm{max},i} = \int_{z_{min}}^{z_{max}} V(z) \, \theta(S,z) \, dz,
\end{equation}
where $V(z)dz$ is the comoving volume across the whole sky between redshift $z$ and $z + dz$, $\theta(S,z)$ is the fractional sky coverage of the LoTSS Deep Fields that accounts for the non-uniform radio-map noise and radio flux density incompleteness, and $S$ is the radio flux density that a source $i$ with a given intrinsic luminosity would have at redshift $z$. In practice, we evaluate this integral numerically, with a step size of $\Delta z = 0.0001$ between $z_{\rm{min}}$ and $z_{\rm{max}}$, the minimum and maximum range of the redshift bin. At each redshift, $\theta(S,z)$ for each source $i$ is given as
\begin{equation}\label{eq:area_comp_full}
    \theta(S,z) = \frac{\Omega[S(z)]}{4\pi} \, \times \, C_{\rm{radio}}[S(z)] 
\end{equation}
where $\Omega[S(z)]$ is the solid angle (in units of sr) of the survey area in which a source $i$ with a flux density $S$ can be detected at 5$\sigma$ given the non-uniform radio image noise, and $C_{\rm{radio}}[S(z)]$ is the radio flux density completeness correction at flux density $S$. The $\Omega[S(z)]$ term is computed by performing a linear interpolation using the cumulative area versus noise plot for each field, shown in Fig.~\ref{fig:noise_area}; these areas are obtained after limiting to the best-available multi-wavelength regions in each field and applying optical-based bright star masking \citep[see][]{Kondapally2021}. Completeness corrections are required to account for the faint undetected radio sources close to the survey detection limit, otherwise leading to an underestimate of the space density; these are discussed in detail in Sect.~\ref{sec:completeness}. 

In constructing the LFs, we consider only the radio sources that have a host galaxy counterpart; this corresponds to 97 per cent of the radio sources in the LoTSS Deep Fields \citep[see][]{Kondapally2021}. For the remaining 3 per cent of sources, the lack of a multi-wavelength counterpart identification means that no photometric redshift was available, and hence SED fitting and subsequent source classification were not possible. \citet{Kondapally2021} performed an optical to mid-IR stacking analysis of the subset of these sources with secure radio positions (which forms the majority) and found that these sources are likely to be predominantly $z > 3$ radio-AGN. Therefore, we expect that this will have a negligible effect on the derived LFs in this paper as we consider only sources with $z \leq 2.5$. 

In the calculation above, we have assumed that the radio flux density sets the limit on the $V_{\rm{max}}$ for a given source. It is, however, possible that the optical/IR dataset may set the limits on the maximum observable volume for a source. The proper application of optical/IR magnitude limits on the LFs is complicated by the use of $\chi^{2}$ detection images, analysis of these would involve detailed modelling of how they would affect both the photometric redshift accuracy and the source classifications; this is beyond the scope of this paper. We have, however, considered the effect of applying an optical/IR magnitude limit in the calculation of $V_{\rm{max}}$ and find that the radio dataset predominantly sets the limits on the $V_{\rm{max}}$. The LFs constructed when also considering an optical/IR magnitude limit agree very well with those based on the radio selection effects alone. Therefore, in this paper, we consider only the radio limits when constructing the LFs.

To compute the uncertainties on the luminosity functions, we performed bootstrap sampling (random sampling by replacement) of the catalogue to generate a distribution of 1000 realisations of the luminosity function. The lower and upper $1\sigma$ uncertainties on our luminosity functions are then determined from the 16\textsuperscript{th} and 84\textsuperscript{th} percentiles of the bootstrap samples. For the faint luminosity bins, where the samples are large and the uncertainties computed from bootstrapping correspondingly small, the true uncertainties are likely to be dominated by other factors such as the photometric redshift errors and source classification uncertainties. Therefore, we set a minimum uncertainty of $0.03$\,dex in the luminosity functions reported, based on the $\sim$ 7 per cent photometric redshift outlier fraction in these fields.

\subsection{Radio completeness corrections}\label{sec:completeness}

\begin{table}
\caption{Completeness corrections determined for the LOFAR Deep Fields calculated using a source-size distribution derived from the observed sources classified as AGN by Best et al. (in prep.) with $1\,\rm{mJy} < S_{150,\rm{tot}} \leq 10\,\rm{mJy}$ in each field. The completeness curves for each field are shown in Fig.~\ref{fig:completeness} (\textit{bottom}).}
\label{tab:comp_corr}
\centering
\begin{tabular}{cccc}
\hline\hline
Flux density & ELAIS-N1 & Lockman Hole & Bo\"{o}tes \\
{[mJy]} & {} & {} & {}\\
\hline
0.09 & 0.143 & - & - \\
0.11 & 0.221 & - & - \\
0.13 & 0.351 & 0.193 & - \\
0.16 & 0.553 & 0.319 & - \\
0.19 & 0.68 & 0.43 & 0.211 \\
0.23 & 0.742 & 0.57 & 0.322 \\
0.28 & 0.778 & 0.664 & 0.486 \\
0.33 & 0.802 & 0.751 & 0.624 \\
0.4 & 0.813 & 0.779 & 0.717 \\
0.63 & 0.849 & 0.841 & 0.816 \\
1.01 & 0.884 & 0.872 & 0.871 \\
1.59 & 0.921 & 0.907 & 0.912 \\
2.52 & 0.951 & 0.941 & 0.952 \\
4.0 & 0.961 & 0.959 & 0.971 \\
6.34 & 0.973 & 0.967 & 0.982 \\
10.05 & 0.975 & 0.974 & 0.981 \\
15.92 & 0.978 & 0.973 & 0.98 \\
25.24 & 0.984 & 0.978 & 0.989 \\
40.0 & 0.987 & 0.978 & 0.985 \\
\hline
\end{tabular}
\end{table}

The radio flux density completeness corrections are generated by performing simulations of inserting mock sources of various intrinsic source-size and flux density distributions into the radio image, and then recovering them using the same \textsc{PyBDSF} parameters as used for the real sources (see \citealt{Sabater2021} for the \textsc{PyBDSF} parameters used).

We simulated mock sources at fixed total flux density intervals separated by $0.2\,\rm{dex}$ in the range $0.4 < S < 40$\,mJy and, using a finer sampling interval of $0.08\,\rm{dex}$ in the range $90 < S < 400\,\rm{\mu}$Jy to better probe the expected sharp decline in completeness at faint flux densities. For the Lockman Hole and Bo\"{o}tes fields, we used the same sampling intervals but only simulating sources down to $\sim 110\,\rm{\mu}$Jy and $\sim 190\,\rm{\mu}$Jy, respectively, due to the slightly shallower depth of the radio data. The flux intervals used are listed in Table~\ref{tab:comp_corr}. 

For each field, we simulated 120\,000 -- 150\,000 mock sources to sample the full range in (total) flux density and source-size parameter space, while also obtaining robust statistics for the bright and extended rare sources. Practically, this was done by inserting 1000 mock sources with convolved sizes between 6 -- 30 arcsec, where 6 arcsec is the size of the LOFAR beam, for a given (total) flux density value into the radio image and extracting the sources using \textsc{PyBDSF}, with this step repeated many times (see Sec.~\ref{sec:comp_size_dist} for details). The injected sources were modelled as Gaussians, although the structure of real sources may be more complex. We ensure that a mock source is placed at least twice its FWHM (along the major axis) away from other mock sources and real radio sources to avoid source overlapping, which can complicate the process of determining if a simulated source has been detected. To define a mock source as being `detected', we require a detection in the mock catalogue within a given separation between the input position and the extracted \textsc{PyBDSF} position. For fainter simulated flux bins, the cross-match radius is set to be three times the expected rms positional uncertainty for a given SNR following \citet{condon1998nvss} and assuming a FWHM of 7 arcsec, the typical for a high-SNR compact source \citep[e.g.][]{Shimwell2022}; this results in a cross-match radius of $\sim$ 3.5 arcsec for the faintest flux bins. At brighter simulated flux densities, this positional uncertainty becomes very small; we therefore set a minimum cross-match radius of 2 arcsec. These angular separation criteria were validated by examining the change in the number of cross-matches as a function of separation.

\subsubsection{Source-size distributions}\label{sec:comp_size_dist}
Completeness depends not only on the flux density but also on the size of the source as source detection is performed based on the peak flux density of a source; therefore, for a given total flux density, the peak flux density for a larger source is more likely to fall below the detection threshold than for a smaller source. However, an accurate determination of the source-size distribution of the sub-mJy radio source population at low frequencies is lacking (see work by \citealt{Mandal2021} at characterising the faint low-frequency source counts) and we must therefore make some assumptions in deriving the corrections.

\begin{figure}
    \centering
    \begin{subfigure}{\columnwidth}
    \centering
    \includegraphics[width=\columnwidth]{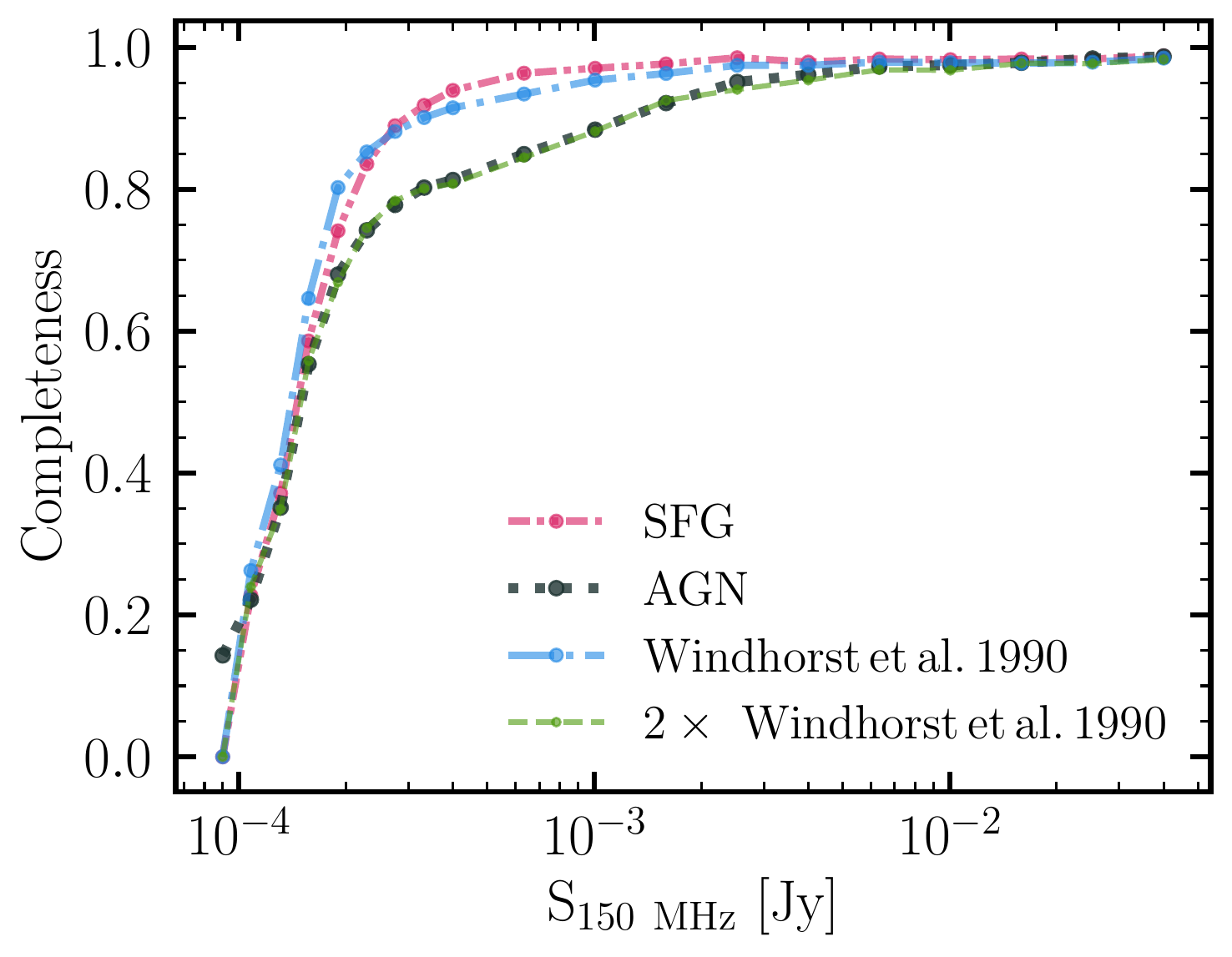}
    \end{subfigure}
    \begin{subfigure}{\columnwidth}
    \includegraphics[width=\columnwidth]{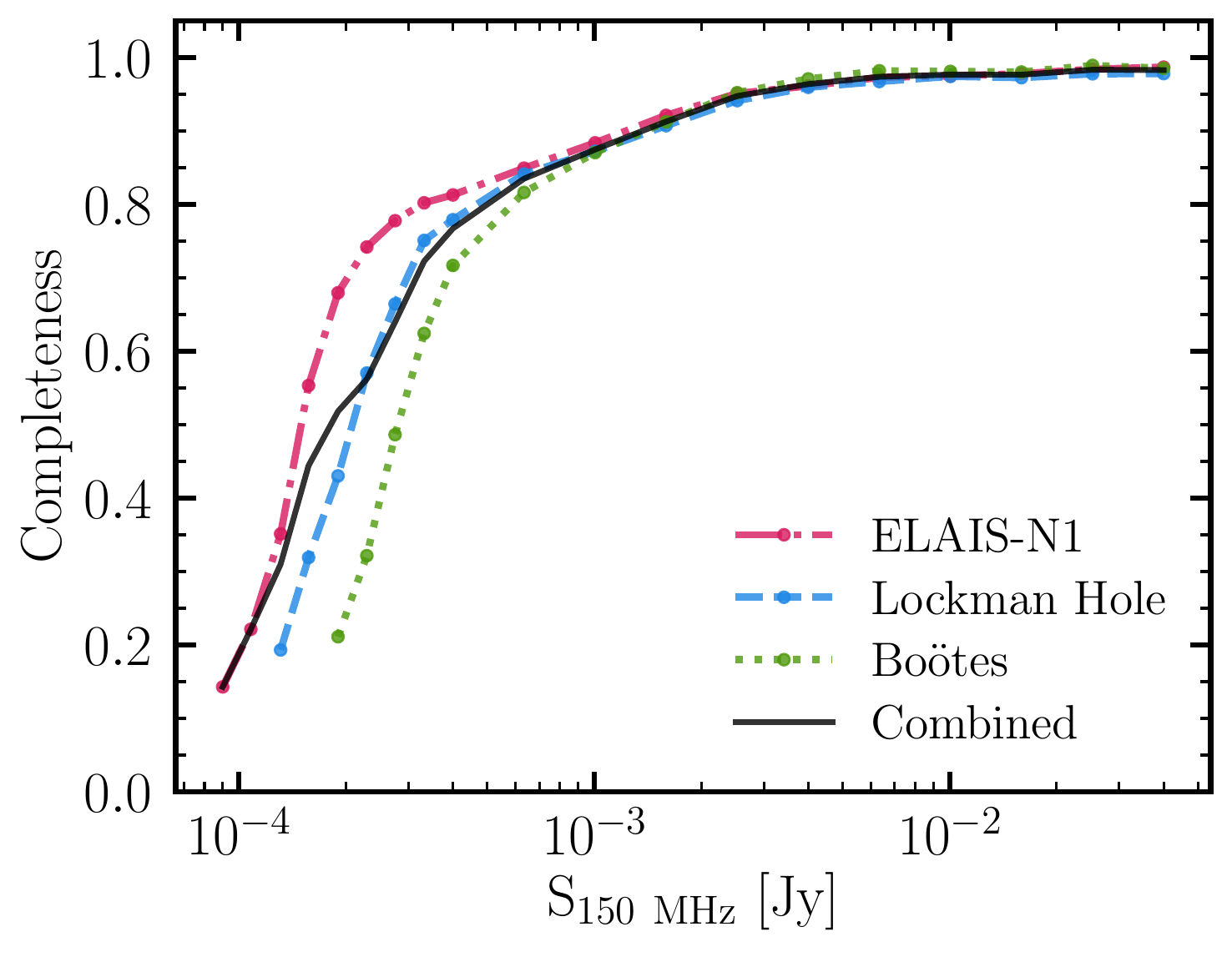}
    \end{subfigure}
    \caption{\label{fig:completeness} Radio flux density completeness corrections computed using the method outlined in Section~\ref{sec:completeness}. \textit{Top:} Completeness curves in the ELAIS-N1 field for different assumed source-size distributions. Curves show completeness based on applying the size-distribution of moderately bright sources (S$_{\rm{tot}} = 1 - 10$\,mJy; i.e. where completeness is high) classified as either SFGs or AGN in Best et al. (in prep.). In addition, the completeness curves resulting from the \protect\cite{windhorst1990} and 2 $\times$ \protect\cite{windhorst1990} integral angular size distribution, often used in the literature, are also shown. \textit{Bottom:} Completeness curves for each of the three LoTSS Deep Fields assuming an `AGN' size-distribution; this is used in the construction of the AGN LFs for each field. A table listing these adopted corrections for each field is shown in Table~\ref{tab:comp_corr}. The `combined' curve is the area-weighted average of the completeness curves in the three fields, used for constructing the combined LFs across the three fields.}
\end{figure}

We start by assuming that the observed size distribution of sources, within a flux density range that is unaffected by completeness, is an accurate description of sizes at fainter flux densities. As our work is focused on generating completeness corrections suited for the AGN subset of the radio population, we generate an `AGN' source-size distribution by selecting all sources classified as radio-excess AGN or SED AGN in the total flux density range $1\,\rm{mJy} < S_{\rm{150,tot}} < 10\,\rm{mJy}$, where we expect the sample to be largely complete. We consider only sources with sizes in the range 6 -- 30 arcsec (along both major and minor axes); larger sizes are not used in our simulations as such sources are poorly represented by a Gaussian surface brightness profile; the small number of sources with larger sizes are all placed at 30 arcsec in our simulations. Within each simulated flux density bin, we weight the simulation output by this size distribution. To determine the completeness correction for each flux density interval, we then considered the subset of mock sources with total flux density $>5\sigma$ based on the local rms, and determined the fraction of these that were detected by \textsc{PyBDSF} with a peak flux density above $5\sigma$ (thus matching the selection criteria applied to the observed source sample, and the limits adopted for the $V_{\rm{max}}$ calculation). The `AGN' completeness corrections obtained in this way in ELAIS-N1 are shown by the grey line in Fig.~\ref{fig:completeness} (top panel). For comparison, we repeat this process using the size distribution of the star-forming galaxy (SFG) subset of the radio population with $1\,\rm{mJy} < S_{\rm{150,tot}} < 10\,\rm{mJy}$, deriving the pink completeness curve in Fig.~\ref{fig:completeness}. The relatively lower completeness for the `AGN' line compared to the `SFG' curve is driven by the sizes of (resolved) radio-AGN being significantly larger than the SFG population; this results in a higher fraction of extended sources (with consequently lower peak flux densities for a given total flux density), which leads to a lower completeness, as expected.

To confirm the robustness of the above approach, we consider also the flux-dependent angular size distribution based on GHz surveys commonly used in the literature. \cite{windhorst1990} describe the integral angular size distribution as
\begin{equation}\label{eq:wh90_htheta}
h\left(>\psi\right) = \exp \left[ -{\ln}\,2\, \left(\frac{\psi}{\psi_{\rm{med}}}\right)^{0.62} \right]
\end{equation}
where, $h(> \psi)$ is the integral angular size distribution for sources with angular sizes larger than $\psi$ (in arcsec) at a given flux density and $\psi_{\rm{med}}$ is the median angular size at the given flux density. \cite{windhorst1990,windhorst1993} proposed a relationship between $\psi_{\rm{med}}$ and the radio flux at 1.4\,GHz, $S_{\rm{1.4\,GHz}}$:
\begin{equation}\label{eq:theta_med}
\psi_{\rm{med}} = 2\, (S_{\rm{1.4\,GHz}} \, / \, \rm{mJy})^{0.3} \rm{arcsec},
\end{equation}
which was converted to 150\,MHz using a spectral index $\alpha = -0.7$. They also considered a potential floor in this relationship at a size of 2 arcsec, however given that this needs to be convolved with the 6 arcsec LOFAR beam, this makes no significant difference to the observed size distribution. Using equations~\ref{eq:wh90_htheta} and~\ref{eq:theta_med}, $h(>\psi)$ was computed for each flux density interval and the simulation outputs were weighted by this to determine the completeness corrections. The resulting correction in the ELAIS-N1 field is shown in Fig.~\ref{fig:completeness} (top panel; blue line). Other low radio frequency studies in the literature \citep[e.g.][]{williams2016facet_bootes,Mahony2016,retanamontenegro2018sourcecounts_bootes} find good agreement between the LOFAR data and the \cite{windhorst1990} size distributions if the median angular size relation (equation~\ref{eq:theta_med}) is scaled by a factor of two. We therefore also compute the completeness corrections for this relation, also shown in the top panel of Fig.~\ref{fig:completeness} (green line), with the larger median sizes resulting in lower completeness. We find a good agreement between the 2$\times$ \citet{windhorst1990} and the `AGN' curves, and similarly between the \citet{windhorst1990} and the `SFG' curves.

In subsequent analysis, we use the `AGN' completeness corrections in the construction of the luminosity functions. The `AGN' completeness corrections for all three LoTSS Deep fields are shown in the bottom panel of Fig.~\ref{fig:completeness} and listed in Table~\ref{tab:comp_corr}. Using `AGN' based corrections, the observations in LoTSS Deep Fields reach a completeness of 50 per cent (and 90 per cent) at $150\,\rm{\mu Jy}$ ($1.3\,\rm{mJy}$), $209\,\rm{\mu Jy}$ ($1.4\,\rm{mJy}$), and $289\,\rm{\mu Jy}$ ($1.4\,\rm{mJy}$), in ELAIS-N1, Lockman Hole, and Bo\"{o}tes, respectively. We note that in all fields, the completeness does not reach 100 per cent; this is largely due to the source finding algorithm struggling to detect simulated sources placed in the higher noise, and lower dynamic range regions near bright genuine sources. We generate a ``combined'' completeness curve for use in constructing the LFs by performing an area-weighted average of the completeness curves in the three fields (black line in Fig.~\ref{fig:completeness}, \textit{bottom}).

\subsubsection{Application of the completeness corrections}\label{sec:comp_application}
The completeness corrections were applied, following equation~\ref{eq:area_comp_full}, by linearly interpolating the corrections at the flux density of each source. We applied a maximum completeness correction of a factor of 10 as any larger corrections are likely not reliable. Then, to determine the point where the completeness corrections to the LFs are too large to be reliable, we recalculated the luminosity functions but this time without applying any completeness corrections (i.e. by setting $C_{\rm{radio}}[S(z)] = 1$ in equation~\ref{eq:area_comp_full}); in our analysis we do not plot or list the space densities for the luminosity bins where the difference between the data points with and without the corrections is larger than 0.3\,dex. To account for uncertainties in the completeness corrections (e.g. the lack of knowledge of the true source-size distribution), we add 25 per cent of the completeness correction in quadrature to the error obtained from bootstrap sampling at each luminosity bin. This value is motivated by experiments of how the completeness correction varies with different bootstrap samples drawn from the simulations.

To take full advantage of the LoTSS Deep Fields, we calculate the LF in each of the three LoTSS Deep Fields separately, and build a combined LF across the three fields covering $\sim 25\,$deg$^{2}$ to obtain more robust number statistics across the full luminosity range and to limit the effects of cosmic variance.

\subsection{The local radio-AGN luminosity function}\label{sec:local_lfs}
\begin{figure}
    \centering
    \includegraphics[width=\columnwidth]{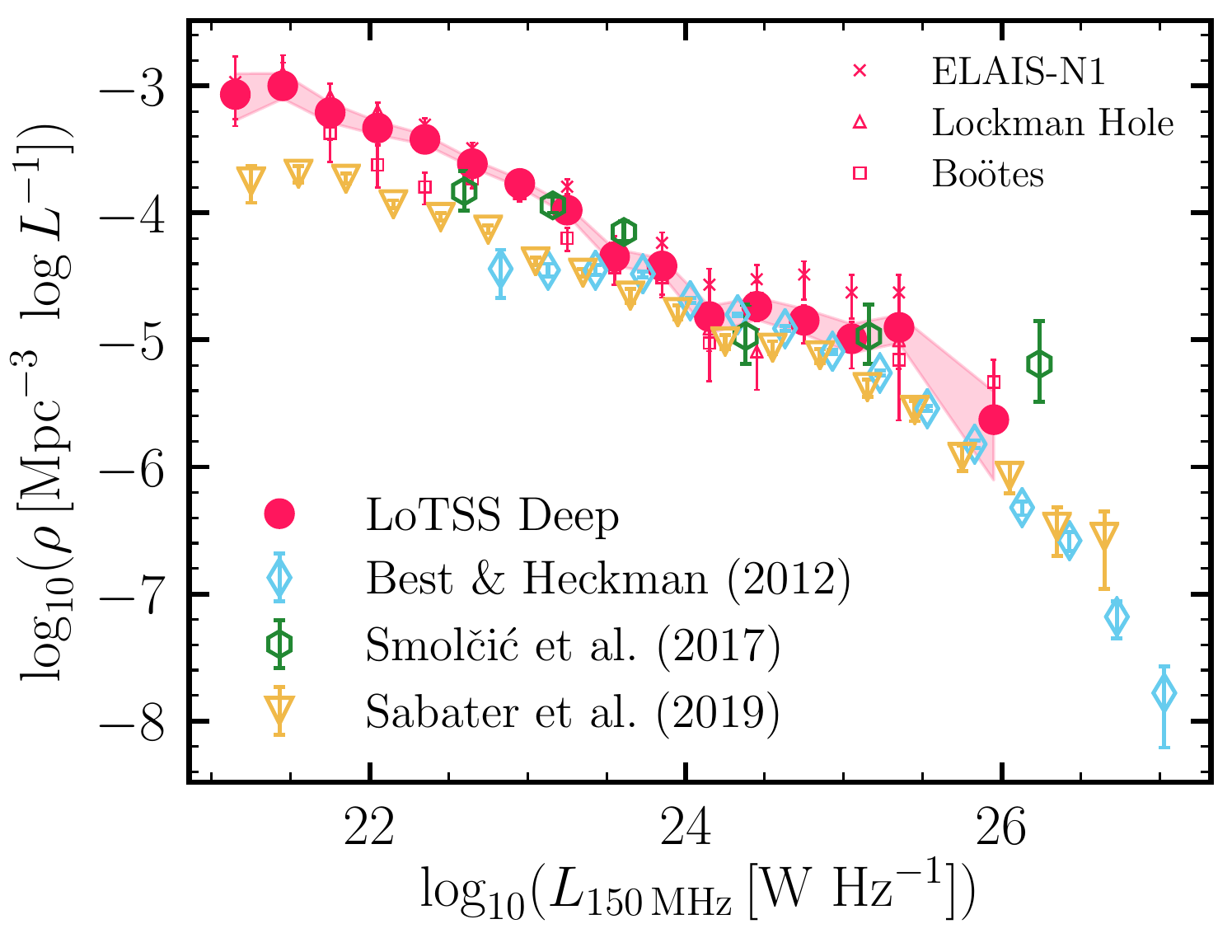}
    \caption{\label{fig:local_lf}Local ($0.03 < z \leq 0.3$) radio-excess AGN luminosity functions for LoTSS Deep Fields built with a bin width of $\Delta\log L = 0.3\,$dex. The combined luminosity function is shown in pink filled circles (for 939 sources) with shaded regions showing the 1$\sigma$ uncertainties computed from bootstrapping. The luminosity functions for each field individually are also shown. We find fairly good agreement with the \citet{2012MNRAS.421.1569B} and \citet{smolcic2017agn_evol_vla} results from GHz surveys, scaled to 150\,MHz using a spectral index $\alpha = -0.7$.}
\end{figure}

Although the LoTSS Deep Fields cover a relatively small volume at low redshifts, comparison of the low redshift LF against previous measurements in the literature is useful. Using the methods outlined above, we have built the local $0.03 < z \leq 0.3$ luminosity function for the radio-excess AGN (939 sources) in the LoTSS Deep Fields, which is shown in Fig.~\ref{fig:local_lf} and listed in Table~\ref{tab:rad_AGN_local_lf}. The combined luminosity function across the three fields is shown by pink filled circles, with the shaded pink region showing the 1$\sigma$ uncertainties. The luminosity functions for each individual deep field are also shown by pink crosses, triangles, and squares (with their respective error bars) for ELAIS-N1, Lockman Hole, and Bo\"{o}tes, respectively. We note that the ELAIS-N1 data points are consistently higher than the other fields, an effect that is also seen in the K-band number counts \citep[see fig.~2 of][]{Kondapally2021}, likely due to large-scale structure within the field as the volume probed at these low redshifts is relatively small. The Bo\"{o}tes data points are offset lower than the other fields, which at high luminosities is likely due to cosmic variance effects, and at low luminosities may also be from incompleteness. Any differences in the radio flux density calibration between the three fields may also introduce an offset in the LFs.

\begin{table}
\caption{The local ($0.03 < z \leq 0.3$) radio-excess AGN LF for LoTSS Deep Fields.\label{tab:rad_AGN_local_lf}}
\centering
\begin{tabular}{ccc}
\hline\hline 
$\log L_{150\, \mathrm{MHz}}$ & $\log \rho$ & N \\
$\log (\mathrm{W~Hz}^{-1})$ & $\log (\mathrm{Mpc}^{-3}~\log L^{-1}$) &  \\
\hline
21.15 & $-3.07_{-0.21}^{+0.18}$ & 6\Tstrut \\[0.05cm]
21.45 & $-3.00_{-0.12}^{+0.11}$ & 20\\[0.05cm]
21.75 & $-3.21_{-0.10}^{+0.09}$ & 34\\[0.05cm]
22.05 & $-3.33_{-0.08}^{+0.07}$ & 67\\[0.05cm]
22.35 & $-3.42_{-0.07}^{+0.06}$ & 135\\[0.05cm]
22.65 & $-3.61_{-0.05}^{+0.05}$ & 187\\[0.05cm]
22.95 & $-3.77_{-0.04}^{+0.04}$ & 176\\[0.05cm]
23.25 & $-3.98_{-0.04}^{+0.04}$ & 119\\[0.05cm]
23.55 & $-4.34_{-0.07}^{+0.06}$ & 54\\[0.05cm]
23.85 & $-4.42_{-0.07}^{+0.06}$ & 47\\[0.05cm]
24.15 & $-4.82_{-0.13}^{+0.08}$ & 19\\[0.05cm]
24.45 & $-4.74_{-0.11}^{+0.07}$ & 23\\[0.05cm]
24.75 & $-4.85_{-0.11}^{+0.09}$ & 18\\[0.05cm]
25.05 & $-4.99_{-0.11}^{+0.12}$ & 13\\[0.05cm]
25.35 & $-4.90_{-0.12}^{+0.10}$ & 16\\[0.05cm]
25.95 & $-5.63_{-0.48}^{+0.22}$ & 3\\[0.05cm]
\hline
\end{tabular}
\end{table}

\begin{figure*}
    \centering
    \includegraphics[width=\textwidth]{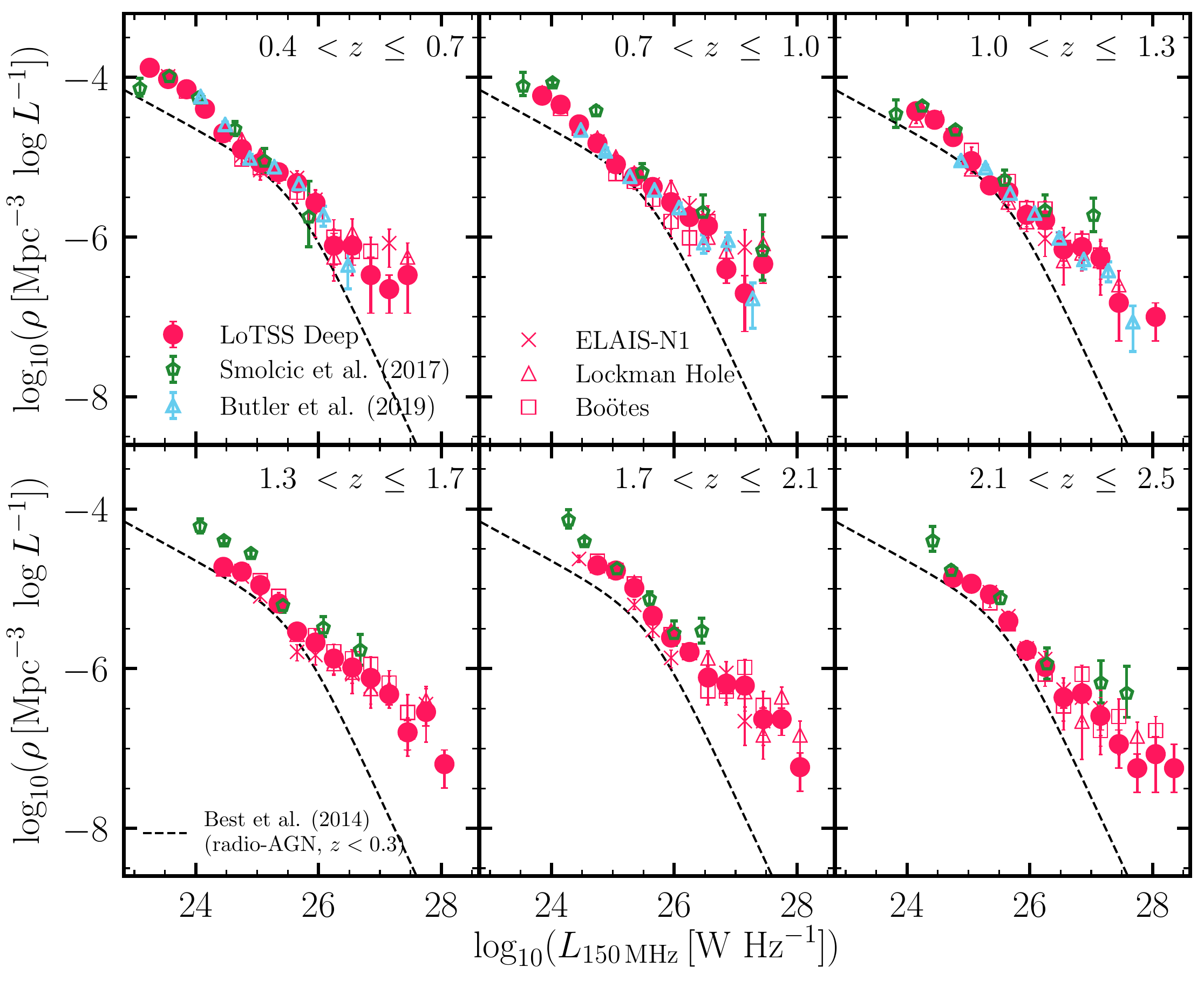}
    \caption{\label{fig:lf_evol_radio_agn} Cosmic evolution (across $0.4 < z \leq 2.5$) of the total radio-AGN luminosity functions at 150\,MHz for the combined LoTSS Deep sample (pink circles). The LFs in individual LoTSS Deep fields are shown in open pink symbols (same symbols as in Fig.~\ref{fig:local_lf}). The dashed line in each panel shows the parameterised form of the local radio-AGN LF by \citet{2014MNRAS.445..955B}, converted to 150\,MHz using a spectral index $\alpha = -0.7$ to guide the eye. The total radio-AGN luminosity functions from \citet{smolcic2017agn_evol_vla} and \citet{Butler2019} in green and cyan open points, respectively, converted to 150\,MHz using $\alpha = -0.7$, are also shown. Overall, we find good agreement with the results from both \citet{Butler2019} and \citet{smolcic2017agn_evol_vla} across redshift.}
\end{figure*}

Fig.~\ref{fig:local_lf} also shows comparison between our local radio-AGN luminosity functions and other previous studies. We find good agreement with the local radio-AGN luminosity function from \citet{smolcic2017agn_evol_vla} built using deep radio imaging from the VLA-COSMOS 3\,GHz Large project \citep{smolcic2017_vla_cosmos_3G}. They identified radio-AGN using a $3\sigma$ redshift-dependent radio-excess selection based on the FIRC following \citet{Delvecchio2017}. Their LFs were computed over $0.1 < z < 0.4$ (105 sources), and for illustration in Fig.~\ref{fig:local_lf} are shifted to 150\,MHz using a spectral index $\alpha = -0.7$. \citet{2012MNRAS.421.1569B} presented a large sample of radio-detected AGN drawn from the combination of FIRST and NVSS with SDSS spectroscopic sample data. We find good agreement with the luminosity function of \citet{2012MNRAS.421.1569B}, shifted to $\rm{L_{150\,MHz}}$ using a spectral index $\alpha = -0.7$, in particular at $\rm{L_{150\,MHz}} \sim 10^{23.5} - 10^{25}\,\rm{W\,Hz^{-1}}$. We are however not able to sample enough volume to probe significantly above the break in the luminosity function. \citet{2012MNRAS.421.1569B} also found their LF to be in good agreement with other determinations \citep[e.g.][]{Mauch2007,Pracy2016}. 

Also shown in Fig.~\ref{fig:local_lf} (green triangles) is the radio-AGN luminosity function from the shallower but wider LoTSS Data Release 1 (LoTSS-DR1) from \citet{sabater2019lotssagn}. Their data, covering $> 400\,$deg$^{2}$, is better suited to sample higher radio luminosities. We find good agreement at moderate luminosities, however our luminosity function is consistently offset to higher space densities by $\sim 0.5\,$dex, especially at $\rm{L_{150\,MHz}} \lesssim 10^{23.5}\,\rm{W\,Hz^{-1}}$. We find that the \citet{2012MNRAS.421.1569B} LFs appear to turn over at around the same point in luminosity. There are a few possible reasons for the difference with both of those studies. First, neither \citet{2012MNRAS.421.1569B} nor \citet{sabater2019lotssagn} apply any completeness corrections; we re-derive the \citeauthor{sabater2019lotssagn} luminosity function by simply applying a 10$\sigma$ radio flux density cut, which increases the space densities by $\sim 0.2\,$dex at $\rm{L_{150\,MHz}} \lesssim 10^{22.3}\,\rm{W\,Hz^{-1}}$. We also re-derived our LF without applying completeness corrections; this reduces the difference with the \citeauthor{sabater2019lotssagn} LF down to the level of field-to-field variations seen across the LoTSS Deep Fields which are a result of cosmic variance and radio flux density calibration differences. Secondly, the \citet{2012MNRAS.421.1569B} and \citet{sabater2019lotssagn} radio-AGN samples were defined by combining their radio data with the SDSS main galaxy spectroscopic sample; sources like quasars or radio-quiet quasars are missing from this sample, likely biasing the luminosity function to lower space densities. Moreover, their classification scheme was tuned to select only `radio-loud AGN', which excludes Seyfert-like AGN, biasing their luminosity function at low luminosities. Finally, the median redshift of the \citet{2012MNRAS.421.1569B} sample is $z_{\rm{med}} \sim 0.16$, and that of \citet{sabater2019lotssagn} is $z_{\rm{med}} = 0.14$, whereas the median redshift for the LoTSS Deep Fields sample (in this redshift bin) is $z_{\rm{med}} = 0.21$; any cosmic evolution of this population can therefore contribute to the difference in space densities observed. We also note that this difference is not simply due to a misclassification of sources between radio-AGN and SFGs, as a similar offset (i.e. higher space densities of SFGs in the LoTSS Deep Fields) is also found by Cochrane et al. (in prep.) when comparing the local SFG luminosity function from the LoTSS Deep Fields with LoTSS-DR1.

\subsection{Evolution of the radio-AGN LFs}\label{sec:radio_agn_evol}
Fig.~\ref{fig:lf_evol_radio_agn} shows the redshift evolution of the total radio-AGN LF in the LoTSS Deep fields over the redshift range $0.4 < z \leq 2.5$. The combined LoTSS Deep LF is shown as pink circles, with the LFs from individual fields shown as open pink symbols (using the same symbols as in Fig.~\ref{fig:local_lf}). For comparison, we show the parametric fit to the local radio-AGN luminosity function by \citet{2014MNRAS.445..955B} in each panel, shifted to 150\,MHz using $\alpha = -0.7$. We again compare our results with the evolution of the total radio-AGN population presented in \citet{smolcic2017agn_evol_vla}, shifted to $\rm{L_{150\,MHz}}$ using $\alpha = -0.7$, shown by green symbols, with our redshift bins chosen to match their analysis. As is evident from Fig.~\ref{fig:lf_evol_radio_agn}, we find excellent agreement with their results at all radio luminosities, out to $z \sim 2.5$; this gives us further confidence that our source classification method for separating radio-AGN from star-forming galaxies is appropriate out to high redshifts. We do however note the disagreement seen at all luminosities in the $1.3 < z \leq 1.7$ bin, which may be driven by the COSMOS field being over-dense at these redshifts \citep[see][]{Duncan2021,McLeod2021}, where the larger volume probed by LoTSS-Deep allows a more robust determination of the LF.

\begin{figure*}
    \centering
    \includegraphics[width=\textwidth]{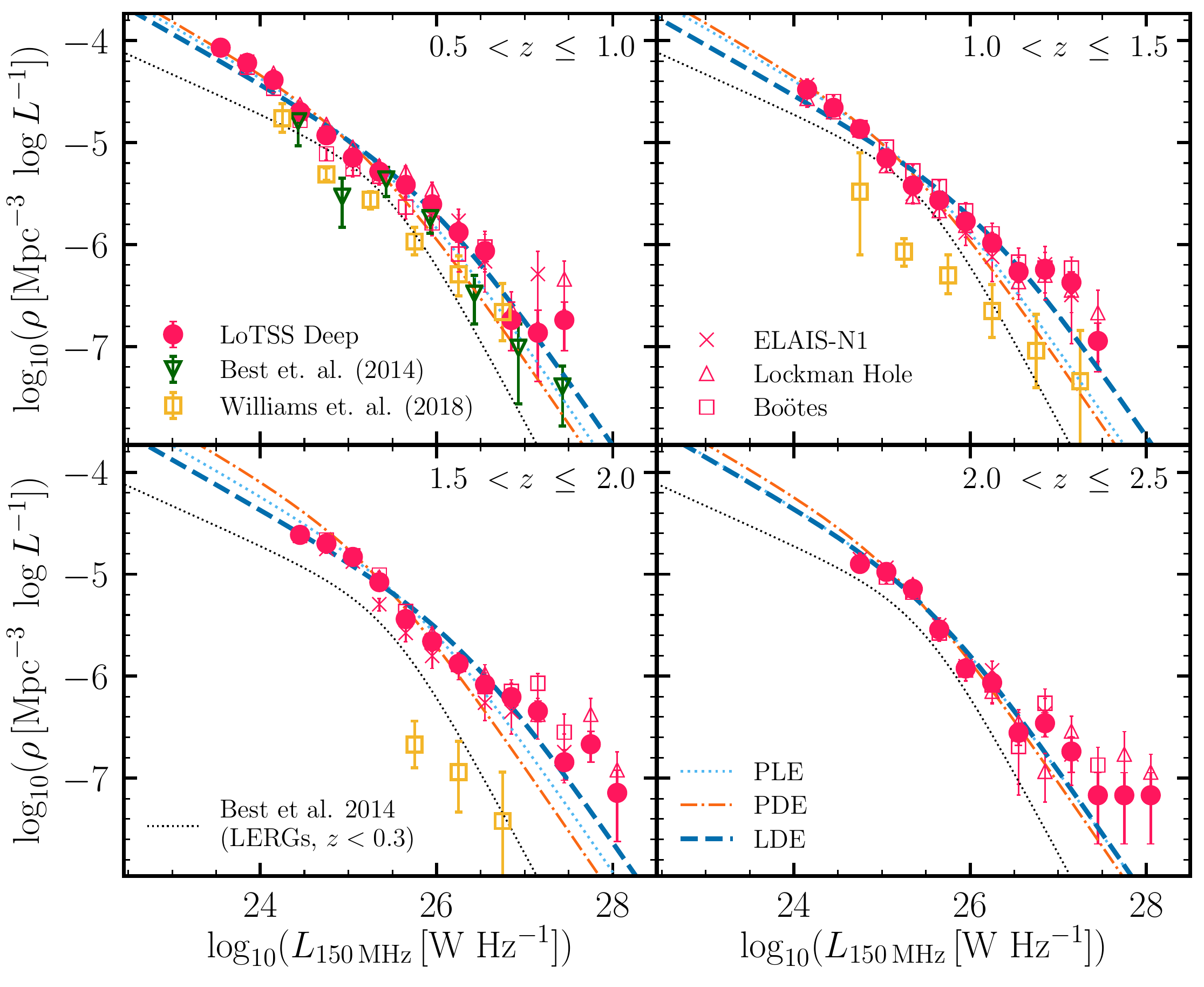}
    \caption{\label{fig:lf_evol_lerg} Cosmic evolution (across $0.5 < z \leq 2.5$) of the LERG luminosity functions at 150\,MHz for the combined LoTSS Deep sample (pink circles), showing relatively mild evolution across the redshift bins examined. Also shown are the LFs in individual LoTSS Deep fields (in pink; same symbols as in Fig.~\ref{fig:local_lf}) and the parametric fit to the local LERG LF (black dotted line) from \citet{2014MNRAS.445..955B}, scaled to 150\,MHz assuming a spectral index $\alpha=-0.7$. The pure density evolution (PDE), pure luminosity evolution (PLE), and combined luminosity and density evolution (LDE) models from fits to individual redshift bins are shown as orange (dash-dotted), blue (dotted), and dark blue (dashed) lines, respectively (see Sect.~\ref{sec:evol_model} for details). In the  $0.5 < z \leq 1.0$ bin, the LERG LF from \citet[][green triangles]{2014MNRAS.445..955B}, scaled to 150\,MHz using a spectral index $\alpha=-0.7$, shows good agreement with our results. The LFs from \citet{Williams2018}, reaching $z \sim 2$ (yellow squares in the three lowest redshift bins), have systematically lower space densities compared to our results. Further tests show that the differences are largely driven by our improvements in the source classification method used.}
\end{figure*}

Also shown in Fig.~\ref{fig:lf_evol_radio_agn} are the radio-AGN LFs of \citet{Butler2019} across $0.3 < z \leq 0.6$, $0.6 < z \leq 0.9$, and $0.9 < z \leq 1.3$ in the closest redshift bins to our LFs. These LFs were compiled using data from the Australian Telescope Compact Array (ATCA) 2.1\,GHz observations of the XXL-S field \citep{Butler2018} and the radio-AGN sample was selected based on radio-source luminosity, morphology, spectral indices, and radio-excess emission based on the FIRC. We also find good agreement with their LFs where available, but our results probe fainter luminosities.

\section{Cosmic evolution of the LERG luminosity functions}\label{sec:cosevol_lerg_lf}
The total radio-AGN luminosity functions presented in Sect.~\ref{sec:radio_agn_evol} contain a mixture of both the LERG and the HERG populations. From studies of the nearby Universe, these two populations are expected to evolve differently with the LERGs dominating the space densities at low luminosities and the HERGs dominating at high radio luminosities \citep[e.g.][]{2012MNRAS.421.1569B}; the LERG population is also particularly interesting for radio-AGN feedback-cycle considerations. Previous studies \citep[e.g.][]{2014MNRAS.445..955B,Pracy2016,Williams2018,Butler2019} have attempted to model the evolution of the LERGs; however, the current LoTSS Deep Fields dataset, with vastly greater numbers of LERGs resulting from the combination of deep radio and multi-wavelength datasets over $\sim 25\,\rm{deg}^{2}$, allows us to study the cosmic evolution of this population in unprecedented detail. The HERGs form a minority of the radio-AGN population in LoTSS-Deep DR1 (see Fig.~\ref{fig:pz_hist}), and moreover, have typically higher radio luminosities than the LERGs \citep[e.g.][]{2014MNRAS.445..955B,Pracy2016}; the present LoTSS-Deep sample therefore does not allow us to robustly constrain the LF and the cosmic evolution of the HERGs. In this paper, we list the HERG LFs in Table~\ref{tab:lerg_evol_table}, which are also plotted in Fig.~\ref{fig:lf_lerg_herg_quies_scvary}, but focus our analysis on the LERG population only in the rest of the paper. Subsequent data releases covering wider areas that better sample the bright end of the luminosity function will enable detailed analysis of the evolution of the HERGs.

In Fig.~\ref{fig:lf_evol_lerg}, we show the evolution of the LERG LF for the LoTSS Deep Fields in four redshift bins ($0.5 < z \leq 1.0$, $1.0 < z \leq 1.5$, $1.5 < z \leq 2.0$, and $2.0 < z \leq 2.5$), each spanning $3 - 4$ decades in luminosity. The combined LoTSS Deep Fields LFs, also tabulated in Table~\ref{tab:lerg_evol_table}, are shown as pink circles. The LFs for individual fields are also shown in pink, using the same symbols for each field as in Fig.~\ref{fig:local_lf}. In each panel, the black dotted line shows the parametrised form of the local ($z < 0.3$) LERG luminosity function determined by \citet{2014MNRAS.445..955B}, scaled to 150\,MHz using a spectral index $\alpha = -0.7$. We find that the LERG population shows modest evolution in the range $0 < z < 0.75$, especially at high luminosities, although the details of this depend on the assumed spectral index. Beyond $z = 0.75$, we see a relatively mild evolution in the LFs with redshift out to $z \sim 2.5$.

We compare our results to the LERG LF derived by \citet{2014MNRAS.445..955B} at $0.5 < z \leq 1$ in Fig.~\ref{fig:lf_evol_lerg}, shifted to $L_{\rm{150\,MHz}}$ using a spectral index $\alpha = -0.7$. \citet{2014MNRAS.445..955B} compiled catalogues of radio-detected AGN from eight surveys, covering different radio-depths and areas, to obtain a sample of 211 radio-loud AGN. They used spectroscopic information to classify their AGN sample into LERGs and HERGs, representing the first study of the evolution of the two modes of AGN, separately, out to $z \sim 1$. The LoTSS-Deep LF shows good agreement with their resulting luminosity function (green triangles), with our data probing fainter luminosities.

\begin{table*}
\caption{Luminosity functions of the LERGs and HERGs in the LoTSS Deep Fields in the range $0.5 < z \leq 2.5$, illustrated in Fig.~\ref{fig:lf_evol_lerg}. Also shown are the LFs in the same redshift bins, for the subsets of the LERG population that are hosted by quiescent galaxies and by star-forming galaxies, respectively; these resulting LFs are discussed in more detail in Sect.~\ref{sec:lerg_lfs_Q_SF}, with the separation of quiescent and star-forming host galaxies of the LERGs described in Sect.~\ref{sec:agn_frac_quies}. Space densities for bins with $N < 2$ are not shown. A machine-readable version of these data is available at \url{https://github.com/rohitk-10/AGN_LF_Kondapally22}.}
\label{tab:lerg_evol_table}
\centering
\begin{tabular}{>{\centering}m{1.47cm}>{\centering}m{1.42cm}>{\centering}m{1.8cm}>{\centering}m{0.92cm}>{\centering}m{1.8cm}>{\centering}m{0.94cm}>{\centering}m{1.8cm}>{\centering}m{0.94cm}>{\centering}m{1.8cm}c} 
\hline\hline
$z$ & $\log L_{150\, \mathrm{MHz}}$ & $\log \rho_{\mathrm{LERG}}$ & $N_{\mathrm{LERG}}$ & $\log \rho_{\mathrm{LERG,Q}}$ & $N_{\mathrm{LERG,Q}}$
 & $\log \rho_{\mathrm{LERG,SF}}$ & $N_{\mathrm{LERG,SF}}$ & $\log \rho_{\mathrm{HERG}}$ & $N_{\mathrm{HERG}}$\\
 & $\log(\mathrm{W~Hz}^{-1})$ & $\log(\mathrm{Mpc}^{-3}\log L^{-1})$ & {} & $\log(\mathrm{Mpc}^{-3}\log L^{-1})$ & {} & $\log(\mathrm{Mpc}^{-3}\log L^{-1})$ & {} & $\log(\mathrm{Mpc}^{-3}\log L^{-1})$ & {}\\
\hline
${0.5} < z \leq {1.0}$ & 23.55 & $-4.07_{-0.07}^{+0.07}$ & 428 & $-4.26_{-0.07}^{+0.07}$ & 279 & $-4.53_{-0.08}^{+0.08}$ & 149 & $-5.15_{-0.11}^{+0.10}$ & 35 \Tstrut\\[0.05cm]
  & 23.85 & $-4.22_{-0.05}^{+0.05}$ & 756 & $-4.48_{-0.05}^{+0.05}$ & 411 & $-4.56_{-0.05}^{+0.05}$ & 345 & $-5.29_{-0.07}^{+0.07}$ & 66 \\[0.05cm]
  & 24.15 & $-4.39_{-0.03}^{+0.03}$ & 719 & $-4.69_{-0.03}^{+0.03}$ & 361 & $-4.69_{-0.04}^{+0.03}$ & 358 & $-5.39_{-0.06}^{+0.06}$ & 72 \\[0.05cm]
  & 24.45 & $-4.70_{-0.03}^{+0.03}$ & 387 & $-5.01_{-0.03}^{+0.03}$ & 190 & $-4.99_{-0.03}^{+0.03}$ & 197 & $-5.41_{-0.05}^{+0.05}$ & 76 \\[0.05cm]
  & 24.75 & $-4.93_{-0.03}^{+0.03}$ & 242 & $-5.15_{-0.04}^{+0.03}$ & 144 & $-5.32_{-0.04}^{+0.04}$ & 98 & $-5.63_{-0.07}^{+0.06}$ & 48 \\[0.05cm]
  & 25.05 & $-5.15_{-0.04}^{+0.03}$ & 151 & $-5.31_{-0.04}^{+0.04}$ & 104 & $-5.65_{-0.07}^{+0.05}$ & 47 & $-5.91_{-0.09}^{+0.08}$ & 26 \\[0.05cm]
  & 25.35 & $-5.28_{-0.04}^{+0.04}$ & 111 & $-5.42_{-0.05}^{+0.04}$ & 82 & $-5.87_{-0.10}^{+0.08}$ & 29 & $-6.33_{-0.15}^{+0.11}$ & 10 \\[0.05cm]
  & 25.65 & $-5.41_{-0.05}^{+0.04}$ & 83 & $-5.51_{-0.06}^{+0.05}$ & 67 & $-6.13_{-0.12}^{+0.10}$ & 16 & $-6.43_{-0.20}^{+0.10}$ & 8 \\[0.05cm]
  & 25.95 & $-5.60_{-0.06}^{+0.06}$ & 54 & $-5.72_{-0.07}^{+0.06}$ & 41 & $-6.22_{-0.15}^{+0.11}$ & 13 & $-6.56_{-0.18}^{+0.12}$ & 6 \\[0.05cm]
  & 26.25 & $-5.88_{-0.10}^{+0.07}$ & 29 & $-5.94_{-0.08}^{+0.07}$ & 25 & $-6.74_{-0.48}^{+0.22}$ & 4 & $-6.86_{-0.48}^{+0.22}$ & 3 \\[0.05cm]
  & 26.55 & $-6.06_{-0.10}^{+0.09}$ & 19 & $-6.20_{-0.15}^{+0.11}$ & 14 & $-6.64_{-0.22}^{+0.15}$ & 5 & $-6.50_{-0.24}^{+0.11}$ & 7 \\[0.05cm]
  & 26.85 & $-6.74_{-0.30}^{+0.18}$ & 4 & $-6.86_{-0.48}^{+0.22}$ & 3 & {} & {} & $-6.74_{-0.30}^{+0.18}$ & 4 \\[0.05cm]
  & 27.15 & $-6.86_{-0.48}^{+0.22}$ & 3 & $-7.04_{-0.30}^{+0.18}$ & 2 & {} & {} & {} & {} \\[0.05cm]
  & 27.45 & $-6.74_{-0.30}^{+0.18}$ & 4 & {} & {} & $-6.86_{-0.48}^{+0.22}$ & 3 & $-6.56_{-0.18}^{+0.18}$ & 6 \\[0.05cm]
\hline
${1.0} < z \leq {1.5}$ & 24.15 & $-4.48_{-0.07}^{+0.07}$ & 396 & $-4.84_{-0.08}^{+0.08}$ & 166 & $-4.73_{-0.07}^{+0.07}$ & 230 & $-6.02_{-0.15}^{+0.12}$ & 14 
\\[0.05cm]
  & 24.45 & $-4.66_{-0.04}^{+0.04}$ & 545 & $-5.16_{-0.05}^{+0.05}$ & 170 & $-4.82_{-0.04}^{+0.04}$ & 375 & $-5.70_{-0.07}^{+0.07}$ & 51 \\[0.05cm]
  & 24.75 & $-4.87_{-0.03}^{+0.03}$ & 403 & $-5.43_{-0.04}^{+0.04}$ & 111 & $-5.01_{-0.03}^{+0.03}$ & 292 & $-5.77_{-0.07}^{+0.06}$ & 50 \\[0.05cm]
  & 25.05 & $-5.15_{-0.03}^{+0.03}$ & 224 & $-5.63_{-0.05}^{+0.05}$ & 74 & $-5.33_{-0.04}^{+0.03}$ & 150 & $-5.92_{-0.08}^{+0.06}$ & 38 \\[0.05cm]
  & 25.35 & $-5.42_{-0.05}^{+0.04}$ & 127 & $-5.80_{-0.06}^{+0.06}$ & 53 & $-5.65_{-0.05}^{+0.05}$ & 74 & $-6.18_{-0.11}^{+0.07}$ & 22 \\[0.05cm]
  & 25.65 & $-5.56_{-0.04}^{+0.04}$ & 94 & $-5.99_{-0.08}^{+0.07}$ & 35 & $-5.77_{-0.05}^{+0.05}$ & 59 & $-6.53_{-0.15}^{+0.11}$ & 10 \\[0.05cm]
  & 25.95 & $-5.77_{-0.06}^{+0.05}$ & 58 & $-6.09_{-0.09}^{+0.07}$ & 28 & $-6.06_{-0.10}^{+0.07}$ & 30 & $-6.58_{-0.18}^{+0.12}$ & 9 \\[0.05cm]
  & 26.25 & $-5.98_{-0.08}^{+0.07}$ & 36 & $-6.20_{-0.10}^{+0.08}$ & 22 & $-6.39_{-0.16}^{+0.09}$ & 14 & $-6.46_{-0.12}^{+0.12}$ & 12 \\[0.05cm]
  & 26.55 & $-6.27_{-0.10}^{+0.08}$ & 19 & $-6.59_{-0.18}^{+0.12}$ & 9 & $-6.54_{-0.18}^{+0.12}$ & 10 & $-6.84_{-0.22}^{+0.15}$ & 5 \\[0.05cm]
  & 26.85 & $-6.25_{-0.12}^{+0.10}$ & 20 & $-6.70_{-0.24}^{+0.11}$ & 7 & $-6.43_{-0.11}^{+0.12}$ & 13 & $-7.07_{-0.48}^{+0.22}$ & 3 \\[0.05cm]
  & 27.15 & $-6.37_{-0.13}^{+0.10}$ & 15 & $-6.94_{-0.30}^{+0.18}$ & 4 & $-6.50_{-0.14}^{+0.11}$ & 11 & $-7.25_{-0.30}^{+0.18}$ & 2 \\[0.05cm]
  & 27.45 & $-6.94_{-0.30}^{+0.18}$ & 4 & $-7.25_{-0.30}^{+0.18}$ & 2 & $-7.25_{-0.30}^{+0.18}$ & 2 & {} & {} \\[0.05cm]
  & 27.75 & {} & {} & {} & {} & {} & {} & $-6.85_{-0.22}^{+0.15}$ & 5 \\[0.05cm]
\hline
${1.5} < z \leq {2.0}$ & 24.45 & $-4.61_{-0.08}^{+0.08}$ & 306 & $-5.31_{-0.09}^{+0.09}$ & 64 & {} & {} & $-6.11_{-0.15}^{+0.13}$ & 13 \\[0.05cm]
  & 24.75 & $-4.70_{-0.04}^{+0.04}$ & 581 & $-5.52_{-0.06}^{+0.06}$ & 85 & $-4.77_{-0.04}^{+0.04}$ & 496 & $-5.83_{-0.07}^{+0.07}$ & 45 \\[0.05cm]
  & 25.05 & $-4.83_{-0.03}^{+0.03}$ & 516 & $-5.79_{-0.06}^{+0.05}$ & 57 & $-4.88_{-0.03}^{+0.03}$ & 459 & $-5.65_{-0.06}^{+0.05}$ & 79 \\[0.05cm]
  & 25.35 & $-5.08_{-0.03}^{+0.03}$ & 314 & $-5.93_{-0.07}^{+0.05}$ & 44 & $-5.14_{-0.03}^{+0.03}$ & 270 & $-5.83_{-0.06}^{+0.05}$ & 55 \\[0.05cm]
  & 25.65 & $-5.44_{-0.04}^{+0.04}$ & 143 & $-6.37_{-0.09}^{+0.09}$ & 17 & $-5.49_{-0.04}^{+0.04}$ & 126 & $-6.08_{-0.09}^{+0.07}$ & 33 \\[0.05cm]
  & 25.95 & $-5.66_{-0.05}^{+0.05}$ & 89 & $-6.35_{-0.11}^{+0.09}$ & 18 & $-5.76_{-0.06}^{+0.05}$ & 71 & $-6.38_{-0.12}^{+0.09}$ & 17 \\[0.05cm]
  & 26.25 & $-5.88_{-0.06}^{+0.05}$ & 54 & $-6.53_{-0.15}^{+0.11}$ & 12 & $-5.99_{-0.07}^{+0.06}$ & 42 & $-6.53_{-0.12}^{+0.10}$ & 12 \\[0.05cm]
  & 26.55 & $-6.08_{-0.08}^{+0.07}$ & 34 & $-6.53_{-0.14}^{+0.10}$ & 12 & $-6.27_{-0.11}^{+0.09}$ & 22 & $-6.66_{-0.18}^{+0.12}$ & 9 \\[0.05cm]
  & 26.85 & $-6.20_{-0.09}^{+0.08}$ & 26 & $-6.66_{-0.20}^{+0.14}$ & 9 & $-6.39_{-0.11}^{+0.09}$ & 17 & $-6.84_{-0.18}^{+0.18}$ & 6 \\[0.05cm]
  & 27.15 & $-6.34_{-0.10}^{+0.10}$ & 19 & $-6.92_{-0.18}^{+0.18}$ & 5 & $-6.48_{-0.11}^{+0.09}$ & 14 & $-6.67_{-0.18}^{+0.12}$ & 9 \\[0.05cm]
  & 27.45 & $-6.84_{-0.18}^{+0.12}$ & 6 & {} & {} & $-6.84_{-0.30}^{+0.12}$ & 6 & $-7.14_{-0.48}^{+0.22}$ & 3 \\[0.05cm]
  & 27.75 & $-6.67_{-0.18}^{+0.12}$ & 9 & $-7.32_{-0.30}^{+0.18}$ & 2 & $-6.78_{-0.24}^{+0.12}$ & 7 & $-7.32_{-0.30}^{+0.18}$ & 2 \\[0.05cm]
  & 28.05 & $-7.14_{-0.48}^{+0.22}$ & 3 & {} & {} & $-7.14_{-0.48}^{+0.22}$ & 3 & {} & {} \\[0.05cm]
\hline
${2.0} < z \leq {2.5}$ & 24.75 & $-4.90_{-0.07}^{+0.07}$ & 210 & $-5.84_{-0.12}^{+0.11}$ & 23 & $-4.95_{-0.07}^{+0.08}$ & 187 & $-6.19_{-0.15}^{+0.13}$ & 13 \\[0.05cm]
  & 25.05 & $-4.98_{-0.04}^{+0.04}$ & 344 & $-6.23_{-0.09}^{+0.08}$ & 19 & $-5.00_{-0.04}^{+0.04}$ & 325 & $-5.93_{-0.07}^{+0.07}$ & 39 \\[0.05cm]
  & 25.35 & $-5.14_{-0.03}^{+0.03}$ & 269 & $-6.67_{-0.18}^{+0.13}$ & 8 & $-5.16_{-0.03}^{+0.03}$ & 261 & $-5.85_{-0.07}^{+0.06}$ & 53 \\[0.05cm]
  & 25.65 & $-5.54_{-0.04}^{+0.04}$ & 115 & $-6.90_{-0.30}^{+0.13}$ & 5 & $-5.56_{-0.04}^{+0.04}$ & 110 & $-6.03_{-0.08}^{+0.06}$ & 38 \\[0.05cm]
  & 25.95 & $-5.92_{-0.07}^{+0.06}$ & 50 & $-7.02_{-0.30}^{+0.18}$ & 4 & $-5.96_{-0.06}^{+0.06}$ & 46 & $-6.39_{-0.12}^{+0.09}$ & 17 \\[0.05cm]
  & 26.25 & $-6.06_{-0.08}^{+0.07}$ & 37 & $-7.15_{-0.24}^{+0.15}$ & 3 & $-6.10_{-0.09}^{+0.07}$ & 34 & $-6.55_{-0.12}^{+0.10}$ & 12 \\[0.05cm]
  & 26.55 & $-6.55_{-0.12}^{+0.10}$ & 12 & $-7.33_{-0.30}^{+0.18}$ & 2 & $-6.63_{-0.15}^{+0.11}$ & 10 & $-6.73_{-0.20}^{+0.14}$ & 8 \\[0.05cm]
  & 26.85 & $-6.46_{-0.13}^{+0.10}$ & 15 & {} & {} & $-6.46_{-0.10}^{+0.11}$ & 15 & $-6.73_{-0.20}^{+0.10}$ & 8 \\[0.05cm]
  & 27.15 & $-6.74_{-0.20}^{+0.14}$ & 8 & {} & {} & $-6.80_{-0.24}^{+0.15}$ & 7 & $-7.17_{-0.48}^{+0.22}$ & 3 \\[0.05cm]
  & 27.45 & $-7.17_{-0.48}^{+0.22}$ & 3 & {} & {} & $-7.17_{-0.48}^{+0.22}$ & 3 & $-7.17_{-0.48}^{+0.22}$ & 3 \\[0.05cm]
  & 27.75 & $-7.17_{-0.48}^{+0.22}$ & 3 & {} & {} & $-7.17_{-0.48}^{+0.22}$ & 3 & {} & {} \\[0.05cm]
  & 28.05 & $-7.17_{-0.48}^{+0.22}$ & 3 & {} & {} & $-7.17_{-0.48}^{+0.22}$ & 3 & {} & {} \\[0.05cm]
  & 28.35 & {} & {} & {} & {} & {} & {} & $-7.34_{-0.30}^{+0.30}$ & 2 \\[0.05cm]
\hline
\end{tabular}
\end{table*}

\begin{table*}
\caption{Results of modelling the evolution of the LERG LFs using a pure luminosity evolution (PLE), pure density evolution (PDE), and the luminosity and density evolution (LDE) models, relative to the local relation from \citet{Mauch2007}, as detailed in Sect.~\ref{sec:evol_model}. The best-fitting LFs for each model are also shown in Fig.~\ref{fig:lf_evol_lerg}.\label{tab:lf_fits}}
\centering
\begin{tabular}{cccccccccccc}
\hline\hline
$z$ & \multicolumn{3}{c}{PLE} & \multicolumn{3}{c}{PDE} & \multicolumn{5}{c}{LDE}\\
{} & \multicolumn{3}{l}{\phantom{}}\dotfill & \multicolumn{3}{c}{\phantom{}} \dotfill & \multicolumn{5}{c}{\phantom{}}\dotfill \\
{} & $\log_{10} L_{\star}(z)$ & $\alpha_{\rm{L}}$ & $\chi^{2}_{\nu}$ & $\log_{10} \rho^{\star}(z)$ & $\alpha_{\rm{D}}$ & $\chi^{2}_{\nu}$ & $\log_{10} \rho^{\star}(z)$ & $\log_{10} L_{\star}(z)$ & $\alpha_{\rm{D}}$ & $\alpha_{\rm{L}}$ & $\chi^{2}_{\nu}$\\
\hline
${0.5} < z \leq {1.0}$ & $25.55_{-0.02}^{+0.02}$ & $1.14_{-0.07}^{+0.07}$ & 6.62 & $-4.92_{-0.01}^{+0.01}$ & $0.76_{-0.05}^{+0.05}$ & 9.47 & $-5.39_{-0.07}^{+
0.07}$ & $25.98_{-0.11}^{+0.12}$ & $-1.19_{-0.30}^{+0.28}$ & $2.93_{-0.49}^{+0.45}$ & 5.59 \Tstrut\\[0.15cm]
${1.0} < z \leq {1.5}$ & $25.50_{-0.02}^{+0.02}$ & $0.66_{-0.05}^{+0.05}$ & 9.19 & $-4.93_{-0.01}^{+0.01}$ & $0.48_{-0.04}^{+0.04}$ & 12.44 & $-5.63_{-0.09}^{
+0.08}$ & $26.24_{-0.13}^{+0.14}$ & $-1.50_{-0.24}^{+0.23}$ & $2.76_{-0.39}^{+0.36}$ & 5.56\\[0.15cm]
${1.5} < z \leq {2.0}$ & $25.79_{-0.02}^{+0.02}$ & $1.18_{-0.04}^{+0.04}$ & 8.28 & $-4.68_{-0.01}^{+0.01}$ & $0.96_{-0.03}^{+0.03}$ & 15.33 & $-5.52_{-0.08}^{
+0.08}$ & $26.35_{-0.11}^{+0.12}$ & $-0.95_{-0.18}^{+0.17}$ & $2.47_{-0.27}^{+0.26}$ & 6.28\\[0.15cm]
${2.0} < z \leq {2.5}$ & $25.58_{-0.02}^{+0.02}$ & $0.60_{-0.04}^{+0.04}$ & 3.14 & $-4.83_{-0.01}^{+0.01}$ & $0.54_{-0.03}^{+0.03}$ & 3.70 & $-5.13_{-0.11}^{+
0.12}$ & $25.61_{-0.14}^{+0.14}$ & $-0.05_{-0.22}^{+0.23}$ & $0.67_{-0.27}^{+0.27}$ & 3.45\\[0.15cm]
\hline
\end{tabular}
\end{table*}

Also in Fig.~\ref{fig:lf_evol_lerg}, we show the LERG LFs computed by \citet{Williams2018}, who studied the cosmic evolution ($0.5 < z \leq 2$) of a sample of 1224 LOFAR-detected sources within the Bo\"{o}tes field (albeit with shallower radio data). They used the combination of the FIR-radio correlation of \citet{CalistroRivera2016} and SED fitting via \textsc{AGNFitter} to perform source classifications. Radio sources that lie $2\sigma$ away from the FIR-radio correlation \citep{CalistroRivera2017} were identified as radio-loud AGN. Then, the various AGN and galaxy models fitted by \textsc{AGNFitter} were used to quantify the fraction of MIR emission arising from the AGN compared to the galaxy component ($f_{\rm{AGN}}$); sources with $f_{\rm{AGN}} > 0.25$, indicative of significant emission from the torus, were classified as HERGs, whereas the remainder are classified as LERGs, which are expected to show little to no torus emission. Their final sample contains 243 LERGs (and 398 HERGs) within $0.5 < z \leq 2$, with their resulting LF shown as yellow squares in Fig.~\ref{fig:lf_evol_lerg}. 

We note that the \citeauthor{Williams2018} LFs appear systematically offset to lower space densities than our dataset in all redshift bins, and also appear offset by $\sim 0.4\,\rm{dex}$ at moderate to faint luminosities ($L_{\rm{150\,MHz}} < 10^{26}\, \rm{W\,Hz^{-1}}$) compared to \citet{2014MNRAS.445..955B}. We have performed tests to investigate the sources of this discrepancy, which are detailed in Appendix~\ref{sec:ap_wendy_lerg}. Firstly, we require a 0.7\,dex ($\sim$ $3\sigma$) radio-excess over the radio luminosity versus SFR relation, whereas \citet{Williams2018} used a $2\sigma$ cut (with $\sigma = 0.529$) on the redshift-dependent FIRC of \citet{CalistroRivera2017}; therefore the radio-excess AGN classification scheme of \citet{Williams2018} is more conservative than our selection and is expected to result in systematically lower space densities of radio-excess objects. Secondly, we find that improvements in the input models for \textsc{AGNFitter} since the analysis of \citet{Williams2018} result in a change in source classification for a significant number of sources, in particular at higher redshifts. Finally, in our analysis, the SFR used for identifying radio-excess AGN is estimated from the \textit{consensus} measurements from four SED fitting codes by Best et al. (in prep), whereas \citeauthor{Williams2018} used estimates from \textsc{AGNFitter} alone. Best et al. (in prep.) show that estimates of SFRs (and infrared luminosities) from \textsc{AGNFitter} are systematically higher than those estimated from other SED fitting routines. As detailed in our analysis in Appendix~\ref{sec:ap_wendy_lerg}, the combination of differences in the SED fitting and source classification criteria results in the apparent discrepancy with \citet{Williams2018}.

\subsection{Modelling the cosmic evolution of the LFs}\label{sec:evol_model}
The radio luminosity function of AGN is generally modelled as a broken power-law of the form
\begin{equation}\label{eq:bpl}
\rho(L) = \frac{\rho^{\star}}{\left( L^{\star}/L\right)^{\beta} + \left( L^{\star}/L\right)^{\gamma}},
\end{equation}
where $\rho^{\star}$ is the characteristic space density, $L^{\star}$ is the characteristic luminosity, and, $\beta$ and $\gamma$ are the bright and faint end slopes, respectively \citep[e.g.][]{1990MNRAS.247...19D}. As evident from comparison with the $0.5 < z \leq 1$ \citet{2014MNRAS.445..955B} LF in Fig.~\ref{fig:lf_evol_lerg}, the area covered by the first data release of the LoTSS Deep Fields is not sufficient to probe bright enough luminosities to constrain the bright-end slope of the LERG LF, at all redshifts. Therefore, we modelled the evolution of the LERG population as the luminosity evolution and density evolution of the local LF such that in equation~\ref{eq:bpl}, $\rho^{\star}(z)$ accounts for the redshift evolution of the normalisation and $L^{\star}(z)$ is the redshift evolution of the characteristic luminosity. For this process, we assumed that the shape of the LF remains constant by fixing the bright-end slope $\beta = -1.27$, and the faint-end slope $\gamma = -0.49$, as found by a broken power-law fit to the local radio-AGN LF by \citet{Mauch2007}. Although the \citeauthor{Mauch2007} LF includes both the LERGs and HERGs, the faint-end slope, which is the key parameter to constrain as our data do not probe the bright end well, will be dominated by the LERGs; the \citet{Mauch2007} faint-end slope provides a better match to our dataset than, for example, the jet-mode AGN LF of \citet{2014MNRAS.445..955B}. We also found that a broken power law fit to the radio-AGN LF of \citet{sabater2019lotssagn} gives a faint-end slope consistent with the \citet{Mauch2007} value, and hence gives similar results in modelling the evolution below.

We first considered a luminosity and density evolution (LDE) model where we allowed $\rho^{\star}(z)$ and $L_{\star}(z)$ to be free parameters: both of these parameters were then fitted for each redshift bin using the \textsc{emcee} Markov Chain Monte Carlo (MCMC) fitting routine \citep{Foreman-Mackey2013}, with the resulting fitted $\rho^{\star}(z)$ and $L_{\star}(z)$ shown in Table~\ref{tab:lf_fits}. This LDE model can also be expressed in terms of a $(1 + z)$ parameterisation such that the evolution of the normalisation is expressed as $\rho^{\star}(z) = \rho^{\star}(0) \times (1 + z)^{\alpha_{\rm{D}}}$, and, the redshift evolution of the characteristic luminosity, $L^{\star}(z)$, is given as
$L^{\star}(z) = L^{\star}(0) \times (1 + z)^{\alpha_{\rm{L}}}$.
Here, $\alpha_{\rm{D}}$ and $\alpha_{\rm{L}}$ correspond to the density and luminosity evolution parameters, respectively. This can be used to both illustrate how strong the evolution is, and whether it follows a simple trend with redshift. For these cases, we used the \citeauthor{Mauch2007} normalisation $\rho^{\star}(0) = \frac{1}{0.4} 10^{-5.5} \log \rm{L^{-1}}\, \rm{Mpc^{-3}}$ (converted to per log luminosity) and the characteristic luminosity $L^{\star}(0) = 10^{25.27}\,\rm{W\,Hz^{-1}}$ (scaled to 150\,MHz assuming $\alpha=-0.7$) along with their two slopes. The corresponding values for the evolution parameters for this LDE model, along with the reduced chi-squared goodness of fit values, $\chi^{2}_{\nu}$, are also listed in Table~\ref{tab:lf_fits}.

\begin{figure*}
    \centering
    \includegraphics[width=\textwidth]{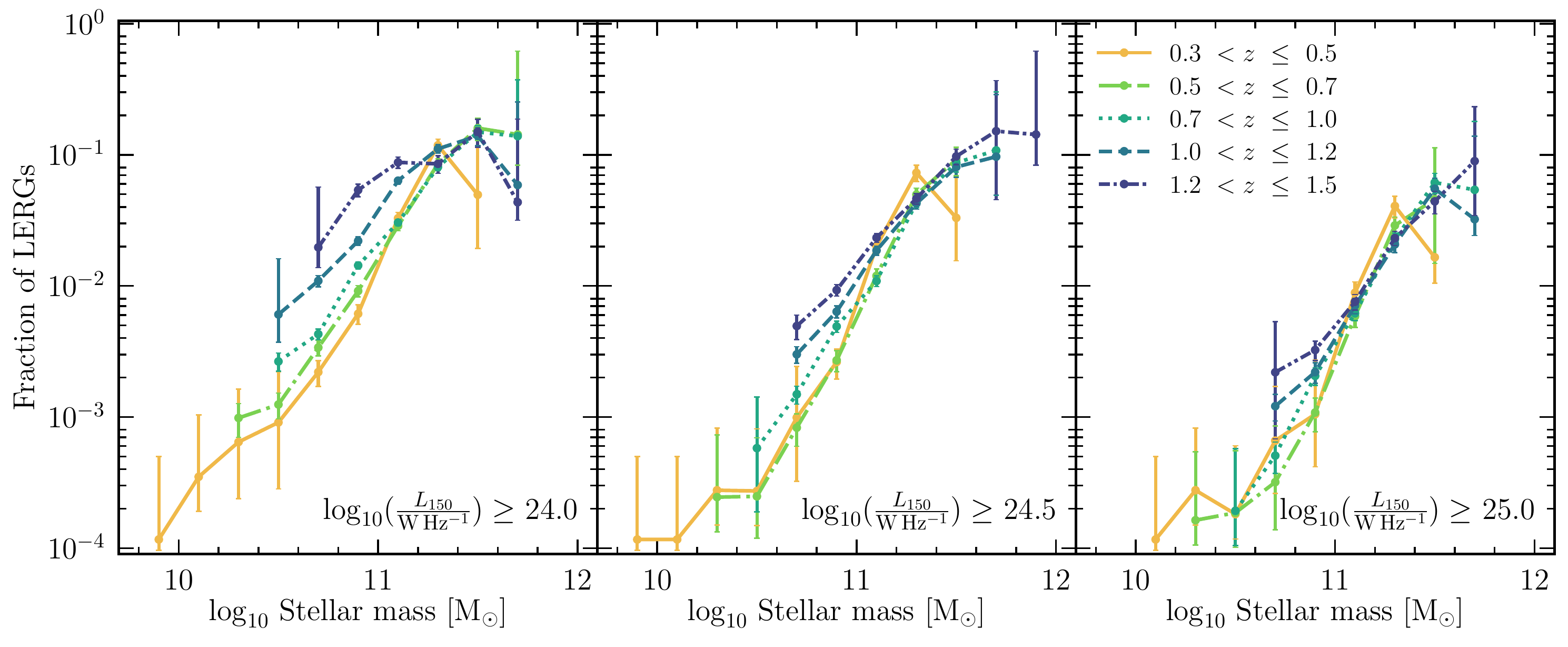}
    \caption{\label{fig:lerg_frac_Lbins}Fraction of galaxies hosting a LERG as a function of stellar mass in five redshift bins between $0.3 < z \leq 1.5$ in the LoTSS Deep Fields. Panels from left to right show this for increasing radio luminosity limits of $L_{\rm{150\,MHz}} \geq 10^{24}\,\rm{W\,Hz^{-1}}$, $L_{\rm{150\,MHz}} \geq 10^{24.5}\,\rm{W\,Hz^{-1}}$, and $L_{\rm{150\,MHz}} \geq 10^{25}\,\rm{W\,Hz^{-1}}$. The error bars represent Poisson uncertainties, following \citet{Gehrels1986} for $\rm{N} < 10$. The fraction of galaxies that host a LERG shows a steep stellar mass dependence, in particular at high radio luminosity limits; this relation remains at least out to $z \sim 1.5$. We also find evidence of this relation flattening at lower radio luminosity limits suggesting the presence of an additional fuelling mechanism which is investigated in Sect.~\ref{sec:agn_frac_quies}.}
\end{figure*}

For each redshift bin, we also considered a pure luminosity evolution (PLE) model (i.e. we set $\alpha_{\rm{D}} = 0$) and a pure density evolution (PDE) model (i.e. we set $\alpha_{\rm{L}} = 0$). The resulting best-fitting $\alpha_{\rm{D}}$ and $\alpha_{\rm{L}}$ parameters (as appropriate for each model) for each redshift bin are listed in Table~\ref{tab:lf_fits}, along with the $\chi^{2}_{\nu}$ values. The resulting best-fitting LFs for all three models at each redshift bin are also shown in Fig.~\ref{fig:lf_evol_lerg}. We find that both the PLE and PDE models are not sufficient at describing the evolution of the LERGs, in particular at high radio luminosities, as also noted by the resulting $\chi^{2}_{\nu}$ values in Table~\ref{tab:lf_fits}. The LDE model is better able to match the high-luminosity end of the LF. The LDE fits indicate that at least out to $z \sim 2$, there is a mild increase in the characteristic luminosity and a corresponding mild decrease in the characteristic space density with increasing redshift. This appears to reverse in the highest redshift bin; however, it should be noted that there is some degeneracy between the fitted $L_{\star}(z)$ and $\rho_{\star}(z)$, especially due to the lack of a prominent break in the LFs.

Finally, we considered how uncertainties in the source classification criteria employed affect our results on the LERG luminosity functions; this is discussed in detail in Appendix~\ref{sec:scvary}. In summary, given the good agreement between our LFs and literature results of the total radio-AGN LFs across redshift, we studied the effects of varying the selection of `SED AGN' on the LERG to HERG classification. We find that the evolution of the LERG population is largely insensitive (at the $\lesssim 0.2\,$dex level) to the exact threshold used for separating the two modes of AGN, even under a very conservative definition of HERGs; this gives us confidence in our source classification criteria adopted and on the resulting evolution of the LERG LFs.

To understand the origin of the mild evolution seen in the luminosity functions of the LERGs, we investigate the incidence of the LERGs as a function of stellar mass and redshift in different types of host galaxies in Sect.~\ref{sec:agn_frac}.

\section{Prevalence of AGN activity with stellar mass and star-formation rate}\label{sec:agn_frac}
It is well-known that the prevalence of radio-AGN increases strongly with stellar mass in the local Universe \citep[e.g.][]{best2005_sdss_xmatch,Smolcic2009,Janssen2012,sabater2019lotssagn} and early studies out to $z \sim 1$ suggest that this trend holds at earlier cosmic time \citep[e.g.][]{Tasse2008,Simpson2013}. In this paper, we study the redshift evolution of LERG activity by measuring how the fraction of galaxies that host a LERG varies with stellar mass across redshift.

For the LERG sample, we used the \textit{consensus} stellar masses derived from SED fitting (see Sect.~\ref{sec:sed_fits}). For the underlying galaxy population, we used the stellar masses (50th percentile of the posterior distribution) computed by \citet{Duncan2021} using a grid-based SED fitting method \citep[see also][]{duncan2014,Duncan2019} for the full multi-wavelength catalogue in each field (resulting in a total of $\sim$1.8\,million sources after satisfying the redshift and stellar mass completeness limits; see below). \citet{Duncan2021} validated their stellar masses for the population as a whole using comparison with literature galaxy stellar mass functions, and also showed them to be in good agreement on a source-by-source basis (in ELAIS-N1, where comparison was possible); Best et al. (in prep.) also find no systematic offset between the \citeauthor{Duncan2021} and the \textit{consensus} stellar masses (although with a scatter of 0.11\,dex for non-AGN, and 0.23\,dex for AGN). As detailed by \citet{Duncan2021}, the photometric redshifts, and hence the derived stellar masses, are found to be reliable at $z < 1.5$ for host-galaxy dominated sources; we therefore restrict our analysis in this section to $z < 1.5$. To avoid biasing our results by stellar mass incompleteness, we restrict analysis to masses above the 90 per cent stellar mass completeness limits estimated by \citet{Duncan2021} for each field separately (given the varying depths of the multi-wavelength data). We estimated the stellar mass completeness limit in each of the five redshift bins as the stellar mass above which a source would be detected over the full redshift bin, and simply removed sources with stellar masses below this completeness limit from the analysis.

\begin{figure*}
    \centering
    \includegraphics[width=0.92\textwidth]{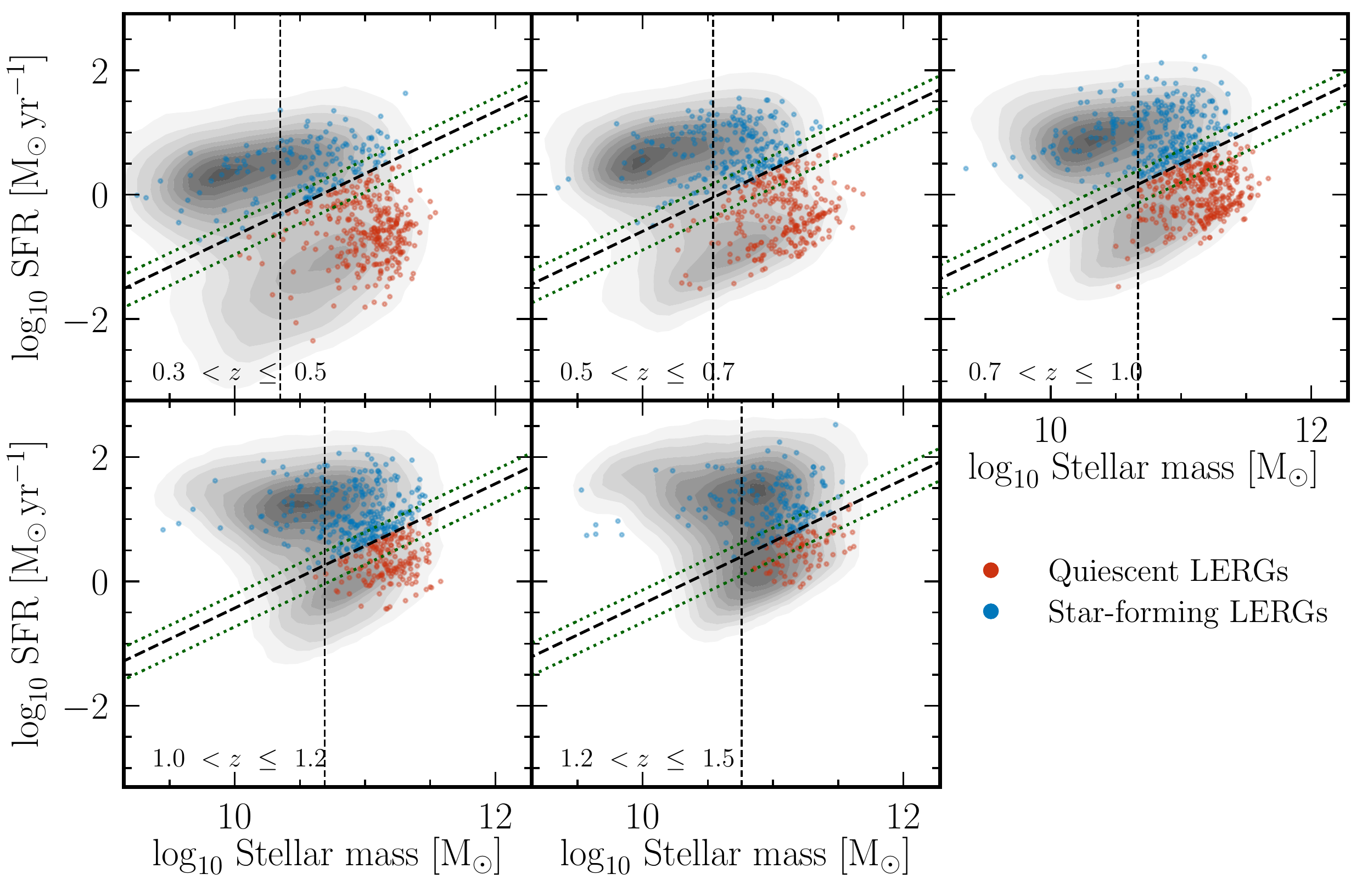}
    \caption{\label{fig:sfr_ms_lerg}The location of LERGs in the ELAIS-N1 field on the SFR-stellar mass (M$_{\rm{\star}}$) plane in five redshift bins ($0.3 < z \leq 1.5)$. The red and blue points correspond to quiescent and star-forming host galaxies of LERGs, respectively. The shaded contours show the SFR-M$_{\rm{\star}}$ distribution (from \textsc{Magphys}) for the underlying IRAC flux-selected sample from \citet{Smith2021} in each redshift bin. The dashed black line corresponds to the sSFR selection threshold adopted for identifying quiescent galaxies (i.e. $f_{\rm{sSFR}} = 1/5$ in equation~\ref{eq:ssfr_q_sel}), as described in Sect.~\ref{sec:agn_frac_quies}. Similarly, the green dotted lines, above and below this line correspond to sSFR thresholds based on $f_{\rm{sSFR}} = 1/3$ and $f_{\rm{sSFR}} = 1/10$, respectively. The vertical lines correspond to the 90 per cent stellar mass completeness limits for each redshift bin in ELAIS-N1; sources below this are removed from the analysis of LERG stellar mass fractions in Fig.~\ref{fig:lerg_frac_quies}.}
\end{figure*}

In Fig.~\ref{fig:lerg_frac_Lbins}, we present the fraction of all galaxies that host a LERG as a function of stellar mass (hereafter; LERG stellar mass fractions) in five redshift bins ($0.3 < z \leq 0.5$, $0.5 < z \leq 0.7$, $0.7 < z \leq 1.0$, $1.0 < z \leq 1.2$, $1.2 < z \leq 1.5$), shown by different coloured lines, for three equally-spaced radio luminosity limits (shown in different panels) from $L_{\rm{150\,MHz}} \geq 10^{24}\,\rm{W\,Hz^{-1}}$ to $L_{\rm{150\,MHz}} \geq 10^{25}\,\rm{W\,Hz^{-1}}$. The lowest radio luminosity limit chosen corresponds to a $5\sigma$ detection (based on noise in the deepest regions) at $z \sim 1.5$, and is broadly comparable to the 1.4\,GHz luminosity limit ($L_{\rm{1.4\,GHz}} = 10^{23}\,\rm{W\,Hz^{-1}}$, assuming $\alpha = -0.7$) typically used in similar studies in the local Universe \citep[e.g.][]{best2005_sdss_xmatch,Janssen2012}. To account for the lack of a volume-limited sample, we weight each source by a 1/V$_{\rm{max}}$ factor based on the maximum volume that a source could be detected in at above $5\sigma$ based on the local radio RMS level. The error bars show Poisson uncertainties (following \citet{Gehrels1986} for stellar mass bins with $\rm{N} < 10$). In total, this results in 2722, 1776, and 835 LERGs (within $0.3 < z \leq 1.5$) with $L_{\rm{150\,MHz}} \geq 10^{24}\,\rm{W\,Hz^{-1}}$, $L_{\rm{150\,MHz}} \geq 10^{24.5}\,\rm{W\,Hz^{-1}}$, and $L_{\rm{150\,MHz}} \geq 10^{25}\,\rm{W\,Hz^{-1}}$, respectively.

For the high radio-power LERGs, corresponding to a radio luminosity limit of $10^{25}\,\rm{W\,Hz^{-1}}$ (right panel of Fig.~\ref{fig:lerg_frac_Lbins}), we find that the fraction of galaxies that host a LERG has a steep dependence on the stellar mass, approaching $\sim$ 10 per cent at the highest stellar masses; these results agree with observations in the local Universe \citep[e.g.][]{Best2005fagn,Janssen2012} and show no signs of redshift evolution. As the radio luminosity limit is lowered (i.e. going from the right panels to left in Fig.~\ref{fig:lerg_frac_Lbins}), we observe an increase in the fraction of LERGs at lower stellar masses at all redshifts, and in particular at higher redshifts, resulting in a shallower dependence of the overall LERG fraction with stellar mass. Physically, the strong dependence on stellar mass is expected from arguments based on the fuelling mechanism of the LERGs \citep[e.g.][]{Best2006,Hardcastle2007}. The flattening of this relation, in particular at lower radio luminosities and stellar masses where star-forming galaxies are expected to dominate the galaxy stellar mass function, suggests the presence of a star-forming LERG population with a potentially different fuelling mechanism. We investigate this in more detail in Sect.~\ref{sec:agn_frac_quies} by considering the dependence of AGN activity on both stellar mass and star-formation rate.

\subsection{Prevalence of LERGs in star-forming and quiescent galaxies}\label{sec:agn_frac_quies}
It is well known that star-forming galaxies occupy a well-defined sequence in the SFR-M$_{\rm{\star}}$ plane, known as the `main-sequence' of star-formation \citep[e.g.][]{Whitaker2012,Speagle2014,Schreiber2015,Tasca2015}, with quiescent galaxies lying below this sequence. Therefore, the ratio of the star-formation rate and the stellar mass of a galaxy, known as the \textit{specific} star-formation rate (sSFR) can be used to identify quiescent galaxies \citep[e.g.][]{Fontana2009,Pacifici2016,Merlin2018}. In this study, using the \textit{consensus} SFRs and stellar masses derived from SED fitting (see Sect.~\ref{sec:sed_fits}), we select sources as quiescent galaxies if they satisfy the condition
\begin{equation}\label{eq:ssfr_q_sel}
    \mathrm{sSFR} < f_{\rm{sSFR}}/t_{\rm{H}}(z),
\end{equation}
where $t_{\rm{H}}(z)$ is the age of the Universe at redshift $z$ of the source, and $f_{\rm{sSFR}} = 1/5$ defines the threshold in sSFR. Galaxy samples selected by this criterion have been found to show good agreement with traditional UVJ rest-frame colour selected samples \citep[e.g.][]{Pacifici2016,Carnall2018,Carnall2020}.

\begin{figure*}
    \centering
    \includegraphics[width=0.92\textwidth]{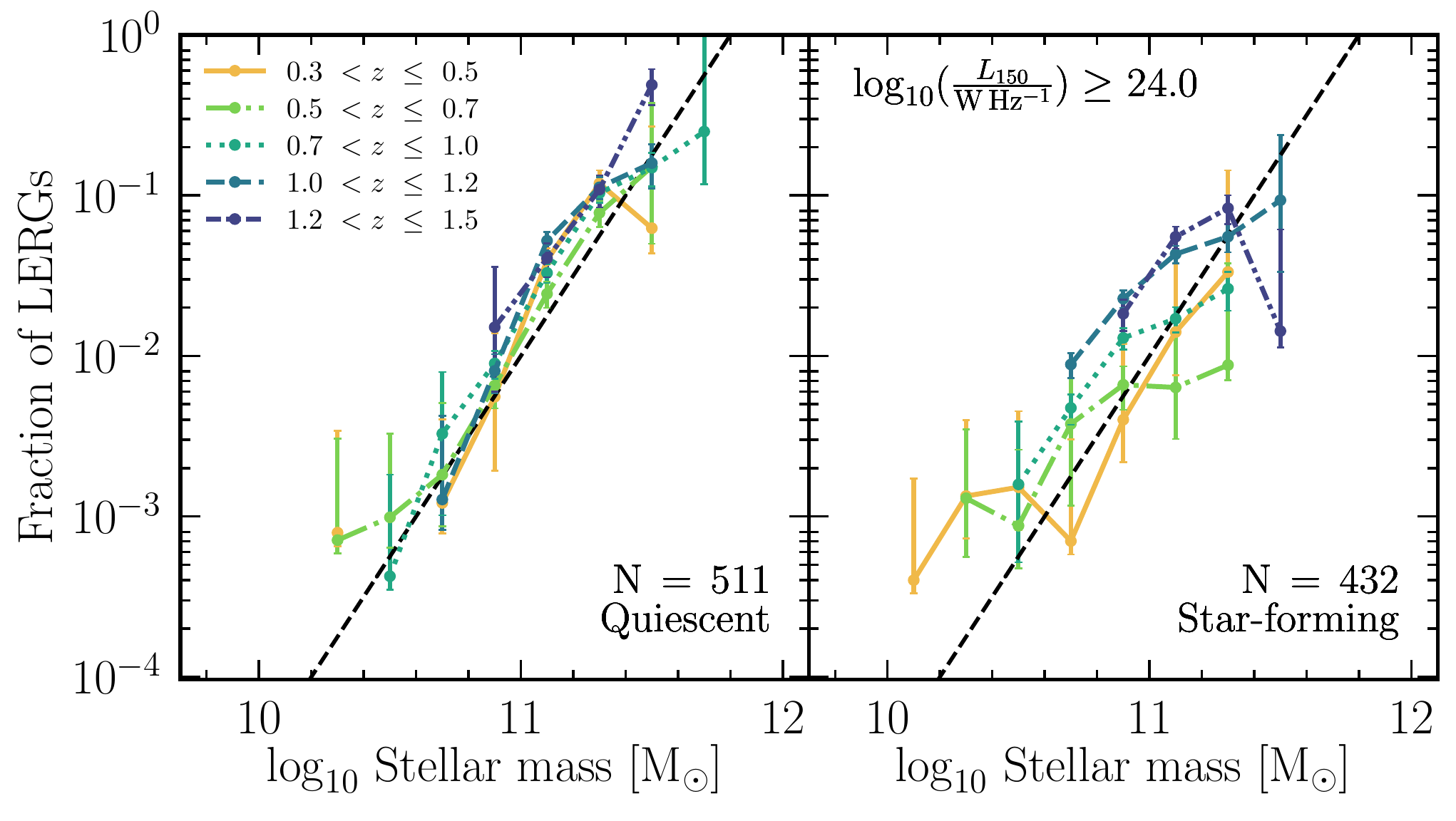}
    \caption{\label{fig:lerg_frac_quies}Fraction of quiescent galaxies that host a LERG (\textit{left}), and the fraction of star-forming galaxies that host a LERG (\textit{right}) as a function of stellar mass with $L_{\rm{150\,MHz}} \geq 10^{24}\,\rm{W\,Hz^{-1}}$ in ELAIS-N1. The number in each panel lists the number of quiescent and star-forming LERGs, respectively, within $0.3 < z \leq 1.5$, for the given radio luminosity limit. The error bars represent Poisson uncertainties, following \citet{Gehrels1986} for $\rm{N} < 10$. The black dashed line shows the $f_{\rm{LERG}} \approx 0.01 (M_{\star}/10^{11}\,\rm{M_{\odot}})^{2.5}$ relation found from studies of LERGs in the local Universe; the normalisation of this relation however depends on the radio luminosity limit, and hence also on the assumed spectral index. We find that the quiescent LERG stellar mass fractions show a steep stellar mass dependence with essentially no evolution with redshift out to $z \sim 1.5$. The star-forming LERGs show a much shallower dependence on stellar mass, with an increase in the fraction at higher redshifts.}
\end{figure*}

\begin{figure*}
    \centering
    \centering
    \begin{subfigure}{0.5\textwidth}
    \centering
    \includegraphics[width=\textwidth]{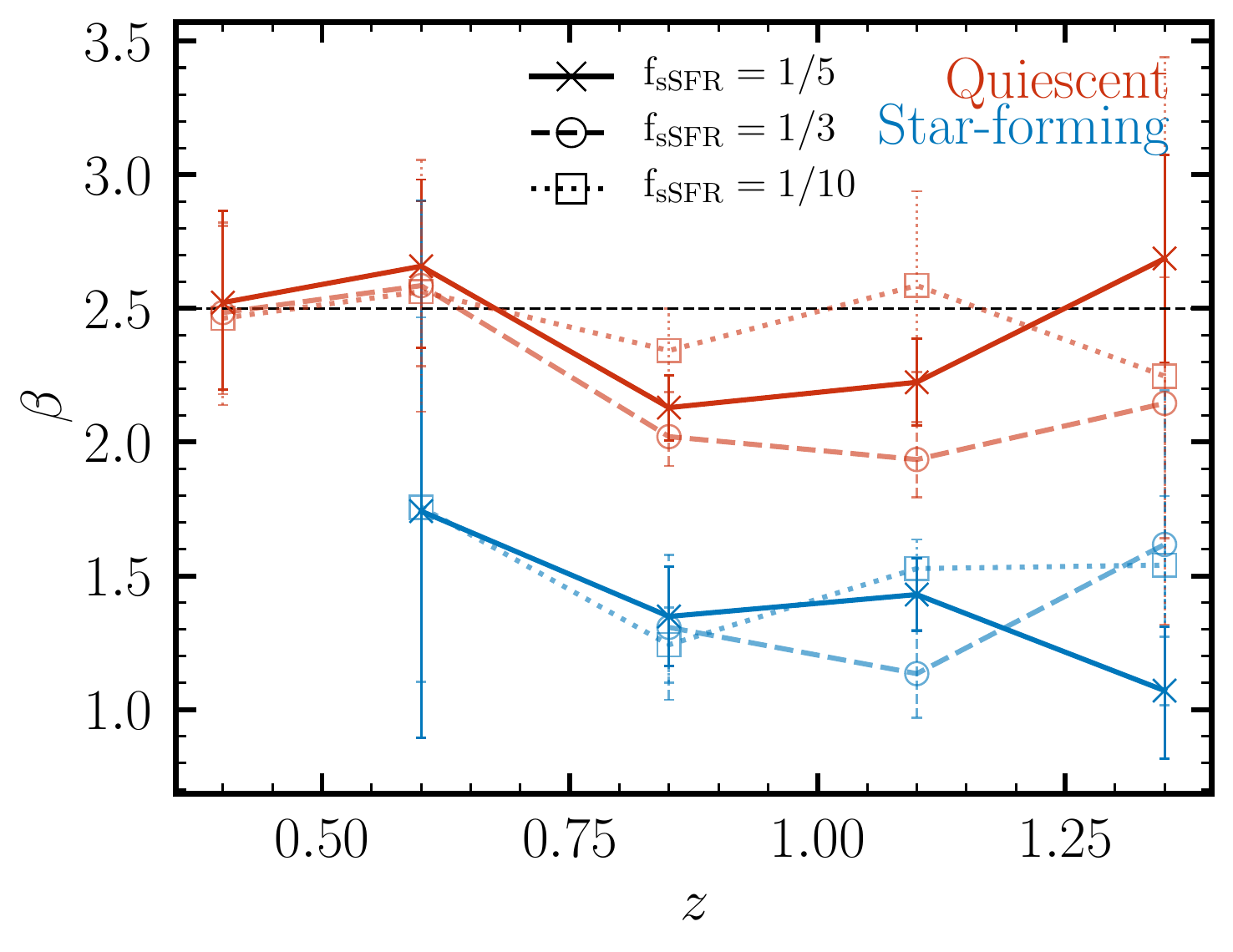}
    \end{subfigure}%
    \begin{subfigure}{0.5\textwidth}
    \includegraphics[width=0.983\textwidth]{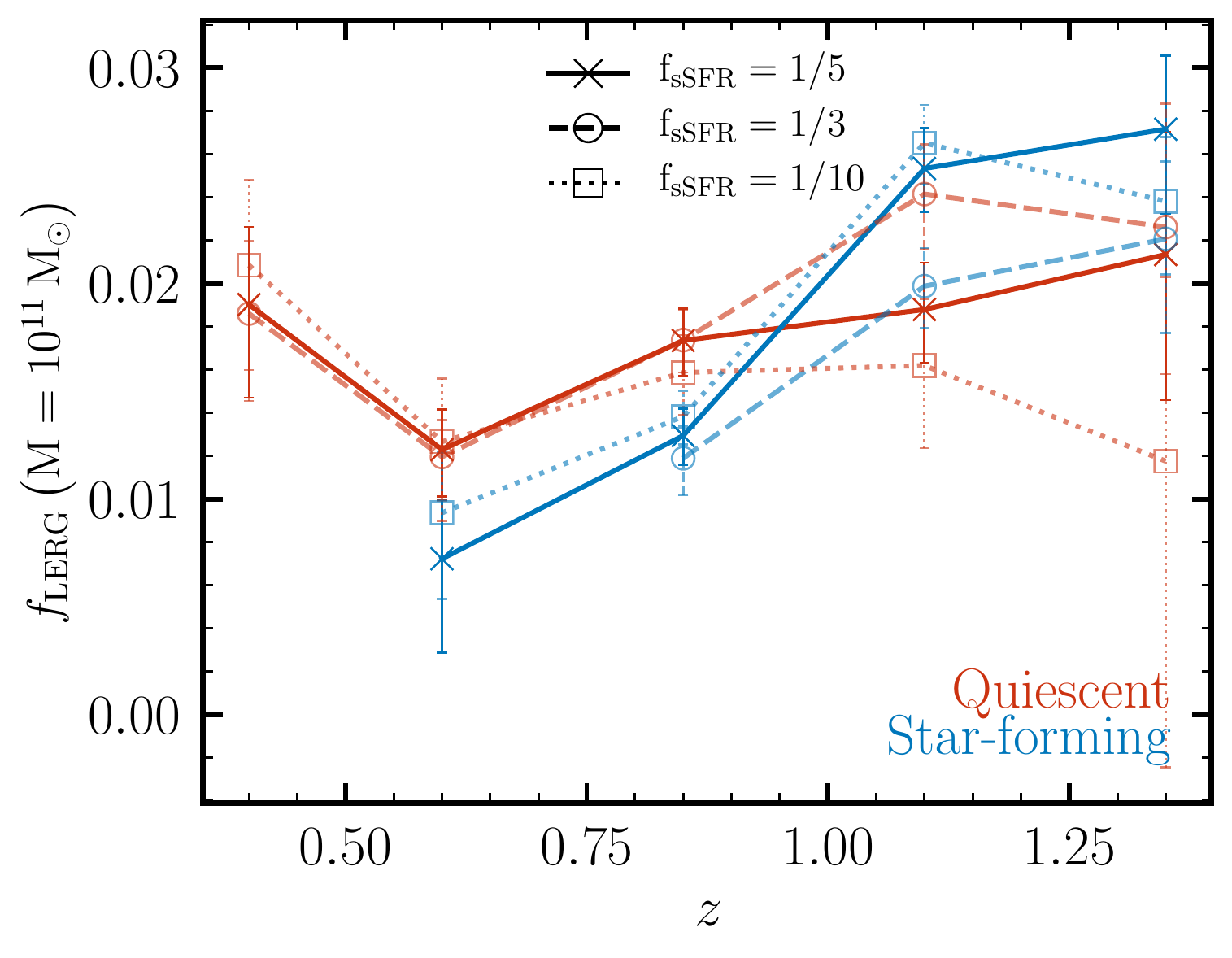}
    \end{subfigure}
    \caption{\label{fig:plaw_fit_agn_frac} Results of power-law fits to the star-forming and quiescent LERG stellar mass fractions with a functional form given by equation~\ref{eq:plaw_agn_frac}. \textit{Left} panel shows the evolution of the power-law slope for quiescent (red lines) and star-forming (blue lines) hosts and the \textit{right} panel similarly shows the evolution in the normalisation of the LERG stellar mass fractions at $M_{\star} = 10^{11}\,\rm{M_{\odot}}$. The solid line (with crosses) corresponds to the results obtained from using the $f_{\rm{sSFR}} = 1/5$ threshold adopted throughout this paper for separating star-forming and quiescent galaxies (see Sect.~\ref{sec:agn_frac_quies}). The dashed and dotted lines correspond to results obtained from using variations of the $f_{\rm{sSFR}}$ threshold adopted. The black horizontal line in the \textit{left} panel corresponds to the slope obtained in the local Universe \citep[e.g.][]{Best2005fagn,Janssen2012}. We find a clear difference in the slope of the LERG stellar mass fractions between the quiescent and star-forming hosts of LERGs. The steep power-law slope of the quiescent LERGs agrees well with the local Universe measurement at all redshifts, consistent with these AGN being fuelled from the hot gas haloes in massive galaxies; the flatter relation for the star-forming LERGs suggests that a different fuelling mechanism, associated with accretion at low levels from the cold gas within these galaxies, may be present. Confidence for this link with the cold gas also comes from the increase in the normalisation of the LERG stellar mass fractions (right panel) for this sub-group at earlier times when gas fractions were higher.}
\end{figure*}

Fig.~\ref{fig:sfr_ms_lerg} shows the SFR-M$_{\rm{\star}}$ relation for the LERG sample in ELAIS-N1 in the same five redshift bins as in Fig.~\ref{fig:lerg_frac_Lbins} ($0.3 < z \leq 1.5$), split into those hosted by quiescent and star-forming galaxies (red and blue points, respectively), separated using the criterion from equation~\ref{eq:ssfr_q_sel}. The black dashed line corresponds to the $\mathrm{sSFR} < 0.2 / t_{\mathrm{H}}(z)$ threshold used to separate the two populations; this results in 1202 quiescent and 1113 star-forming LERG host galaxies within $0.3 < z < 1.5$. We note that the vast majority of the star-forming LERGs lie on or below the main-sequence. An investigation of the rest-frame (\textit{u - r}) colours (uncorrected for dust reddening) reveals that the median colours for both the star-forming and quiescent LERGs are consistent with being close to the green valley \citep[e.g.][]{Schawinski2014}, as is also evident from Fig.~\ref{fig:sfr_ms_lerg}. Recent work by \citet{Mingo2022} found a tail of LERGs with high sSFRs based on a study of the resolved radio-loud AGN within LoTSS-Deep; this is consistent with our study that finds the existence of a significant population of LERGs hosted by star-forming galaxies.

The green dotted lines above and below this dividing line correspond to $f_{\rm{sSFR}} = 1/3$ and $f_{\rm{sSFR}} = 1/10$, respectively, in equation~\ref{eq:ssfr_q_sel}; we use these variations on the standard sSFR threshold to test the robustness of our results on the prevalence of LERGs (see below). Vertical dashed lines show the stellar mass completeness limits in ELAIS-N1 in each redshift bin that are applied when generating the LERG stellar mass fraction plot. The grey shaded contours in each redshift bin show the SFR-M$_{\star}$ relation for the underlying parent galaxy population in ELAIS-N1; these are drawn such that the outermost contour encompasses 99 per cent of all the underlying sources. For this underlying population, we use the F$_{\rm{3.6\,{\mu}m}} > 10\,\mu$Jy IRAC selected sample of \citet{Smith2021} in ELAIS-N1, consisting of 183\,399 sources for which SED fitting was performed using \textsc{magphys}. We limit the following analysis to the ELAIS-N1 field as a similar \textsc{magphys} SED fitting output for the underlying population is not available in the other two fields. Limiting to the sources with the best available multi-wavelength coverage, good $\chi^{2}$ SED fits, and within the chosen redshift range (i.e. $0.3 < z \leq 1.5$) results in a final sample of 140\,754 sources in the underlying population that are used in this analysis. We separate this underlying population into star-forming and quiescent galaxies using the same sSFR threshold as that used for the LERG population. As evident from the panels in Fig.~\ref{fig:sfr_ms_lerg}, the $f_{\rm{sSFR}} = 1/5$ threshold is found to be an appropriate division between star-forming and quiescent galaxies at all redshifts.

Fig.~\ref{fig:lerg_frac_quies} shows the fraction of quiescent galaxies (of a given mass) that host a LERG (\textit{left}), and similarly, the fraction of star-forming galaxies hosting a LERG (\textit{right}), for the same five redshift bins as in Fig.~\ref{fig:lerg_frac_Lbins}, with a radio luminosity limit of $L_{\rm{150\,MHz}} \geq 10^{24}\,\rm{W\,Hz^{-1}}$. We note that this radio luminosity limit is not used to define the LERG sample, but rather the luminosity range over which the LERG stellar mass fraction analysis is carried out. The black dashed line also plotted in each panel shows the $f_{\rm{LERG}} \approx 0.01 (M_{\star}/10^{11}\,\rm{M_{\odot}})^{2.5}$ relationship found for the prevalence of jet-mode AGN with stellar mass in the local Universe \citep{Best2005fagn}. The exact value of the normalisation in this relationship depends on the luminosity limit and hence on the spectral index adopted; the line drawn in Fig.~\ref{fig:lerg_frac_quies} is for illustration purposes only. The fraction of quiescent galaxies hosting a LERG agrees well with this steep stellar mass dependence found in the local Universe, showing essentially no evolution with redshift. In contrast, the LERG stellar mass fractions for the star-forming host galaxies show a weaker dependence on stellar mass, with this relation evolving with redshift such that the fraction increases with increasing redshift at a fixed stellar mass, with the prevalence of these LERGs in star-forming hosts reaching comparable levels to the quiescent hosts at high redshifts.

To quantitatively investigate the trends in the evolution of the LERG stellar mass fractions seen in Fig.~\ref{fig:lerg_frac_quies}, we parametrised the LERG stellar mass fractions as a power-law of the form
\begin{equation}\label{eq:plaw_agn_frac}
    f_{\rm{LERG}}(M_{\star}) = c \left( \frac{M_{\star}}{10^{11}\,\rm{M_{\odot}}}\right)^{\beta}
\end{equation}
such that $f_{\rm{LERG}}(M_{\star})$ is the LERG fraction at mass $M_{\star}$, $c$ is the normalisation at $M_{\star} = 10^{11}\,\rm{M_{\odot}}$, and $\beta$ is the power law slope. We then fitted this power law form to the LERG stellar mass fractions shown in Fig.~\ref{fig:lerg_frac_quies} at each redshift individually, for both the star-forming and quiescent LERG populations.

Fig.~\ref{fig:plaw_fit_agn_frac} shows the results of this power-law fitting process for the quiescent (red) and star-forming (blue) LERG populations in the same five redshift bins used in Fig.~\ref{fig:lerg_frac_quies}. The \textit{left} panel of Fig.~\ref{fig:plaw_fit_agn_frac} shows the redshift evolution of the power-law slope, $\beta$, and \textit{right} panel shows the redshift evolution of the normalisation at $M_{\star} = 10^{11}\,\rm{M_{\odot}}$ for both populations. The error bars show the $1\sigma$ uncertainties in the fitted parameters. We omit data points where the uncertainties are of order or larger than the magnitude of the fitted parameters due to a poor fit; this in particular affected the lowest redshift bin of the star-forming subset due to large uncertainties in the LERG stellar mass fractions. In each panel, the solid lines correspond to $f_{\rm{sSFR}} = 1/5$, our adopted threshold (see Sect.~\ref{sec:agn_frac_quies} for details) for separating star-forming (blue) and quiescent (red) host-galaxies used throughout this study. Also shown are the results from fitting the LERG stellar mass fractions that were derived based on variations of the $f_{\rm{sSFR}}$ selection thresholds equal to $1/3$ and $1/10$ (see also Fig.~\ref{fig:sfr_ms_lerg}). The black dashed line in the \textit{left} panel corresponds to the power-law slope ($\beta = 2.5$) found in studies of jet-mode AGN population in the local Universe \citep[e.g.][]{Best2005fagn,Janssen2012}.

We find that the quiescent and star-forming LERG populations show distinct power-law slopes, with the quiescent LERG stellar mass fractions having a steep slope, close to the local redshift relation, that remains roughly constant with redshift. Moreover, if a stricter definition of `quiescence' is used (i.e. $f_{\rm{sSFR}} = 1/10$; dotted line), we find that the slope of our quiescent LERG population at higher redshifts agrees even better with the local relation. The star-forming LERG stellar mass fractions show a much shallower slope, which is interestingly similar to the relation of the radiative-mode AGN population in the local Universe \citep[e.g.][]{Janssen2012}. As is evident from Fig.~\ref{fig:plaw_fit_agn_frac}, the LERG stellar mass fractions resulting from the two alternative selection thresholds for quiescent/star-forming galaxies agree well with the values derived from our adopted `quiescence' criteria (within $1\sigma$ based on our uncertainties); this demonstrates that our results are robust to changes in how quiescent and star-forming host-galaxies are selected.

Looking at the evolution of normalisation of the power-law at $M = 10^{11}\,\rm{M_{\odot}}$ in Fig.~\ref{fig:plaw_fit_agn_frac} (\textit{right}), we find that for the quiescent LERGs, the normalisation stays roughly constant out to $z \sim 1.5$. In contrast, the star-forming subset shows a strong increase in the normalisation of the LERG stellar mass fraction at higher redshifts, increasing by a factor of $\sim$ 4 by $z \sim 1$, showing hints of higher prevalence at $M = 10^{11}\,\rm{M_{\odot}}$ compared to the quiescent hosts at these redshifts.

We note that our use of a radio-excess $> 3\sigma$ criterion may miss low-luminosity radio-AGN, hosted in star-forming galaxies in particular, if the star-formation rate is sufficiently high that a jet with a power close to our adopted radio luminosity limit does not produce a significant radio excess. This may be particularly relevant in massive galaxies, due to the correlation between mass and star-formation rate for star-forming galaxies. Therefore this may impact the results determined in Fig.~\ref{fig:lerg_frac_quies}, leading to an artificial flattening of the relation with stellar mass for the star-forming LERGs. We have investigated applying a higher radio-luminosity limit, $L_{\rm{150\,MHz}} \geq 10^{25}\,\rm{W\,Hz^{-1}}$ (with broader redshift bins $0.5 < z \leq 1$ and $1 < z \leq 1.5$ to account for fewer sources); this ensures a complete sample as radio jets with such high powers will dominate over any star-formation, and thus be selected as radio-excess AGN. Using this higher radio luminosity limit does result in a steeper relation with stellar mass for the star-forming LERGs. We attempted to fit a power law to these star-forming LERG stellar mass fractions, however, such a high radio luminosity limit removes a large fraction of the LERGs from the sample, leading to poorer statistics and hence we were unable to converge on a fit for the lowest redshift bin. For the $1 < z \leq 1.5$ bin, we find a power law slope $\beta = 1.37 \pm 0.57$, which is $\sim 2\sigma$ away from the power law slope of $\beta = 2.5$ found for the quiescent LERGs in the local Universe. This gives us further confidence that the shallower stellar mass dependence observed for the star-forming LERGs is not purely driven by a luminosity selection effect. It is also important to note that this limit only selects LERGs with the most powerful jets, rather than the ``typical'' LERG population which we are more interested in. Furthermore, in future work, more robust measurements of the star-forming LERG population will be enabled by upcoming datasets from the WEAVE-LOFAR survey \citep{2016sf2a.conf..271S}, which will allow a selection of radio-AGN using emission line diagnostics, overcoming the limitations of the radio-excess selection, and data from sub-arcsecond LOFAR imaging \citep[e.g.][]{Sweijen2022} which will enable more robust identification of jetted AGN hosted by star-forming galaxies.

\subsection{The nature of star-forming and quiescent LERGs}\label{sec:agn_frac_disc}
Physically, the differences observed between the quiescent and star-forming hosts of LERGs suggests that the LERGs in these galaxies are fuelled by different mechanisms. For the quiescent LERGs, we observe a steep stellar mass dependence on the LERG fraction, similar to that found in the local Universe \citep[e.g.][]{Best2005fagn,Janssen2012}, where the LERG population is dominated by red, quiescent host galaxies. This strong dependence of radio-AGN activity with stellar mass for the LERG population has been shown to be coupled to the cooling rate of hot gas from haloes in massive elliptical galaxies, supporting the idea that these LERGs are fuelled by the hot phase of the intergalactic medium \citep[e.g.][]{Allen2006,Best2006,Hardcastle2007}. Since the observed power-law slope (and normalisation) for this quiescent LERG population remains roughly constant out to $z \sim 1.5$, this suggests that the higher-redshift quiescent LERG population is also fuelled by accretion from the hot medium, as in the local Universe, and that this mechanism has been in place at least since $z \sim 1.5$.

In contrast, the LERG activity in star-forming galaxies shows a much flatter dependence with stellar mass, suggesting that a different physical mechanism may be responsible for fuelling the AGN in these galaxies. Given that this population is undergoing a recent episode of star-formation, likely fuelled by the availability of cold gas (e.g. via mergers or secular processes), we may expect this cold gas to provide a fuel source for the black hole as well. Cold gas accretion is often thought to lead to a radiative-mode AGN (which are widely associated with star-forming galaxies), but these star-forming LERGs are understood to be fuelled by a radiatively-inefficient accretion flow; hence, the cold gas associated with the on-going star-formation must also be capable of fuelling the black hole. This is consistent with the idea that it is the Eddington-scaled accretion rate onto the black hole that determines the accretion mode (i.e. radiatively efficient/inefficient mode) and not necessarily the fuel source, such that it is possible to achieve accretion of cold material at low Eddington rates, resulting in a radiatively inefficient mode of AGN \citep[e.g. see discussion by][]{2014MNRAS.445..955B,Hardcastle2018,Hardcastle2020}. In such a scenario, we would expect the HERGs to be the higher accretion rate extension of this star-forming LERG population.

Other evidence in support of this picture comes from the observation of enhanced LERG activity in interacting galaxies \citep[e.g.][]{Sabater2013,Gordon2019} and that in lower-mass galaxies  ($M_{\star} \lesssim 10^{11}\,\rm{M_{\odot}}$), the merger fraction of LERGs is higher than that of a control sample (by $\sim 20$ per cent; \citealt{Gao2020}); these results could be due to the supply of cold gas brought in to the galaxy due to the interaction/merger events. Indeed there is evidence for large reservoirs of cold molecular gas ($\sim 10^{8} - 10^{10}\,\rm{M_{\odot}}$) in the host galaxies of LERGs \citep[e.g.][]{Okuda2005,OcanaFlaquer2010,Ruffa2019}, along with evidence for the cold gas playing a role in the fuelling of the AGN in some systems \citep[e.g.][]{Tremblay2016,Maccagni2018,Ruffa2019b}. However, it is also worth noting that while the presence of cold gas within the host galaxy is associated with increased AGN activity \citep[e.g.][]{Vito2014,Aird2019,Carraro2020}, it does not necessarily imply that this gas reaches and accretes on to the central engine; the cold gas in some systems is mostly found in a relaxed rotating disc \citep[e.g.][]{North2019} and as such would be expected to lead to only relatively low accretion rates.

\begin{figure*}
    \centering
    \includegraphics[width=\textwidth]{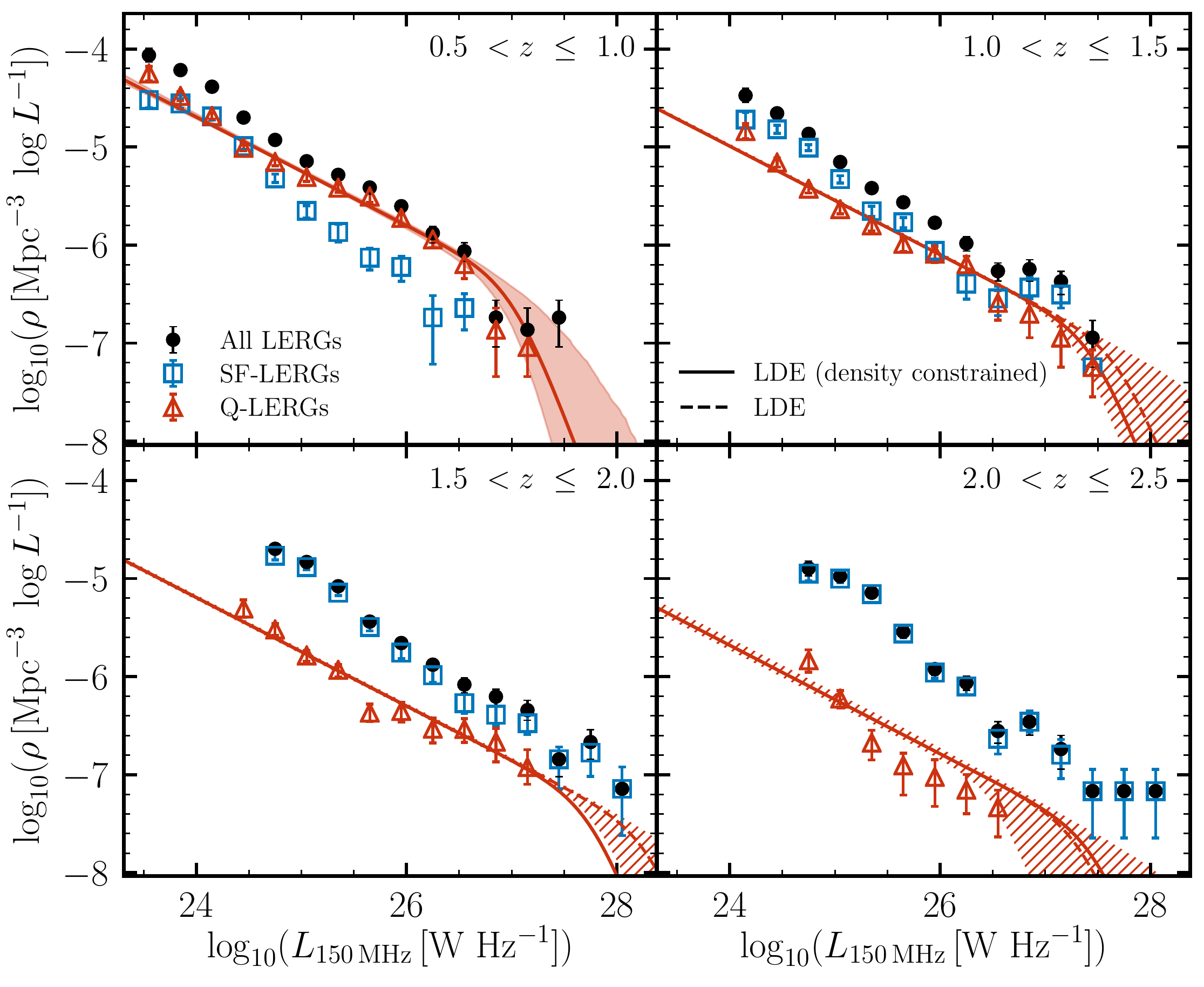}
    \caption{\label{fig:lf_evol_lerg_quies}The evolving LFs for the LERGs in the combined LoTSS Deep Fields split into the subset hosted by quiescent and star-forming galaxies for the same redshift bins as in Fig.~\ref{fig:lf_evol_lerg}. The space densities of the quiescent LERGs dominate the total LERG population at $z < 1$ and decrease steadily with redshift, however the star-forming LERGs begin to dominate beyond $z \sim 1$; the combination of the evolution of these two sub-groups explains the evolution of the total LERG population seen in Sect.~\ref{sec:cosevol_lerg_lf}. For the quiescent LERGs, we show a broken power law fit to the data in the $0.5 < z \leq 1$ bin, with the shaded region forming the $1\sigma$ uncertainty on the best fit. At higher redshifts, the dashed line and hatched regions represent the best-fit and $1\sigma$ uncertainties from a 2-parameter LDE fit, whereas, the solid line shows an LDE fit where the density evolution is fixed based on the evolution of the expected host galaxies, and can be seen to provide a good match to the data within the $1\sigma$ region at each redshift.}
\end{figure*}

The present data are not sufficient for us to determine the physical processes underpinning the AGN fuelling and the direct link with the cold gas; instead detailed characterisation of the multi-phase medium and its kinematics for matched samples of the three sub-groups of the radio-excess AGN are required. Nonetheless, the trends of shallower stellar mass dependence of the star-forming LERG stellar mass fractions (compared to the quiescent LERGs) and the increase in their normalisation at earlier times is consistent with the idea that a different mechanism is responsible for fuelling the LERG activity in these more star-forming galaxies, likely associated with the cold gas component.

\section{The luminosity functions of the quiescent LERGs}\label{sec:lerg_lfs_Q_SF}
Given that we see this significant population of LERGs being hosted in star-forming galaxies at higher redshifts in Sect.~\ref{sec:agn_frac}, it is informative to study how the luminosity functions of both of these populations evolve with redshift in order to understand the evolution of the LERG population.

Fig.~\ref{fig:lf_evol_lerg_quies} shows the evolution of the total LERG LFs (black points, as shown in Fig.~\ref{fig:lf_evol_lerg}), but also split into the subsets that are hosted by quiescent (red triangles) and star-forming (blue squares) galaxies. The error bars represent the $1\sigma$ uncertainties with the LFs also tabulated in Table~\ref{tab:lerg_evol_table}. Across the four redshift bins, the LFs are constructed in total for 3406 quiescent and 5024 star-forming LERGs. We find that LERGs hosted by star-forming and quiescent galaxies, are both found across all radio luminosities. At $z < 1$, the total LERG population is primarily hosted in quiescent galaxies at almost all radio luminosities, with the star-forming population reaching comparable space densities only at the faint end. At higher redshifts, we find a steady decline in the space density of the quiescent LERGs; simultaneously, the space density of the star-forming subset increases steadily with redshift, with a switch-over in the dominant population occurring by $z \sim 1$. This star-forming LERG population dominates over the quiescent LERGs across all radio luminosities at $z > 1.5$. We note that this trend is not simply due to an increase in the number of star-forming galaxies at early epochs as our analysis of the LERG stellar mass fractions (see Fig.~\ref{fig:lerg_frac_quies} and Fig.~\ref{fig:plaw_fit_agn_frac}), which accounts for this effect, shows an increase in the prevalence of this population too. The combination of the evolution of these two sub-groups can thus explain the little evolution seen in the total LERG LFs (see Sect.~\ref{sec:cosevol_lerg_lf}).

We also considered how uncertainties in the separation of LERGs and HERGs affect the LFs of the quiescent and star-forming LERGs; this analysis is detailed in Appendix~\ref{sec:scvary}. In summary, we modified and applied additional criteria for selecting `SED AGN' (i.e. those that show signs of MIR emission from an AGN, typical for HERGs) and reconstructed the LFs of the HERGs and the two sub-groups of LERGs. We found that at all redshifts, the quiescent LERGs are a robust population, largely unaffected by the exact threshold used to separate HERGs from the LERGs. Similarly, the star-forming LERGs also show little change in their LFs, even for quite extreme changes in classification criteria, with changes being largely consistent within $2\sigma$ of the errors.

Here, we focus on modelling the evolution of the quiescent LERG LFs as this population is particularly important for radio-mode feedback considerations. To do this, we first fitted the $0.5 < z \leq 1.0$ quiescent LERG LFs as a broken power law (with all four free parameters), with the best-fitting model shown as a solid red line in the $0.5 < z \leq 1$ panel of Fig.~\ref{fig:lf_evol_lerg_quies} and the shaded region representing the $1\sigma$ uncertainties on the best fit. The fitted parameters are also tabulated in Table~\ref{tab:lf_fits_quies}.

\begin{figure}
    \centering
    \includegraphics[width=\columnwidth]{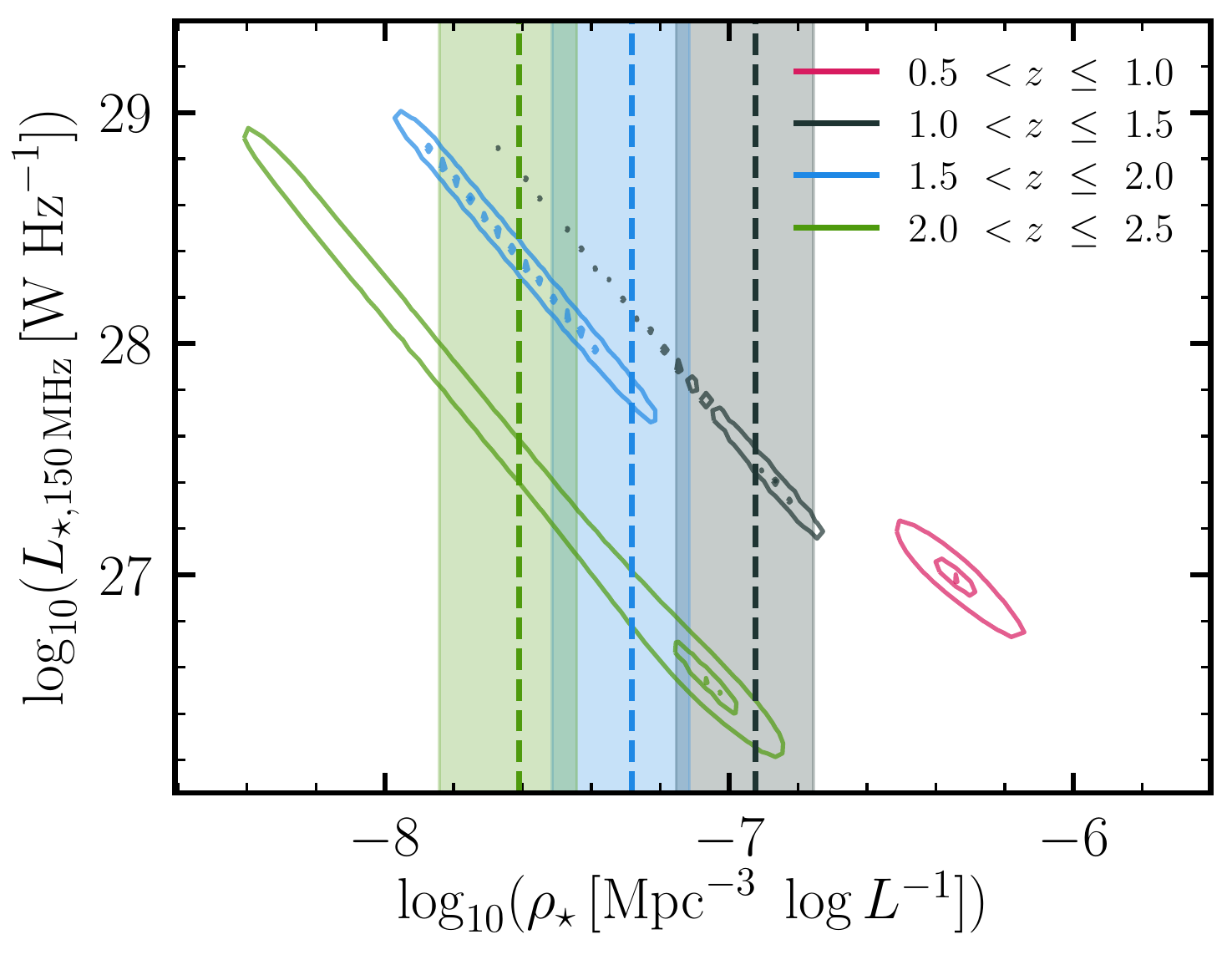}
    \caption{\label{fig:qlerg_lde_contour}2D contour plot showing the posterior distribution of $\rho_{\star}(z)$ and $L_{\star}(z)$ for the quiescent LERGs. For $0.5 < z \leq 1$, the contours are based on a broken power law fit, while at higher redshifts, the contours show the posterior distributions from a 2-parameter LDE fit to the LFs. The large degeneracy in these parameters can be better constrained when the evolution of $\rho_{\star}(z)$ is fixed based on the evolution of the expected host galaxies of quiescent LERGs as shown by the dashed lines (see Fig.~\ref{fig:host_density_evol}); based on this estimate, the shaded vertical regions correspond to the $1\sigma$ uncertainties on the evolved $\rho_{\star}(z)$.}
\end{figure}

\begin{table}
\caption{Results of modelling the evolution of the quiescent LERGs. We fitted a broken power law to the $0.5 < z \leq 1$ LF. Using this broken power law fit (i.e. constant slopes), the LF at higher redshifts is modelled with a luminosity and density evolution. To reduce the degeneracy between parameters, the evolution of $\rho_{\star}(z)$ was fixed based on the evolution of the expected host galaxies of the quiescent LERGs (see Fig.~\ref{fig:host_density_evol}), with $L_{\star}(z)$ fitted as a free parameter. The best-fitting models are also shown in Fig.~\ref{fig:lf_evol_lerg_quies}.\label{tab:lf_fits_quies}}
\centering
\resizebox{0.48\textwidth}{!}{
\begin{tabular}{cccccc}
\hline\hline
$z$ & $\log_{10} \rho^{\star}(z)$ & $\log_{10} L_{\star}(z)$ & $\beta$ & $\gamma$ & $\chi^{2}_{\nu}$\Bstrut\\
\hline
${0.5} < z \leq {1.0}$ & $-6.37_{-0.23}^{+0.17}$ & $27.03_{-0.22}^{+0.39}$ & $-2.88_{-1.76}^{+1.11}$ & $-0.55_{-0.03}^{+0.03}$ & 3.57 \Tstrut\\[0.15cm]
${1.0} < z \leq {1.5}$ & -6.92 & $27.33_{-0.04}^{+0.04}$ & {...} & {...} & 3.39\\[0.15cm]
${1.5} < z \leq {2.0}$ & -7.28 & $27.91_{-0.06}^{+0.06}$ & {...} & {...} & 2.28\\[0.15cm]
${2.0} < z \leq {2.5}$ & -7.61 & $27.93_{-0.12}^{+0.12}$ & {...} & {...} & 5.89\\[0.15cm]
\hline
\end{tabular}
}
\end{table}

Fitting all parameters for a broken power law independently at higher redshifts for the quiescent LERGs is not possible due to the lack of any strong break in the LFs\footnote{Interestingly, in Fig.~\ref{fig:lf_evol_lerg_quies}, it can be seen that all redshifts, the LF could potentially be fit better with a shallower faint-end slope, with then an upturn in the space densities of the quiescent LERGs at the faintest luminosities. At the lowest redshift bin ($0.5 < z \leq 1$), the upturn starts to appear at $L_{\rm{150\,MHz}} \approx 10^{24.5}\,\rm{W\,Hz^{-1}}$; this is typically below the limit probed by previous surveys. If this upturn is real, then it might suggest that a double Schechter function may be more appropriate than a broken power law for modelling the luminosity functions of the quiescent LERGs in particular. However, the present dataset does not have enough data points at the faint end beyond the upturn to provide constraints on such a model; deeper radio data, for example from subsequent data releases of the LoTSS Deep Fields, will allow us to better sample the faint end of the quiescent LERGs.}. Instead, we fixed the two power law slopes at the values derived from the fit at $0.5 < z \leq 1$, and adopted an LDE model by fitting directly for $\rho_{\star}$ and $L_{\star}$ at each redshift\footnote{We also examined the PLE and PDE models, but neither is able to match the data well: the PLE model cannot explain the strong evolution in space densities seen at low luminosities, whereas the PDE model fails to match the evolution at high luminosities.}. The resulting fitted model and $1\sigma$ uncertainties are shown by the dashed lines and hatched regions in Fig.~\ref{fig:lf_evol_lerg_quies}. Due to the lack of an obvious break in the LFs, there is a large degeneracy between the evolution of $\rho_{\star}$ and $L_{\star}$, as can be visualised in Fig.~\ref{fig:qlerg_lde_contour} which shows a 2D contour plot of the posterior distribution of the LDE model fit in $L_{\star}(z)$ and $\rho_{\star}(z)$ parameter space for different redshift bins. For the $0.5 < z \leq 1$ redshift bin, the contours shown are from the broken power law fit described above.

Under a simple model, we can consider the evolution of the quiescent LERG LFs to be dictated by the availability of their expected host galaxies (i.e. the massive quiescent galaxies; e.g. \citealt{2014MNRAS.445..955B}), which can help reduce the degeneracy of the two-parameter LDE fit. To determine the evolution of the space densities of the hosts of the quiescent LERGs, we follow the method outlined by \citet{2014MNRAS.445..955B}. \citeauthor{2014MNRAS.445..955B} used various determinations of the quiescent stellar mass functions from the literature \citep[from][]{DominguezSanchez2011,baldry2012gama_smf,Moustakas2013,Ilbert2013,Muzzin2013} and combined this with the prevalence of LERGs as a function of stellar mass to predict the total space density of the expected LERG hosts. We use these data compiled by \citet{2014MNRAS.445..955B}, but scale these by a factor of $0.018/0.01$ based on the normalisation of quiescent LERG stellar mass fractions found in this study (where $0.018$ is the mean normalisation at $10^{11}\,\rm{M_{\odot}}$ across $0.3 < z \leq 1.5$, and $0.01$ is the normalisation derived by \citealt{Best2005fagn}, at their 1.4\,GHz radio luminosity limit). We also remove the small scale-factors that \citet{2014MNRAS.445..955B} introduced to line up the different datasets. To this analysis, we add additional points based on a more recent measurement of the quiescent stellar mass function by \citet{McLeod2021}. 

\begin{figure}
    \centering
    \includegraphics[width=\columnwidth]{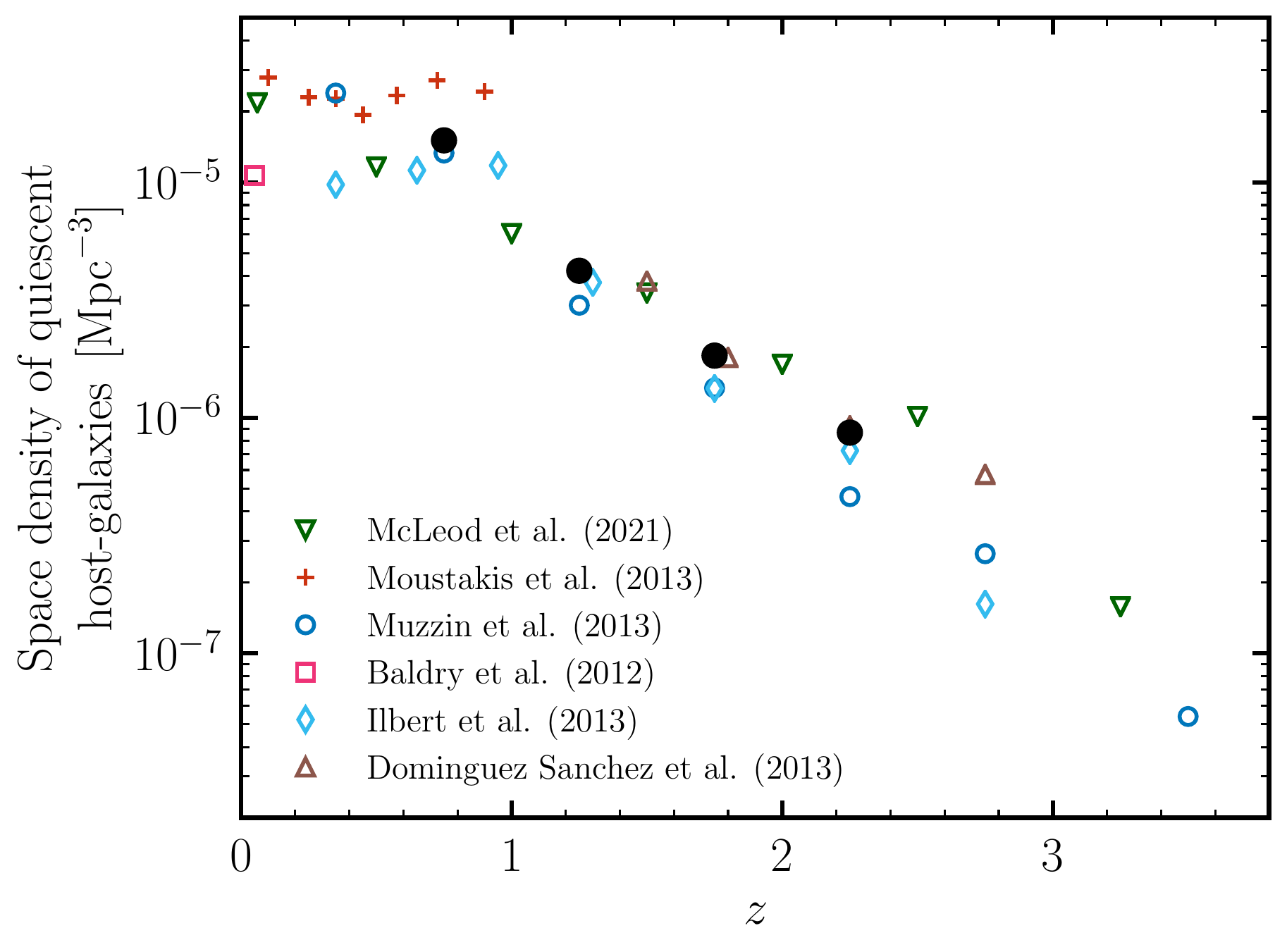}
    \caption{\label{fig:host_density_evol}The evolution of the space density of the expected host-galaxies of the quiescent LERGs from literature datasets at $z < 4$. The \citet{McLeod2021} data points were determined by combining the stellar mass function of quiescent galaxies with the prevalence of quiescent LERGs with stellar mass ($f_{\rm{AGN(M_{\star})}} = 0.018 (M_{\star}/10^{11}\,\rm{M_{\odot}})$; see text) and integrating in stellar mass (down to $M_{\star} = 10^{9}\,\rm{M_{\odot}}$) to obtain the space density of the expected host galaxies. The other data points \citep[from][]{DominguezSanchez2011,baldry2012gama_smf,Moustakas2013,Ilbert2013,Muzzin2013} were complied by \citet[][taken from their fig.~6]{2014MNRAS.445..955B} and scaled to match the normalisation in $f_{\rm{AGN}}\rm{(M_{\star})}$ found in this study (see text). The black filled points show the mean of the datasets obtained from interpolating at the midpoints of our four LF redshift bins.}
\end{figure}

Fig.~\ref{fig:host_density_evol} shows the results of this analysis, where we find that the space density of the available hosts remains relatively flat out to $z \sim 0.75$, beyond which there is a sharp decline with redshift. We then determined the evolution of the space density of the expected hosts at the midpoint of our LF redshift bins (i.e. at $z = 0.75, 1.25, 1.75, 2.25$) by interpolating and computing the mean of the six datasets (shown by black solid points in Fig.~\ref{fig:host_density_evol}). This evolution of the hosts was then used constrain the LDE model by scaling the $\rho_{\star}(z=0.75)$ from the broken power law fit to predict $\rho_{\star}(z)$ at higher redshifts, as shown by the dashed lines in the 2D contour plot in Fig.~\ref{fig:qlerg_lde_contour} for each redshift bin, with the shaded regions representing the uncertainties in $\rho_{\star}(z=0.75)$ from the broken power law fit. To test this explicitly, we used the host density evolution determined above to fix the value of $\rho_{\star}(z)$ at higher redshifts, and fitted directly for $L_{\star}(z)$ only for each of the three $1.0 < z \leq 2.5$ bins. The resulting best-fitting models are shown as red solid lines for the corresponding redshift bins in Fig.~\ref{fig:lf_evol_lerg_quies}; these are found to show a good match to the data, lying within the $1\sigma$ hatched region at each redshift.

These results agree well with studies of the total LERG population out to $z \lesssim 1$ that find a mild evolution \citep[e.g.][]{2014MNRAS.445..955B,Pracy2016}; at these redshifts, the total LERG population is dominated by the quiescent hosts which show little evolution (see Fig.~\ref{fig:host_density_evol}). At higher redshifts, while the total LERG population still shows weak evolution, as also found by \citet[][out to $z < 1.3$]{Butler2019}, the quiescent LERG population declines strongly with increasing redshift in line with the expected host galaxies, while the characteristic luminosity of the sources increases; this is counter-balanced by an increasing star-forming LERG population at higher redshifts.

\vspace{-0.6cm}
\section{Conclusions}\label{sec:conclusions}
In this paper, we have used the first data release of the LoTSS Deep Fields to study the evolution of the radio-AGN population out to $z \sim 2.5$. The LoTSS Deep Fields represents the deepest low-frequency radio continuum survey to date, covering 25\,deg$^{2}$ across the ELAIS-N1, Lockman Hole, and Bo\"{o}tes fields. The depth of the radio and multi-wavelength dataset, in combination with the wide area coverage, makes this an ideal survey to constrain the evolution of the low-luminosity radio-AGN out to high redshifts.

The total radio-AGN luminosity functions were constructed out to $z \leq 2.5$ using a sample of 11\,783 radio-AGN; this shows good agreement across all redshifts with the literature LFs determined from GHz surveys. We separated the radio-AGN into LERGs and HERGs via the identification of accretion disc and torus features in galaxy SEDs; the presence of such signatures is indicative of a HERG. This results in a sample of 10\,481 LERGs and 1302 HERGs across the deep fields (within $z \leq 2.5$), with the LERGs dominating over the HERGs across all luminosities probed by this study. Using this sample, we have then presented the first robust measurement of the cosmic evolution of the LERG luminosity functions out to $z \sim 2.5$. We find relatively mild evolution within the redshift bins ($0.5 < z \leq 2.5$) examined in this study; this evolution is found to be driven by a very different evolution of those LERGs hosted by quiescent galaxies and those hosted by star-forming galaxies. We therefore investigated these two populations separately.

The main results of our study for the quiescent LERGs are as follows:

\begin{itemize}
\item The quiescent LERG LFs dominate the space densities of LERGs across essentially all luminosities at $z \lesssim 0.75$.
\item The incidence of LERG activity in quiescent galaxies shows a strong stellar mass dependence, which can be well-described at all redshifts as $f_{\rm{LERG,Q}} = c (M_{\star} / 10^{11}\,\rm{M_{\odot}})^{\beta}$, with a power-law slope of $\beta \approx 2.5$ and a normalisation of $c \approx 0.018$ (for a luminosity limit of $L_{\rm{150\,MHz}} \geq 10^{24}\,\rm{W\,Hz^{-1}}$). This relation for the quiescent LERGs also shows no evolution with redshift (between $0.3 < z \leq 1.5$).
\item The steep dependence of LERG activity with stellar mass, and the lack of a redshift evolution suggest that the LERGs in these quiescent galaxies are fuelled by cooling of hot gas from haloes, consistent with local Universe observations, at least over the past $\sim 9\,\rm{Gyrs}$.
\item The quiescent LERGs show a strong negative evolution in their LFs beyond $z \sim 0.75$, which is consistent with the characteristic space density evolving in accordance with the availability of the expected host galaxies, while there is also an increase in the characteristic luminosity with increasing redshift.
\end{itemize}

The main results of our study for the star-forming LERGs are as follows:

\begin{itemize}
\item The incidence of LERG activity in star-forming galaxies shows a much weaker stellar mass dependence ($\beta \approx 1.5$ consistently across redshift) than that found for the quiescent LERGs in this study, and for LERGs in the local Universe, and is instead comparable to that found for the HERGs locally.
\item The incidence of LERGs in star-forming galaxies (at fixed stellar mass; i.e. the normalisation $c$) increases by nearly a factor of four over the range $0.3 < z \leq 1.5$, reaching a higher prevalence than that of the quiescent LERGs at the highest redshifts.
\item The star-forming LERGs are a minority of the overall LERG population at $z < 1$, but space densities of the star-forming LERGs increase with increasing redshift, such that they become the dominant population at $z \gtrsim 1.5$ over all radio luminosities.
\end{itemize}

In the local Universe, LERGs have been argued to be primarily fuelled by cooling of hot gas, typically at low accretion rates, from their massive host galaxies, whereas the incidence of HERGs show a weaker stellar mass dependence, attributed to fuelling via cold gas, typically at high accretion rates. In our study, we find a flatter stellar mass dependence of LERGs hosted by star-forming galaxies (broadly similar to the local HERGs), which increase in prevalence at earlier cosmic times, when gas fractions were higher; this suggests that a different fuelling mechanism, likely associated with the cold gas, is responsible for triggering LERG activity within these star-forming galaxies compared to the LERGs in quiescent galaxies. Therefore, it is possible that the radio-AGN activity in the star-forming LERGs are triggered by a similar mechanism to the HERGs in the local Universe, however, the present dataset is not sufficient to fully understand the nature of fuelling of the star-forming LERGs. Detailed characterisation of the molecular gas and other host galaxy properties of matched samples of the different subsets of the radio-AGN population are required to understand the physical processes driving the observed differences, and whether the cold gas does indeed play a role in fuelling the LERGs in star-forming galaxies.
 
The upcoming WEAVE spectrograph in the near future will obtain spectra for all of the radio sources detected in these LOFAR Deep Fields as part of the WEAVE-LOFAR survey \citep{2016sf2a.conf..271S}; this will not only enable confirmation of the photometric redshifts, but also more robust host galaxy properties and source classification using emission-line diagnostics out to at least $z \sim 1$. Combined with the large samples of AGN detected in these deep fields, this will allow for detailed characterisation of the radio-AGN population as a function of multiple parameters simultaneously. Future data releases of the LoTSS Deep Fields will present deeper and wider radio and multi-wavelength data, providing yet further constraints on the faint-end of the AGN luminosity functions.

\vspace{-0.7cm}
\section*{Acknowledgements}
We thank the anonymous referee for their helpful comments that have improved the paper. This paper is based (in part) on data obtained with the International LOFAR Telescope (ILT) under project codes LC0\_015, LC2\_024, LC2\_038, LC3\_008, LC4\_008, LC4\_034 and LT10\_01. LOFAR \citep{2013A&A...556A...2V} is the Low Frequency Array designed and constructed by ASTRON. It has observing, data processing, and data storage facilities in several countries, which are owned by various parties (each with their own funding sources), and which are collectively operated by the ILT foundation under a joint scientific policy. The ILT resources have benefitted from the following recent major funding sources: CNRS-INSU, Observatoire de Paris and Université d'Orléans, France; BMBF, MIWF-NRW, MPG, Germany; Science Foundation Ireland (SFI), Department of Business, Enterprise and Innovation (DBEI), Ireland; NWO, The Netherlands; The Science and Technology Facilities Council, UK; Ministry of Science and Higher Education, Poland. For the purpose of open access, the author has applied a Creative Commons Attribution (CC BY) licence to any Author Accepted Manuscript version arising from this submission.

RK acknowledges support from the Science and Technology Facilities Council (STFC) through an STFC studentship via grant ST/R504737/1. PNB and JS are grateful for support from the UK STFC via grants ST/R000972/1 and ST/V000594/1. We thank James Aird and Seb Oliver for useful discussions on the paper. RKC acknowledges support from the Flatiron Institute which is supported by the Simons Foundation. MJH and DJBS acknowledge support from STFC grant ST/V000624/1. MB acknowledges support from INAF under PRIN SKA/CTA FORECaST and from the Ministero degli Affari Esteri della Cooperazione Internazionale - Direzione Generale per la Promozione del Sistema Paese Progetto di Grande Rilevanza ZA18GR02. KJD acknowledges funding from the European Union’s Horizon 2020 research and innovation programme under the Marie Sk\l{}odowska-Curie grant agreement No. 892117 (HIZRAD). KM has been supported by the National Science Centre (UMO-2018/30/E/ST9/00082). CLH acknowledges support from the Leverhulme Trust through an Early Career Research Fellowship. BM acknowledges support from the UK STFC under grants ST/R00109X/1, ST/R000794/1, and ST/T000295/1. IP acknowledges support from INAF under the SKA/CTA PRIN “FORECaST” and the PRIN MAIN STREAM “SAuROS” projects. HJAR acknowledges support from the ERC Advanced Investigator programme NewClusters 321271. WLW acknowledges support from the CAS-NWO programme for radio astronomy with project number 629.001.024, which is financed by the Netherlands Organisation for Scientific Research (NWO). This research made use of {\sc Astropy}, a community-developed core Python package for astronomy \citep{astropy:2013, astropy:2018} hosted at \url{http://www.astropy.org/}, and of {\sc Matplotlib} \citep{hunter2007matplotlib}.

\section*{Data Availability}
The dataset used in this paper is derived from the LoTSS Deep Fields Data Release 1 \citep{Tasse2021,Sabater2021,Kondapally2021,Duncan2021}. The images and catalogues are publicly available from \url{https://lofar-surveys.org/deepfields.html}. The luminosity function data presented are available in a machine-readable format at the following repository: \url{https://github.com/rohitk-10/AGN_LF_Kondapally22}. Other results presented in the paper are available upon reasonable request to the corresponding author.



\bibliographystyle{mnras}
\bibliography{lofar_deepfields_agn} 



\appendix
\section{Comparison with Williams et al. (2018)}\label{sec:ap_wendy_lerg}
\begin{figure}
    \centering
    \includegraphics[width=0.8\columnwidth]{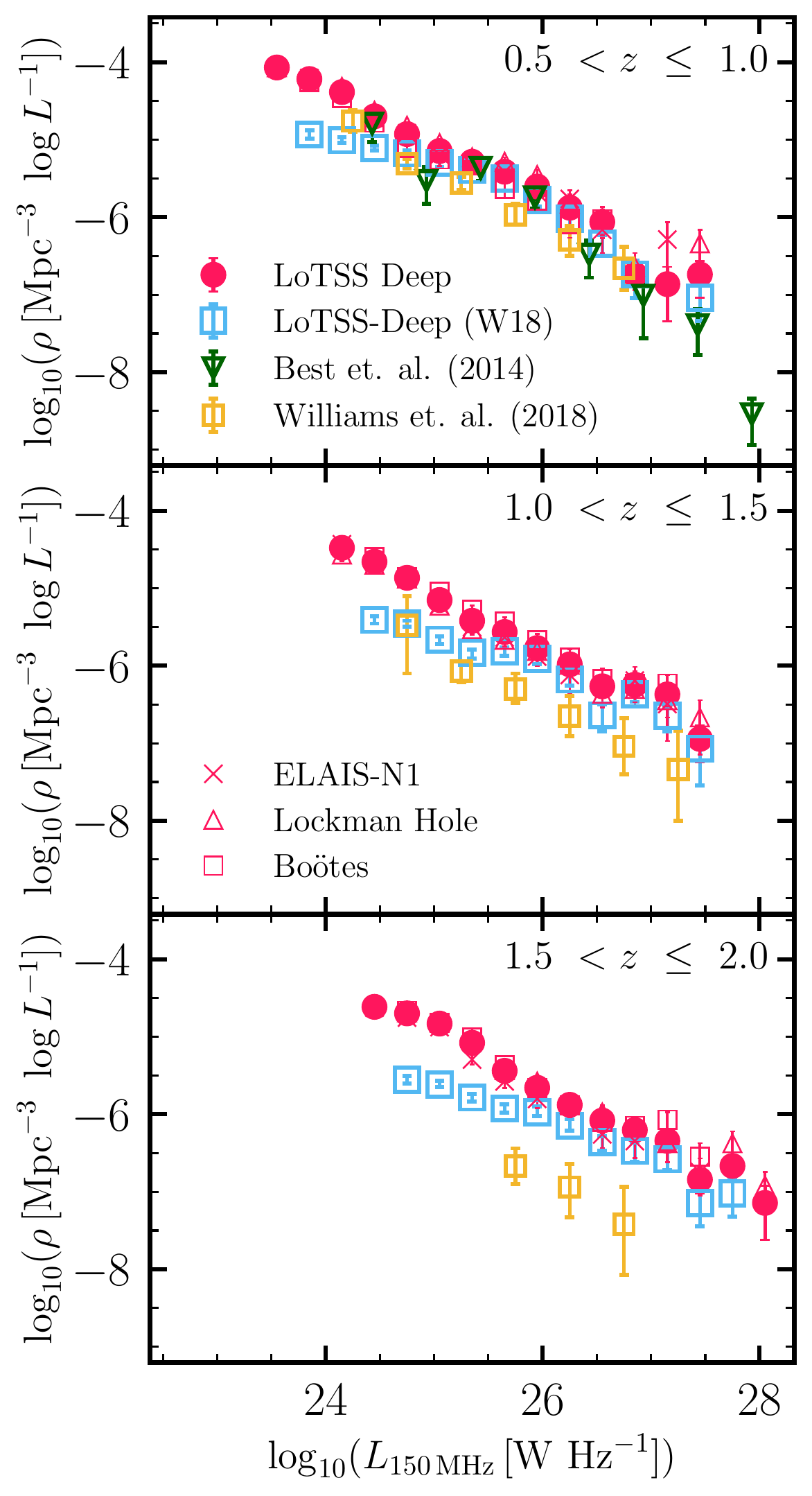}
    \caption{\label{fig:lf_w18_comp}Comparison of the cosmic evolution of LERG LF with \citet{Williams2018} using different source classification schemes. We show the LoTSS-Deep LERG LF resulting from the Best et al. (in prep.) classification scheme (pink circles), and the resultant LoTSS-Deep LF when the source classification method of \citet{Williams2018} is applied to our sample (cyan squares; see also Sec.~\ref{sec:cosevol_lerg_lf}). The literature results from \citet{Williams2018} (yellow squares) and \citet{2014MNRAS.445..955B} (green triangles) are also shown.}
\end{figure}
In Sect.~\ref{sec:cosevol_lerg_lf}, we compared the cosmic evolution of LERGs in LoTSS-Deep with other literature results and found that the \citeauthor{Williams2018} LFs are systematically offset to lower space densities; here, we investigate the potential causes for this apparent discrepancy.

As detailed in Sect.~\ref{sec:cosevol_lerg_lf}, \citeauthor{Williams2018} constructed their LFs using a sample of 243 LERGs which were selected using a radio-excess selection based on the FIRC and SED fitting from \textsc{AGNFitter}. In principle, this selection method is similar to the one used by Best et al. (in prep.), however, in practice, the \citet{Williams2018} radio-excess criterion is more conservative, increasingly so with increasing redshift, and their criteria for separating HERGs and LERGs is also different (see Sect.~\ref{sec:agn_sel}). Therefore we reconstructed the LERG LFs with the LoTSS-Deep data using the classification scheme adopted by \citet{Williams2018}. Specifically, we identified radio-excess sources making use of the FIRC of \citet{CalistroRivera2017} using properties derived from \textsc{AGNFitter} only instead of the \textit{consensus} values, and we classified these sources as HERGs or LERGs using the \citet{Williams2018} $f_{\rm{AGN}}$ parameter.

Fig.~\ref{fig:lf_w18_comp} shows comparison with \citet{Williams2018} when applying their source classification scheme to our LoTSS Deep dataset (cyan squares; hereafter LoTSS-Deep W18). We find that the derived LFs now show better agreement at all redshift with the LFs derived in this work (within $2\sigma$) based on Best et al. (in prep.) classifications than with the \citet{Williams2018} results, especially at high radio luminosities and higher redshifts. This appears to be largely because we used an updated version of \textsc{AGNFitter} compared to the one used in \citeauthor{Williams2018} with improved models being used in the fitting process resulting in better fits in general. This has a significant effect on the classification of sources as HERGs or LERGs.

The reconstructed LoTSS-Deep W18 LFs do not agree well with ours at faint radio luminosities, at all redshifts. We find that this is due to a significant number of sources classified as radio-excess AGN (and subsequently as LERGs) by our criteria, but as SFGs under the \citet{Williams2018} criteria. One reason for this discrepancy is that the two studies use different estimates of SFRs: \textit{consensus} SFRs versus \textsc{AGNFitter} based IR luminosities (and hence SFRs) used by \citet{Williams2018}. Best et al. (in prep.) show that the SFRs reported by \textsc{AGNFitter} are found to be systematically offset to higher values than those reported by the other three SED fitting routines used. This may be caused by a combination of a lack of energy balance in \textsc{AGNFitter} unlike other routines, and differences between the various SED fitting routines used in this study. Another reason for the discrepancy is that the radio-excess selection used by \citet{Williams2018} is more conservative as their redshift-dependent FIRC correlation has a $\sigma = 0.527$, and so their $2\sigma$ cut corresponds to a larger radio-excess than that of Best et al. (in prep.). Furthermore, the use of an evolving FIRC by \citet{Williams2018} means that their selection becomes relatively more conservative with increasing redshift. The combination of these two factors can help explain the observed differences with \citet{Williams2018}; a detailed SED fitting comparison to investigate the cause of this discrepancy is outside the scope of this paper.

\vspace{-0.6cm}
\section{Effects of varying the source classification criteria}\label{sec:scvary}
\begin{figure*}
    \centering
    \includegraphics[width=0.8\textwidth]{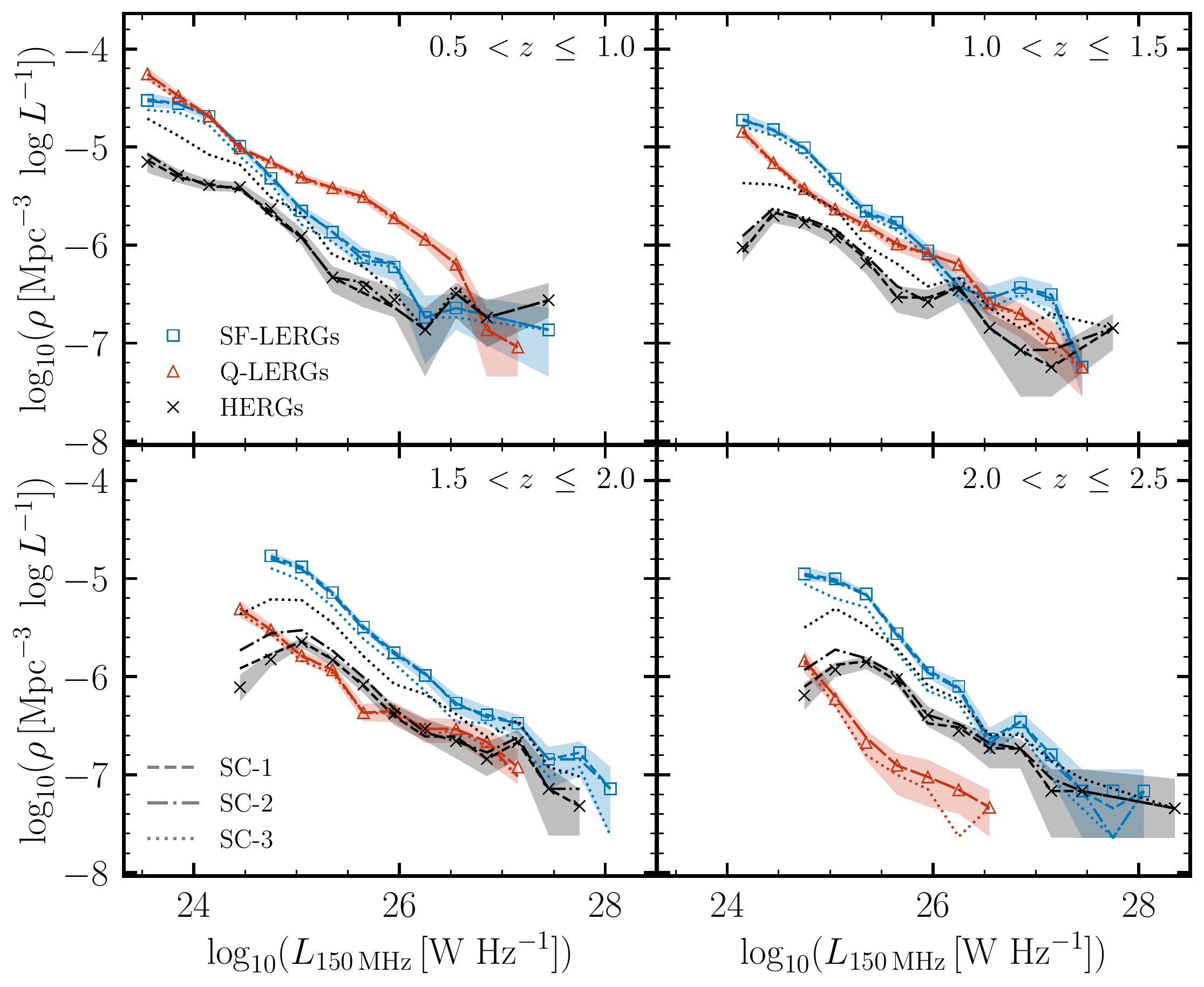}
    \caption{\label{fig:lf_lerg_herg_quies_scvary}Evolution of the luminosity functions of star-forming and quiescent LERGs, and the HERGs in the LoTSS-Deep fields using different source classification criteria. The separation of quiescent and star-forming LERGs is performed based on the sSFR criteria defined in Sect.~\ref{sec:agn_frac_quies}. The markers and associated shaded regions correspond to the LFs and $1\sigma$ uncertainties resulting from the adopted source classification criteria from Best et al. (in prep.), whereas the lines show the LFs obtained under three variations of the source classification criteria (see text). In general, we find that the adopted quiescent and star-forming LERGs, which are the focus of this paper, show little change in space densities with changes in source classification selection, suggesting that the results in the main paper are robust against uncertainties in the source classification scheme adopted.}
\end{figure*}

In this section we explore how deviations from our adopted criteria for source classification affect our main results on the evolution of the LERG LFs presented in Sects.~\ref{sec:cosevol_lerg_lf}~--~\ref{sec:lerg_lfs_Q_SF}.

As discussed in Sects.~\ref{sec:local_lfs}~--~\ref{sec:radio_agn_evol}, the local and evolving LFs for the total radio-AGN population derived in LoTSS-Deep agree well with literature results (as do those of the SFGs by Cochrane et al. in prep.), giving us confidence in our separation of star-forming galaxies from radio-excess AGN across redshift. Furthermore, we have investigated increasing and decreasing the radio-excess criteria by $0.1\,\rm{dex}$ to find that both the star-forming and quiescent LERG LFs remain largely unchanged, with typical changes in space densities of $< 0.15\,\rm{dex}$. In this section, we only investigate the modification of the selection of the so-called `SED AGN', which will result in a change between sources classified as HERGs and LERGs. We note that it can be challenging to identify (optical) AGN of moderate to low luminosities using photometric data; this may be particularly important in star-forming galaxies as both AGN activity and star-formation contribute to the MIR emission, making it difficult to disentangle contributions associated with the different emission processes. If such sources also show radio-loud emission and the SED AGN component is missed, then based on our source classification criteria, these would be misclassified as LERGs (and likely as star-forming LERGs) instead of HERGs. Although this issue is inevitable in the absence of emission line diagnostics that allow a more robust selection of optical AGN, we consider the impact of this potential contamination on our results.

Fig.~\ref{fig:lf_lerg_herg_quies_scvary} shows the cosmic evolution of the quiescent LERGs, star-forming LERGs and HERGs between $0.5 < z \leq 2.5$ based on our adopted source classification criteria from Best et al. (in prep.), plotted with triangle (red), square (blue), and cross (black) symbols, respectively. The shaded regions correspond to the $1\sigma$ uncertainties on the LFs. Also shown in Fig.~\ref{fig:lf_lerg_herg_quies_scvary} are the LFs for these three classes of AGN obtained using three variations of the `SED AGN' selection criteria from Best et al. (in prep.): SC-1, SC-2, and SC-3. As outlined in Sect.~\ref{sec:agn_sel}, based on the output from each of \textsc{AGNFitter} and \textsc{cigale}, Best et al. (in prep.) defined a value $f_{\rm{AGN}}$ for each code, as the fraction of the mid-infrared luminosity associated with AGN emission compared to the host galaxy. This diagnostic (in particular the 16\textsuperscript{th} percentile of $f_{\rm{AGN}}$) forms a key part in the selection of `SED AGN', and hence the separation of `radio-excess AGN' into LERGs and HERGs. We arbitrarily reduced the thresholds for these parameters from both \textsc{AGNFitter} and \textsc{cigale} by a factor of two (i.e. leading to an increase in the number of HERGs) and reconstructed the LFs, as shown by the SC-1 (dotted) lines.

\begin{figure}
    \centering
    \includegraphics[width=0.9\columnwidth]{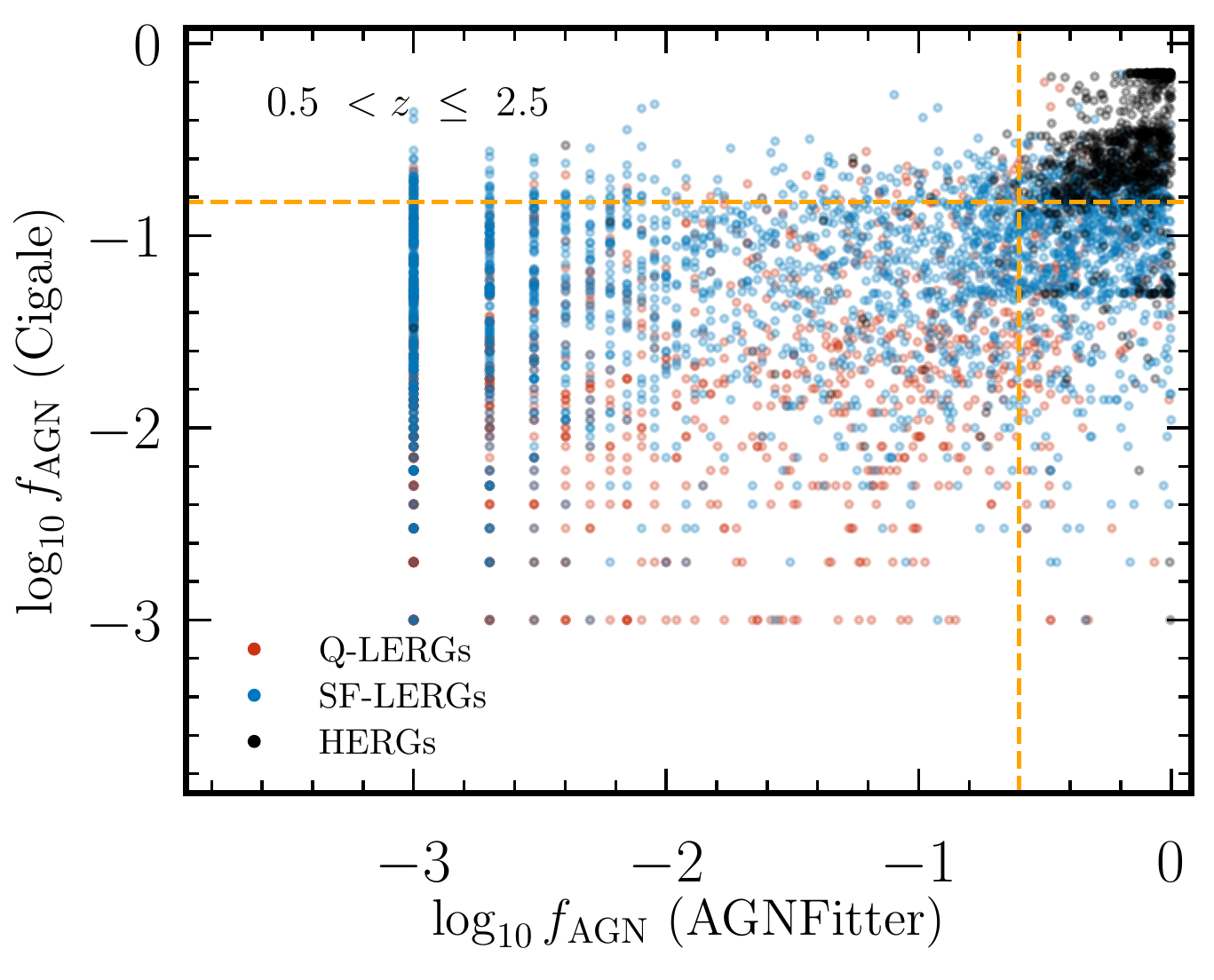}
    \caption{\label{fig:agn_frac_af_cg_scatter}2D scatter plot showing the location of the quiescent LERGs, star-forming LERGs, and HERGs (in ELAIS-N1), in AGN fraction (\textsc{cigale}) - AGN fraction (\textsc{AGNFitter}) parameter space over a broad redshift bin. The dashed lines along the x and y-axes show the $f_{\rm{AGN}} \rm{(\textsc{AGNFitter})} = 0.25$ or $f_{\rm{AGN}} \rm{(\textsc{cigale})} = 0.15$ thresholds used to select additional `SED AGN' for testing the source classification criteria.}
\end{figure}

We also implemented additional criteria to that adopted by Best et al. (in prep.) in the selection of `SED AGN', using the best-fitting $f_{\rm{AGN}}$ values from \textsc{AGNFitter} and \textsc{cigale} in addition to the 16\textsuperscript{th} percentile values. We used $f_{\rm{AGN}}$ as we wanted to select objects with high $f_{\rm{AGN}}$ values, but with large uncertainties (e.g. due to low SNR), that may be missed by the Best et al. (in prep.) criteria; this allowed us to test a more complete but possibly contaminated sample, compared to the clean but possibly incomplete `SED AGN' sample of Best et al. (in prep.). In order to find an appropriate threshold in $f_{\rm{AGN}}$, we considered first the distribution of sources in the $f_{\rm{AGN}} \rm{(\textsc{AGNFitter})}$ versus $f_{\rm{AGN}} \rm{(\textsc{Cigale})}$ 2D parameter space. Fig.~\ref{fig:agn_frac_af_cg_scatter} shows a scatter plot in this 2D space, with the red, blue, and black points corresponding to the quiescent LERGs, star-forming LERGs, and HERGs, respectively, as classified based on the Best et al. (in prep.) criteria. It is reassuring to see that although the best-fitting $f_{\rm{AGN}}$ parameters were not used in the selection of the `SED AGN', the HERGs occupy a well-defined region of this 2D plane, corresponding to high $f_{\rm{AGN}}$ values from both the SED fitting codes, albeit with some overlap with the star-forming LERG population. Moreover, the quiescent and star-forming LERGs also show very similar distributions in best-fitting $f_{\rm{AGN}}$ values (from both \textsc{AGNFitter} and \textsc{cigale}) -- and distinct from the HERGs -- suggesting that these are both drawn from the same parent population and that the star-forming LERG population is not significantly affected by contamination from HERGs. Based on the location of the HERGs, and that of X-ray detected AGN (indicative of radiative-mode AGN) from the X-Bo\"{o}tes survey in the 2D scatter plot in Fig.~\ref{fig:agn_frac_af_cg_scatter}, we defined an extra criteria of `SED AGN' as sources with $f_{\rm{AGN}} \rm{(\textsc{Cigale})} > 0.15$ and $f_{\rm{AGN}} \rm{(\textsc{AGNFitter})} > 0.25$. We note that this selection criterion is applied in addition to the Best et al. criteria, such that there will always be an increase in the number of sources classified as HERGs; in the case adopted here, this leads to 258 more sources being classified as HERGs. The LFs resulting from this criteria (SC-2) are shown as dashed lines in Fig.~\ref{fig:lf_lerg_herg_quies_scvary}.

Finally, we considered adopting a more extreme approach such that sources with either $f_{\rm{AGN}} \rm{(\textsc{Cigale})} > 0.15$ or $f_{\rm{AGN}} \rm{(\textsc{AGNFitter})} > 0.25$ are classed as `SED AGN'. As seen from Fig.~\ref{fig:agn_frac_af_cg_scatter}, while this selection may result in a higher completeness of `SED AGN', it will suffer from even more contamination; this is particularly the case at higher redshifts. This is due to a significant population of sources (mostly star-forming LERGs based on the Best et al. classification) with $f_{\rm{AGN}} \rm{(\textsc{AGNFitter})} > 0.25$ but significantly lower (up to a factor of 10 lower) values of $f_{\rm{AGN}} \rm{(\textsc{cigale})}$, which is a result of the significantly higher uncertainties on the \textsc{AGNFitter} values. Nevertheless, the SC-3 criteria allows us to examine the extent to which even an extreme classification could potentially affect the LERG LFs (dash-dotted lines in Fig.~\ref{fig:lf_lerg_herg_quies_scvary}).

As evident from Fig.~\ref{fig:lf_lerg_herg_quies_scvary}, given both the small number of HERGs in the LoTSS Deep Fields (see also Fig.~\ref{fig:pz_hist}) and the small numbers of sources with a change in their classification based on SC-1 and SC-2 criteria, there is only a very small effect on the LERG LFs across all redshifts (though slightly more significant for the HERG LFs). The SC-1 and SC-2 lines agree with the originally determined Best et al. HERG LFs within $1-2\sigma$ of our best estimate LFs at all but the faintest few luminosity bins; at the faint end, the change in the LFs is comparable to the field-to-field variation found in the LFs. For the quiescent and star-forming LERGs, both SC-1 and SC-2 variations result in very similar LFs across redshift. The quiescent LERG population in particular remains robust against changes to the source classification method adopted within the uncertainties across all of the redshift bins; this also holds even when they become the sub-dominant population of all LERGs at $z > 1.5$. The star-forming LERG LFs at $z \leq 1$ based on SC-1 and SC-2 are lower by $\sim 0.1-0.2\,$dex, however this has little to no effect on the total LERG LFs given that the quiescent LERGs dominate the space densities over most of the radio luminosities studied here; at higher redshifts, the differences with the SC-1 and SC-2 based LFs are consistent within $1-2\sigma$ of the uncertainties. 

For SC-3, the space density of the HERGs (2734 sources) now increases substantially across all luminosities, and especially at the faint end at $z > 1.5$, as might be expected. For the LERG population however, which is the focus of this paper, the reconstructed LFs at $z \leq 1.5$ show good agreement with our best estimate of the LERG space densities; at higher redshifts, the space densities decrease, in particular at faint luminosities, reaching $\sim 0.2\,$dex lower in the faintest bins. 

We therefore conclude that the LERG LFs presented in the main sections of the paper based on the Best et al. (in prep.) source classification -- and subsequent results on the LERG stellar mass fractions -- are robust to uncertainties in the source classification criteria used.


\bsp	
\label{lastpage}
\end{document}